\documentclass[aps,prd,twocolumn,amssymb,eqsecnum,showpacs,showkeyes,secnumarabic,graphics,floatfix,nofootinbib,tightenlines,longbibliography]{revtex4-1} 
 
\usepackage{graphicx}
\usepackage{epsfig}
\usepackage{bm}
\usepackage{cancel}
\usepackage{dcolumn}
\usepackage{amsmath}
\usepackage{hhline}
\usepackage{enumerate}
\usepackage[utf8]{inputenc}
\usepackage{cancel}
\usepackage[dvipsnames]{xcolor}
\usepackage{float}
\usepackage{longtable}
\usepackage[breaklinks=true,colorlinks=true,
linkcolor=blue,urlcolor=blue,citecolor=MidnightBlue,% PDF VIEW
bookmarks=true,bookmarksopenlevel=2]{hyperref}
\usepackage{mathrsfs}

\def \beq  {\begin{equation}}
\def \eeq  {\end{equation}}
\def \ber  {\begin{eqnarray}}
\def \eer  {\end{eqnarray}}

\begin{document}
\newcommand{\newc}{\newcommand}
\newc{\be}{\begin{equation}}
\newc{\ee}{\end{equation}}
\newc{\ba}{\begin{eqnarray}}
\newc{\ea}{\end{eqnarray}}
\newc{\bea}{\begin{eqnarray*}}
\newc{\eea}{\end{eqnarray*}}
\newc{\D}{\partial}
\newc{\ie}{{\it i.e.} }
\newc{\eg}{{\it e.g.} }
\newc{\etc}{{\it etc.} }
\newc{\etal}{{\it et al.}}
\newc{\lcdm}{$\Lambda$CDM }
\newc{\wcdm}{wCDM }
\newc{\lcdmnospace}{$\Lambda$CDM}
\newc{\plcdm}{Planck15/$\Lambda$CDM }
\newc{\plcdmnospace}{Planck15/$\Lambda$CDM}
\newc{\geffz}{$G_{\rm eff}(z)$ }
\newc{\geffznospace}{$G_{\rm eff}(z)$}
\newcommand{\nn}{\nonumber}
\newc{\ra}{\Rightarrow}
\newc{\la}{\label}
\newc{\fs}{{\rm{\it f\sigma}}_8}
\newc{\omm}{$\Omega_m$ }
\newc{\wlcdm}{WMAP7/$\Lambda$CDM }
\newc{\wlcdmnospace}{WMAP7/$\Lambda$CDM}
\newc{\ommnospace}{$\Omega_m$}
\newcommand{\gsim}{\raisebox{-0.13cm}{~\shortstack{$>$ \\[-0.07cm]
      $\sim$}}~}
\newcommand{\lsim}{\lesssim}

\title{Hints of Modified Gravity in Cosmos and in the Lab?}

\author{Leandros Perivolaropoulos}\email{leandros@uoi.gr}
\author{Lavrentios Kazantzidis}\email{lkazantzi@cc.uoi.gr}
\affiliation{Department of Physics, University of Ioannina, GR-45110, Ioannina, Greece}

\date{\today}

\begin{abstract}
General Relativity (GR) is consistent with a wide range of experiments/observations from millimeter scales up to galactic scales and beyond. However, there are reasons to believe that GR may need to be modified  because it includes singularities (it is an incomplete theory) and also it requires fine-tuning to explain  the accelerating expansion of the universe through the cosmological constant. Therefore, it is important to check various experiments and observations beyond the above
 range of scales for possible hints of deviations from the predictions of GR. If  such hints  are found it is important to understand which classes of modified gravity theories are consistent with them. The goal of this review is to summarize recent progress on these issues.   On sub millimeter scales we show an analysis of the data of the Washington experiment \cite{Kapner:2006si-washington3} searching for modifications of Newton's Law on sub-millimeter scales and demonstrate that a spatially oscillating signal is hidden in this dataset. We demonstrate that even though this signal cannot be explained in the context of standard modified theories (viable scalar tensor and $f(R)$ theories), it is a rather generic prediction of nonlocal gravity theories. On cosmological scales we review recent analyses of Redshift Space Distortion (RSD) data which measure the growth rate of cosmological perturbations at various redshifts and show that these data are in some tension with the \lcdm parameter values indicated by Planck/2015 CMB data at about 3$\sigma$  level. This tension can be reduced by allowing for an evolution of the effective Newton constant that determines the growth rate of cosmological perturbations. We conclude that even though this tension between the data and the predictions of GR could be due to systematic/statistical uncertainties of the data, it could also constitute early hints pointing towards a new gravitational theory. 
\end{abstract}
\maketitle

%\keywords{Modified gravity; growth cosmological data; newton's constant; small scale gravity experiments.}

%\ccode{PACS numbers:}

%\tableofcontents

\section{Introduction}
\label{sec:Introduction}

General Relativity (GR) has been tested in a wide range of scales starting from sub-mm scales out to supercluster $O(100Mpc)$ scales. Even though no statistically significant evidence has been found so far indicating deviations from GR, there are theoretical arguments and experimental/observational hints that indicate that GR may need to be modified on both the smallest and the largest probed scales.

From the theoretical point of view, it is clear that GR has to face the following challenges:
\begin{itemize}
\item
It predicts the existence of unphysical singularities which indicate that it is a physically incomplete theory.
\item
It is nonrenormalizable and inconsistent with Quantum Field Theory (QFT) at high energies due to the prediction of black hole formation when small scales are probed via scattering experiments.
\item
It can not explain the observed accelerating expansion of the universe unless extreme fine tuning is assumed.
\end{itemize}

From the experimental/observational point of view GR has been well tested on solar system scales where the PPN parameters measuring deviations from GR have been shown to reduce to the values predicted by GR at an accuracy level of about $10^{-5}$ \cite{Will:2014kxa}. On larger and smaller scales however the constraints on deviations from GR are not as strong. In fact there have been claims for hints of deviations from GR predictions even on solar system scales (e.g. Pioneer anomaly \cite{Anderson:2001ks}) and on galactic scales (e.g. the formation of black holes at discrete values of mass \cite{Sokolov}). 

On galactic scales, the deviation of star velocities from the velocities expected in the presence of visible matter in the context of GR indicates that the Einstein equation $G_{\mu \nu}=T_{\mu\nu}^{lum}$ (where $T_{\mu\nu}^{lum}$ is the energy momentum tensor of luminous matter) is violated. The usual approach is restoring consistency between the two sides of the Einstein equation has been to modify the right side of the Einstein equation and write it in the form $G_{\mu \nu}=T_{\mu\nu}^{lum}+T_{\mu\nu}^{dm}$ where $T_{\mu\nu}^{dm}$ is the energy momentum tensor of matter that interacts only gravitationally (dark matter \cite{Jungman:1995df}).  An alternative approach is to modify the left side of the Einstein equation as $G_{\mu \nu}+G_{\mu \nu}^{TeVeS}=T_{\mu\nu}^{lum}$ leading to a modified version of GR: Tensor Vector Scalar theory (TeVeS) \cite{Bekenstein:2004ne}. Clearly, a combination of the above solutions is also possible leading to $G_{\mu \nu}+G_{\mu \nu}^{TeVeS}=T_{\mu\nu}^{lum}+T_{\mu\nu}^{dm}$. The recent detection of gravitational waves coming from the collision of two neutron stars (GW170817) \cite{TheLIGOScientific:2017qsa}, however, seems to exclude \cite{Boran:2017rdn} all types of dark matter emulator theories such as TeVeS theory. An exception to this exclusion may be \cite{Green:2017qcv} the alternative Scalar-Tensor-Vector Gravity theory (STVG) \cite{Moffat:2005si}, which seems to remain viable after the GW170817 event since the photon and graviton geodesics are identical in this theory. Another modified dark matter emulator approach similar to the Modified Newtonian Dynamics (MOND) approach \cite{Milgrom:1983ca,Milgrom:1983pn,Milgrom:1983zz} is based on the assumption that the gravitational constant varies with acceleration \cite{Christodoulou:2018xxw,Christodoulou:2018xzg,Christodoulou:2019ixr} or equivalently with the dimensionless surface density of a spherical mass distribution $s$ \cite{Christodoulou:2019atk}.

The frontiers of current gravitational research lie on the two extreme scales that gravitational experiments/observations can currently probe: sub-mm scales where a wide range of experiments \cite{Murata:2014nra-good-review-ofexperiments} search for new types of forces and cosmological scales of a few $Mpc$ or larger where observations of the growth rate of cosmological perturbations through Redsift Space Distortions \cite{Nesseris:2017vor,Macaulay:2013swa,Tsujikawa:2015mga,Johnson:2015aaa,Basilakos:2017rgc,Kazantzidis:2018rnb} or Weak Lensing \cite{Joudaki:2017zdt,Hildebrandt:2016iqg,Troxel:2017xyo,Kohlinger:2017sxk} can probe the gravitational laws and the consistency of GR with data. Current research on these frontier scales is the focus of the present review.

\begin{figure}[!h]
\centering
\includegraphics[width = 0.42\textwidth]{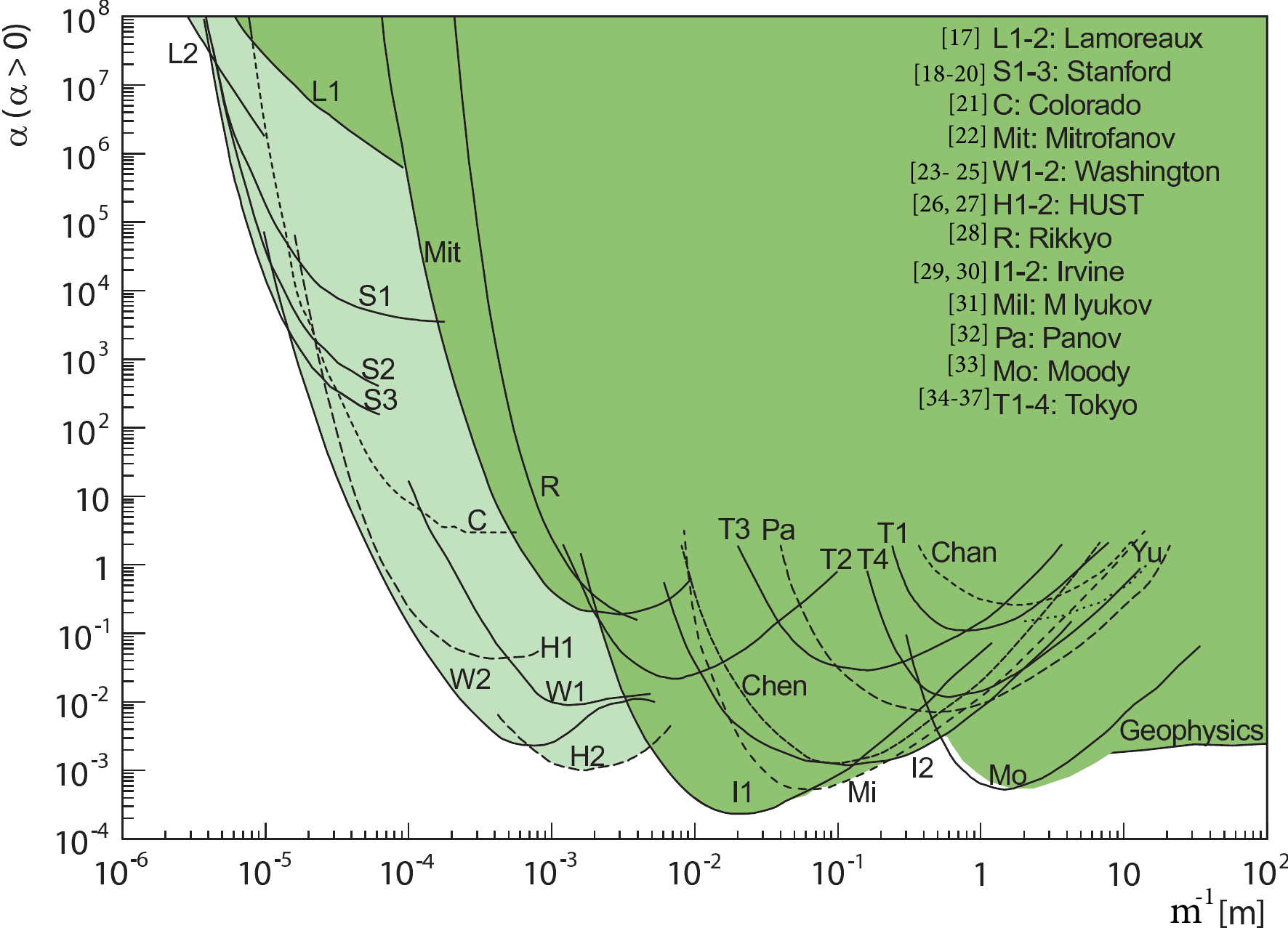}
\caption{A review of current constraints based on the Yukawa parametrization \eqref{yukawaanz} for deviation from Newton's law. From Ref. \cite{Murata:2014nra-good-review-ofexperiments}.}
\label{fig:experiments}
\end{figure}

Small scale gravity experiments \cite{Lamoreaux:1996wh,Chiaverini:2002cb,Smullin:2005iv,Geraci:2008hb,Long:2002wn,Mitrofanov1,Hoyle:2004cw,Hoyle:2000cv-washington1,Kapner:2006si,Tu:2007zz,Yang:2012zzb,RikkyoMurata1,Hoskins:1985tn,Spero:1980zz,Milyukov1,Panov1,Moody:1993ir,Hirakawa1,Ogawa1,Kuroda1,Mio1} probe sub-mm scales searching for new forces on these scales. The forces between test masses are measured at various distances and compared with the expected forces on the basis of known physics. Deviations from the null result corresponding to Newtonian gravitational interaction are fit to specific parametrizations that are well motivated based on theoretical arguments.
 The most commonly used parametrization for fitting the above deviations of gravitational experiment data is the Yukawa parametrization, where the effective gravitational interaction potential is expressed as 
\be 
V_{\rm eff}= -G \frac{M}{r}(1+\alpha e^{- m r})
\label{yukawaanz}
\ee
corresponding to a spatially varying effective Newton's constant of the form
\be 
G_{\rm eff}(r)= G (1+\alpha e^{-mr})
\label{geffyukawaanz}
\ee
Eq. \eqref{geffyukawaanz} depends on the parameters $\alpha$ and $m$, which denote the amplitude and the range of Yukawa force. Fig. \ref{fig:experiments} shows current constraints from small scale gravity experiments. For $\alpha \approx 1$, the range of this Yukawa exponential is constrained to be less than about $0.1mm$ \cite{Murata:2014nra-good-review-ofexperiments} (see also Ref. \cite{Brax:2019iut} which constrains $m$ to be in the range $4 \mu m\lesssim m^{-1} \lesssim 68 \mu m$ by using results from the Washington experiment on the modification of the inverse-square law, the observations of the hot gas of galaxy clusters and the Planck satellite data on the neutrino masses.

The Yukawa parametrization is the most commonly used parametrization for testing for deviations from Newton's law on sub-mm scales. It is generic and well motivated theoretically as it is a natural prediction in the context of a wide range of modified gravity theories including Brans-Dicke \cite{Perivolaropoulos:2009ak-massive-bd-ppn,Hohmann:2013rba,Jarv:2014hma}, scalar-tensor \cite{EspositoFarese:2000ij,Gannouji:2006jm,Faraoni:2004pi,Chiba:2003ir} and $f(R)$ theories \cite{Berry:2011pb-weak-field-fr-incl-osc,Capozziello:2009vr,Schellstede:2016ldu}. It is also a natural prediction of theories involving compactified extra dimensions such as Kaluza-Klein theories \cite{Perivolaropoulos:2003we}.

For example, consider a generic form of $f(R)$ theories with an $R^2$ correction of the form \cite{Chiba:2003ir}
\be 
f(R)=R+\frac{1}{6 m^2} R^2 
\label{frform}
\ee 
The generalized  Einstein-Hilbert action is of the form
\be 
S_R=\frac{1}{16 \pi G}\int d^4 x \sqrt{-g} f(R)+S_{matter}
\label{fraction}
\ee
 Varying action \eqref{fraction} with respect to the metric leads to the dynamical equations
\be  f'(R)R_{\mu\nu} - \frac{1}{2}g_{\mu\nu}f(R) = 8\pi G T_{\mu\nu} + \nabla_{\mu}\nabla_{\nu}f'(R) - g_{\mu\nu}\Box f'(R)
\label{frdyneqs}
\ee
Assuming that $f(R)$ has the form of Eq. \eqref{frform}, in the weak field limit  $g_{\mu\nu}=\eta_{\mu\nu}+h_{\mu\nu}$, it is straightforward to show that a solution for the metric perturbation $h_{00}$ in the presence of a point mass $M$ takes the form \cite{Perivolaropoulos:2016ucs}
\be 
h_{00} = \frac{2G M}{r} \left(1 + \frac{1}{3}e^{-m r} \right)
\label{h00sol1}
\ee
which compared to the usual Newtonian case has the correction factor $\frac{1}{3}e^{-m r}$. A comparison with the Yukawa ansatz \eqref{geffyukawaanz}  implies that $\alpha=\frac{1}{3}$. A similar form of modified Newtonian force is obtained for massive Brans-Dicke (BD) theories \cite{Perivolaropoulos:2016ucs} and for Kaluza-Klein theories \cite{Perivolaropoulos:2002pn,Perivolaropoulos:2003we}. In those cases the phenomenological parameters $\alpha$, $m$ depend on the fundamental parameters of the theories (e.g. mass of scalar field, Brans-Dicke parameter $\omega$, size and number of extra dimensions). 

The above Yukawa parametrization is well motivated theoretically and is currently the standard parametrization used to fit experimental residuals of the Newtonian force. However, alternative parametrizations may also be theoretically motivated in the context of other theoretical models and they may in fact provide better fits to experimental residuals with respect to the Newtonian force on sub-mm scales. For example some brane theories favor a power law residual parametrization \cite{Donini:2016kgu,Benichou:2011dx,Bronnikov:2006jy,Nojiri:2002wn-newtonptl-brane}. A purely phenomenological approach could also consider arbitrary parametrizations (e.g. spatially oscillating parametrizations) of residual forces designed so that they provide the best fit to residual force data.

Stability of the theories that lead to a Yukawa type of modified Newton's law usually implies that $m^2>0$ \cite{Perivolaropoulos:2016ucs}. The case $m^2<0$ is usually associated with instabilities \cite{Faraoni:2006sy,Dolgov:2003px} of the underlying theories and also with an oscillating behaviour of the additional term modifying the Newtonian gravitational force. Despite the fact that in these cases we may have no Newtonian limit, such a spatially oscillating term can escape detection if its spatial wavelength is smaller than a fraction of a $mm$ \cite{Perivolaropoulos:2016ucs}. This case will be discussed in Section \ref{sec:submmscales} along with an example of a healthy theory (nonlocal gravity \cite{Edholm:2016hbt-nonlocal-potential-stable-spatial-oscillations,Kehagias:2014sda-nonlocal-oscillations,Frolov:2015usa-newton-potential-nonlocal}) that predicts such spatial oscillations without the presence of ghosts/instabilities \cite{Tomboulis:1997gg,Siegel:2003vt,Biswas:2013cha}.

On the other frontier of testing GR, cosmological scales, the properties of the gravitational theory can be probed by measuring the growth rate of cosmological perturbations through  the measurement of peculiar velocities of galaxies (obtained using  Redshift Space Distortion (RSD) data \cite{Macaulay:2013swa,Alam:2016hwk}) and through weak gravitational lensing \cite{Joudaki:2017zdt,Amon:2017lia,Abbott:2018jhe}. In the presence of perturbations, the perturbed metric in the Newtonian gauge takes the form 
\be
ds^2= -(1 + 2 \phi) dt^2 + a^2 (1 - 2\psi) d{\vec{x}}\,^2
\ee
where $\phi$ and $\psi$ are potentials with $\phi$ corresponding to the Newtonian potential.  These two potentials in general obey modified Poisson equations of the following form
\ba
\nabla^2 \phi &=& 4 \pi G_{\rm eff} a^2 \rho \, \delta_m,  \label{potphi}\\
\nabla^2 (\phi+\psi) &=& 8 \pi G_{L}  a^2 \rho \, \delta_m, \label{potphipsi}
\ea
where $\delta_m$ in the linear matter overdensity, $\rho$ is the mean matter density and $a$ is the cosmic scale factor.
The potential $\phi$ can be probed using growth of density  perturbations observations through RSD data \cite{Macaulay:2013swa,Alam:2016hwk} and $\phi+\psi$ is usually probed using weak lensing data \cite{Joudaki:2017zdt,Amon:2017lia,Abbott:2018jhe}. In Eq. \eqref{potphi} and Eq. \eqref{potphipsi} we also have the parameters $G_{\rm eff}$ and $G_L$ which in GR are equal and constant
\be 
G_{\rm eff}=G_L=G_N
\ee
while in modified gravity theories they can be spacetime dependent. Therefore a basic question arises. ``How can the actual data constrain possible scale or redshift dependence of these parameters?". Here we focus on the $G_{\rm eff}$ that is associated with the Newtonian potential $\phi$ and can be constrained using RSD data measuring the growth of density perturbations. 

Early  hints of modifications of GR are most likely to come from experiments/observations at the frontier scales: sub-mm and cosmological scales. Important questions that need to be addressed in this context are the following: 
\begin{itemize}
 \item Is GR consistent with currently available data on each scale?
 \item Even if it is consistent what is the optimum parametrization of the effective Newton's constant $G_{\rm eff}$ in providing the best quality of fit to the data? 
\item If there is such  parametrization providing a better fit to the data, then what are the theoretical models that support it?
\end{itemize}
These questions will be the focus of the present brief review.

The structure of this review is the following: In Section \ref{sec:cosmoscales}, we focus on cosmological scales and review the phenomenological predictions of modified gravity theories on the observable growth rate of matter density perturbations which can be used as a probe of gravitational physics on cosmological scales. We also focus on Redshift Space Distortions (RSD) as a probe of the growth of matter density perturbations and use an extended compilation of RSD data to identify the tension level between the \lcdm parameter values favoured by Planck 2015 \cite{Ade:2015xua} and the corresponding parameter values favoured by the RSD growth data. The effect of an evolving with redshift $z$ effective Newton's constant $G_{\rm eff}(z)$ on the level of this tension is reviewed and the qualitative features of the best fit form of $G_{\rm eff}(z)$ are identified. The consistency of these qualitative features with specific modified gravity theories is also discussed. In Section \ref{sec:submmscales} we focus on sub-mm scales and identify the quality of fit of a novel oscillating residual force parametrizations on the data of the Washington small scale gravity experiment. The consistency of this parametrization with specific modified gravity models ($f(R)$ theories and nonlocal gravity) is also discussed. Finally, in Section \ref{sec:concl} we conclude, summarize and discuss interesting extensions of the reviewed research.

\section{Hints of Modified Gravity on Cosmological Scales}
\label{sec:cosmoscales}

\subsection{RSD Data: Analysis and Phenomenological Implications}
A particularly useful probe of the growth rate of density perturbations is weak lensing \cite{Joudaki:2017zdt,Amon:2017lia,Abbott:2018jhe}. Recent \lcdm parameter constraints emerging from a tomographic weak gravitational lensing analyses indicates a 2-3 $\sigma$ tension in the $\sigma_8 - \Omega_m$ parameter space between the parameter values favoured by Planck 2015 \cite{Ade:2015xua} (which can be seen in Table \ref{tab:planck1}) and specific weak lensing survey data \cite{Joudaki:2017zdt,Abbott:2017wau}.

\begin{table}[h]
\caption{The \plcdm  parameters as reported in Ref. \cite{Ade:2015xua}}
\label{tab:planck1}
\begin{centering}
\begin{tabular}{|c|c|}
 \hline
Parameter & \plcdm Values \cite{Ade:2015xua} \\ 
\hline
\rule{0pt}{3ex} 
$\Omega_b h^2$ & $0.02225\pm0.00016$ \\
$\Omega_c h^2$ & $0.1198\pm0.0015$ \\
$n_s$ & $0.9645\pm0.0049$ \\
$H_0$ & $67.27\pm0.66$ \\
$\Omega_{m}$ & $0.3156\pm0.0091$ \\
$w$ & $-1$ \\
$\sigma_8$ & $0.831\pm0.013$ \\
\hline
\end{tabular}
\end{centering}
\end{table}

\begin{figure*}[ht!]
\centering
\includegraphics[width = 0.74\textwidth]{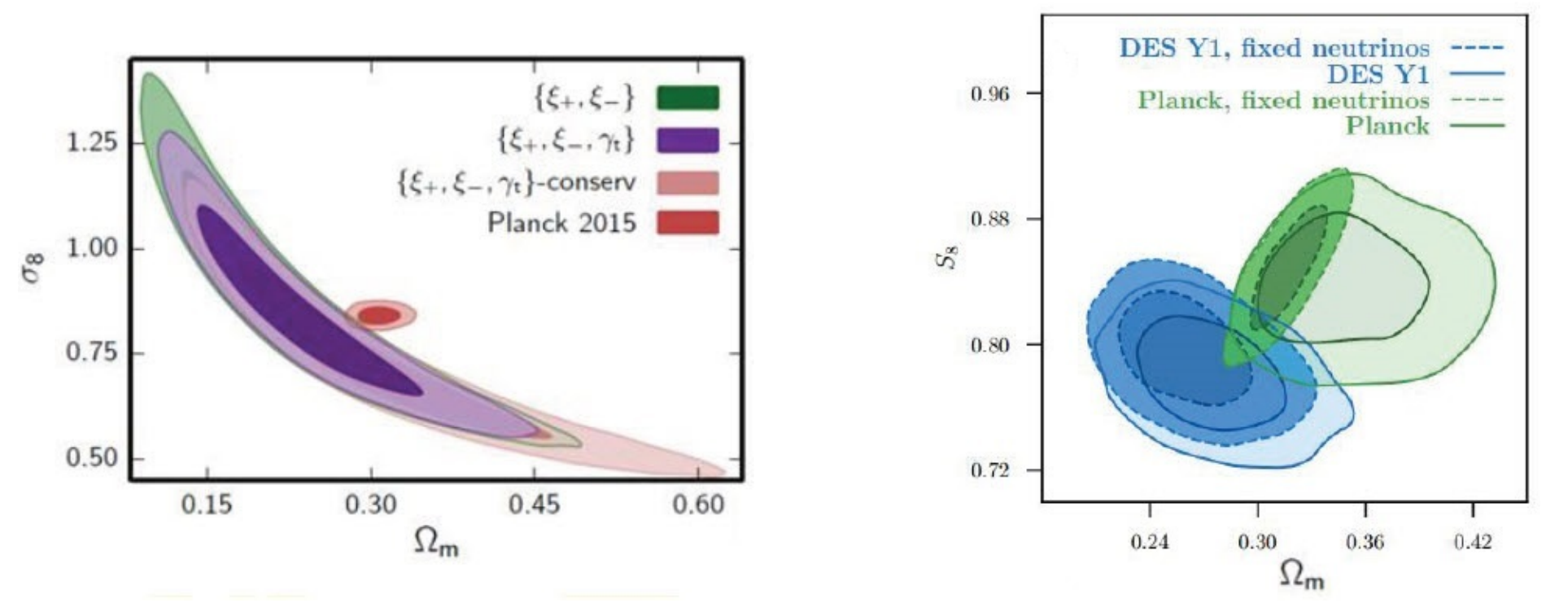}
\caption{A review of current constraints of weak lensing data with \plcdmnospace. The left figure corresponds to the Kilo Degree Survey (KiDS) \cite{Joudaki:2017zdt} and the right one to the Dark Energy Survey (DES) \cite{Abbott:2017wau} where $S_8 \equiv \sigma_8 \left( \frac{\Omega_m}{0.3}\right)^{0.5}$}
\label{fig:weaklens}
\end{figure*}

This tension is demonstrated in Fig. \ref{fig:weaklens}. On the left panel we show the $1-2\sigma$, $\sigma_8 - \Omega_m$ best fit parameter contours obtained by the Kilo Degree Survey (KiDS) \cite{Joudaki:2017zdt} superposed with the corresponding Planck 2015 \cite{Ade:2015xua} contours.  On the right panel we show the $1-2\sigma$, $S_8 - \Omega_m$ ($S_8 \equiv \sigma_8 \left( \frac{\Omega_m}{0.3}\right)^{0.5}$) best fit parameter contours obtained by the Dark Energy Survey (DES) \cite{Abbott:2017wau} superposed with the corresponding Planck 2015 \cite{Ade:2015xua} contours.  In both cases a $2-3\sigma$ tension between the \plcdm best fit and the weak lensing best fit parameter values is evident. This tension may be either due to systematics of the weak lensing or \plcdm data or could be an early hint of new gravitational physics since the weak lensing data are much more sensitive to the growth of cosmological perturbations (gravitational physics) than the CMB data which only probe this growth rate through the ISW effect on very large scales (low $l$).

As is clearly seen from Fig. \ref{fig:weaklens} weak lensing data appear to favour a lower value for \omm compared to the value of \omm favoured by \plcdmnospace. The requirement of lower \omm favoured by the weak lensing data may also be viewed as a requirement of weaker gravity than implied by GR (\plcdmnospace) at low redshifts. An interesting question therefore emerges: ``Is the same trend for weaker gravity at low redshifts and tension with \plcdm  also favoured by other probes of the growth rate of density perturbation like the RSD data?"

The RSD surveys probe the growth of matter density perturbations by detecting the distortion of the power spectrum of perturbations which are induced by peculiar velocities. This distortion probes the peculiar velocities of galaxies on large scales which in turn can be used to obtain the growth rate of perturbations  $f(a)=\frac{d ln\delta}{dlna}$, where $a$ is the scale factor and  $\delta(a)\equiv \delta \rho/\rho$ is the linear matter overdensity growth factor. Combined with density rms fluctuations within spheres of radius $R=8 h^{-1} Mpc$ which may be wriiten as $\sigma_8(a)=\sigma_8 \frac{\delta(a)}{\delta(1)}$, the observable product $f\sigma_8(a)$ measured by RSD surveys at various redshifts $z$ (or values of the scale factor $a$) may be expressed in terms of the present value of $\sigma_8(a=1)\equiv \sigma_8$ and the derivative of $\delta(a)$ with respect to the scale factor $a$ as
\ba
\fs(a)&\equiv& f(a)\cdot \sigma(a)=\frac{\sigma_8}{\delta(1)}~a~\delta'(a) \label{fs8}
\ea
This combination, \ie Eq.\eqref{fs8}, at various redshifts is published by various surveys as a probe of the growth of matter density perturbations.

Given the background expansion rate $H(z)$ which can be parametrized as  $wCDM$ 
\ba
E(a)^2&\equiv & \frac{H(a)^2}{H_0^2}= \Omega_{m}a^{-3}+\left(1-\Omega_{m}\right)a^{-3(1+w)}
\label{Hubblew}
\ea
the theoretically predicted functional form of $\delta(a)$ and therefore of $f\sigma_8(a)$ can be obtained on sub-Hubble scales
by solving the dynamical growth equation \cite{Nesseris:2017vor}
\be 
\delta''(a)+\left(\frac{3}{a}+\frac{H'(a)}{H(a)}\right)\delta'(a)
-\frac{3}{2}\frac{\Omega_{m} G_{\rm eff}(a,k)/G_{\textrm{N}}}{a^5 H(a)^2/H_0^2}~\delta(a)=0 \label{odedeltaa}
\ee
or in redshift space
\be 
\delta'' + \left[\frac{(H^2)'}{2~H^2} -
{\frac{1}{1+z}}\right]\delta'
= {\frac{3}{2}} (1+z)\frac{ H_0^2}{H^2} {\frac{G_{\rm eff}(z,k)}
{G_{N}}}~\Omega_m\delta
\label{odedeltaz}
\ee

In Eqs. (\ref{odedeltaa}), (\ref{odedeltaz}) possible deviations from GR are expressed by allowing for a scale and redshift-dependent effective Newton's constant $G_{\rm eff}=G_{\rm eff}(a,k)$. It should be stressed that an observed value of $G_{\rm eff}$ that is not constant and/or differs from the Newton's constant value $G_N$ on solar system scales does not necessarily mean that GR is violated. It could also mean that dark energy clusters on sub-Hubble scales and/or that there is a coupling between dark matter and dark energy. Both of these effects would lead to a modification of Eq. \eqref{odedeltaa} from its standard form with $G_{\rm eff}=G_N$.

In the context of standard GR ($G_{\rm eff}=G_N$) and assuming a $wCDM$ background \eqref{Hubblew} it is straightforward to solve Eq. \eqref{odedeltaa} numerically with initial conditions deep in the matter era ($\delta(a)\sim a$) and obtain the solution $\delta(a,w,\Omega_m)$ and then use \eqref{fs8} to obtain the theoretically predicted form of $f\sigma_8(a,\sigma_8,w,\Omega_m)$ in the context of GR. A fit of this theoretical prediction to the observed RSD datapoints $f\sigma_8(z_i)$ can lead to constraints on the parameters $\sigma_8,w,\Omega_m$. The comparison of these constraints with the corresponding \plcdm constraints can be a measure of the consistency of the RSD data with \plcdm in the context of GR. 

A fit along the above lines has been implemented in Refs. \cite{Nesseris:2017vor,Kazantzidis:2018rnb} where $\chi^2$ was constructed by defining the vector 
\ba
V^i(z_i,\Omega_{m},\sigma_8,g_a)\equiv f\sigma_{8i} -f\sigma_8(z_i,\sigma_8,w,\Omega_m)
\label{vecvidef}
\ea
where  $f\sigma_{8i}$ are the RSD datapoints and $f\sigma_8(z_i,\sigma_8,w,\Omega_m)$ is the theoretical prediction at the same redshift $z_i$.
The best fit $\sigma_8,w,\Omega_m$ parameter values were obtained \cite{Nesseris:2017vor,Kazantzidis:2018rnb} by minimizing 
\be
\chi^2(\sigma_8,w,\Omega_m)= V^i C_{ij}^{-1}V^j 
\label{chi2}
\ee
where $C_{ij}$ is the covariance matrix assumed to be diagonal except of the WiggleZ survey $3\times 3$ subset \cite{Blake:2012pj}. Thus, the covariance matrix may be written as
\be
C_{ij}^{\textrm{growth,total}}=\left(
         \begin{array}{cccc}
           \sigma_1^2 & 0 & 0 & \cdots \\
           0 & C_{ij}^{WiggleZ} & 0& \cdots \\
           0 & 0 & \cdots &   \sigma_N^2 \\
         \end{array}
       \right) \label{eq:totalcij}
\ee
where \cite{Blake:2012pj}
\be
C_{ij}^{\text{WiggleZ}}=10^{-3}\left(
         \begin{array}{ccc}
           6.400 & 2.570 & 0.000 \\
           2.570 & 3.969 & 2.540 \\
           0.000 & 2.540 & 5.184 \\
         \end{array}
       \right) \label{eq:wigglez}
\ee
The rest of the non-diagonal terms are assumed to be 0, implying no correlation among the corresponding datapoints. This assumption is an approximation which as discussed below using Monte Carlo simulations has a relatively small effect on the derived best fit parameter values \cite{Kazantzidis:2018rnb}.

A wide range of $\fs$ datasets have been used to constrain cosmological model parameters. Three of the largest such compilations have been constructed in Refs. \cite{Nesseris:2017vor,Kazantzidis:2018rnb}.
In Ref. \cite{Nesseris:2017vor} a compilation of 34 $\fs$ datapoints was constructed including datapoints published until 2016. In an attempt to minimize correlations among datapoints a second compilation consisting of 18 $\fs$ datapoints was constructed which included those datapoints that appeared to have minimal levels of correlation (originating from different redshift surveys and different patches in the sky). This more robust compilation is shown in Table \ref{tab:fs8-data-gold} in Appendix \ref{sec:appA}. The third more recent compilation \cite{Kazantzidis:2018rnb} is the largest $\fs$ dataset published to date consisting of $63$ distinct datapoints (Table \ref{tab:fs8-data-kazan} in Appendix \ref{sec:appA}). Despite the possible correlations among the datapoints of this compilation, it contains interesting useful information which has been extracted in the detailed analysis of Ref. \cite{Kazantzidis:2018rnb}.

The growth rate of cosmological perturbations is obtained from the RSD data by comparing the observed power spectrum of large scale structures in redshift space $P_{obs}(k_{ref\perp},k_{ref\parallel},z)$ with the expected isotropic (due to the cosmological principle) true underlying spectrum $P_{matter}(k,z)\sim \delta \rho/\rho(k,z)^2$ where $k_{ref\parallel}$ is the Fourier scale wavevector component parallel to the line of sight and $k_{ref\perp}$ is the corresponding wavevector perpendicular to the line of sight in the context of a given reference (fiducial) cosmology used to convert the measured angles and redshifts to distances. The true statistically isotropic power spectrum depends only on the magnitude of the true Fourier scale wavevector. The observed spectrum of perturbations is distorted for two reasons:
\begin{itemize}
\item
{\bf Incorrect Fiducial Cosmology:} The redshift surveys measure galaxy redshifts and angles of galaxies. In order to construct the correlation function and thus the power spectrum, these angles and redshifts need to be converted to comoving coordinates. This conversion requires the assumption of a particular form of $H_{ref}(z)$ (a reference or fiducial cosmology) which is not necessarily identical with the true cosmology $H(z)$.  The use of an incorrect fiducial cosmology $H_{ref}(z)$ would lead to an incorrect distorted nonisotropic power spectrum $P_{obs}(k_{ref\perp},k_{ref\parallel},z)$ which may be shown \cite{Alam:2015rsa,Hinton:2016jfq} to be connected with the galaxy power spectrum $P_g(k_{ref\perp},k_{ref\parallel},z)$ obtained with the correct cosmology $H(z)$ with the relation \cite{Alam:2015rsa}
\be 
P_{obs}(k_{ref\perp},k_{ref\parallel},z)=\frac{d_A(z)^2_{_{ref}} H(z)}{d_A(z)^2 H_{ref}(z)}   P_g(k_{ref\perp},k_{ref\parallel},z)
\label{apeffect}
\ee
where $d_A(z)$ is the angular diameter distance.  This geometric distortion of the correlation function and the power spectrum due to the use of the incorrect fiducial cosmology is known as the Alcock-Paczynski (AP) effect. 

Even if the correct cosmology was used for the conversion of angles-redshifts to distances, the power spectrum $P_g$ is still nonisotropic. The reason for this remaining distortion are the peculiar velocities of galaxies which encapsulate the information for the gravitational growth of perturbation. Thus, the second effect  that distorts the observed power spectrum is the peculiar velocity effect.

\item
{\bf Peculiar Velocities:} Peculiar velocities add an extra component to the cosmological redshifts thus perturbing the real positions of galaxies $x_r$ along the line of sight to a new position $x_p$ of the form \cite{Ballinger:1996cd,Amendola:2016saw}
\be
x_p=x_r+(1+z)\frac{{\hat x}\cdot {\vec v}}{H(z)}
\label{pecveldist}
\ee
This distortion of galaxy positions due to their peculiar velocities leads to an additional distortion of the observed power spectrum of the form
\begin{widetext}
\be
P_g(k_{ref\perp},k_{ref\parallel},z)=b(z)^2 \left[1+\beta(z)\frac{k_{ref\parallel}^2}{k_{ref\parallel}^2+k_{ref\perp}^2}\right]^2\;P_{matter}(k,z)
 \label{pvelspecdist}
 \ee
 \end{widetext}
where $b(z)$ is the bias factor (the ratio of the galaxy overdensities over the underlying matter overdensities) and $\beta(z)\equiv \frac{f(z)}{b(z)}$ is the linear redshift space distortion parameter. The wavenumbers $k_{ref\perp}$ and $k_{ref\parallel}$ obtained using the fiducial cosmology are connected to the wavenumbers $k_{\parallel}$ and $k_{\perp}$ in the true cosmology as 
$k_{ref\parallel}=\frac{H_{ref}(z)}{H(z)} k_{\parallel}$, $k_{ref\perp}=\frac{d_A(z)}{d_{Afid}(z)} k_{\perp}$. Using Eq. (\ref{pvelspecdist}), the measured distorted power spectrum $P_g(k_{ref\perp},k_{ref\parallel},z)$ and the isotropy of the true power spectrum, the parameter $\beta(z)$ can be inferred and from it the bias free product $\fs$ can be derived.
\end{itemize}
Each one of the datapoints of Tables \ref{tab:fs8-data-kazan} and \ref{tab:fs8-data-gold} is constructed under the assumption of a particular fiducial cosmology. Thus an Alcock-Paczynski  correction factor needs to be imposed to each one of the datapoints converting them to the values corresponding to the true cosmology $H(z)$. 
If an  $\fs'$ measurement has been obtained assuming a fiducial \lcdm cosmology $H'(z)$, the corresponding $\fs$ obtained with the true cosmology $H(z)$ is approximated as \cite{Macaulay:2013swa} 
\be 
f\sigma_8(z)\simeq\frac{H(z) d_A(z)}{H'(z)d_A'(z)} f\sigma_8'(z)\equiv q(z,\Omega_m,\Omega_m')\; f\sigma_8'(z)
\label{eq:fs8corr}
\ee
This equation should be taken as a rough order of magnitude estimate of the AP effect as it appears in somewhat different forms in the literature \cite{Saito16,Wilson:2016ggz,Alam:2015rsa}.  

\begin{figure}[!h]
\centering
\includegraphics[width = 0.47\textwidth]{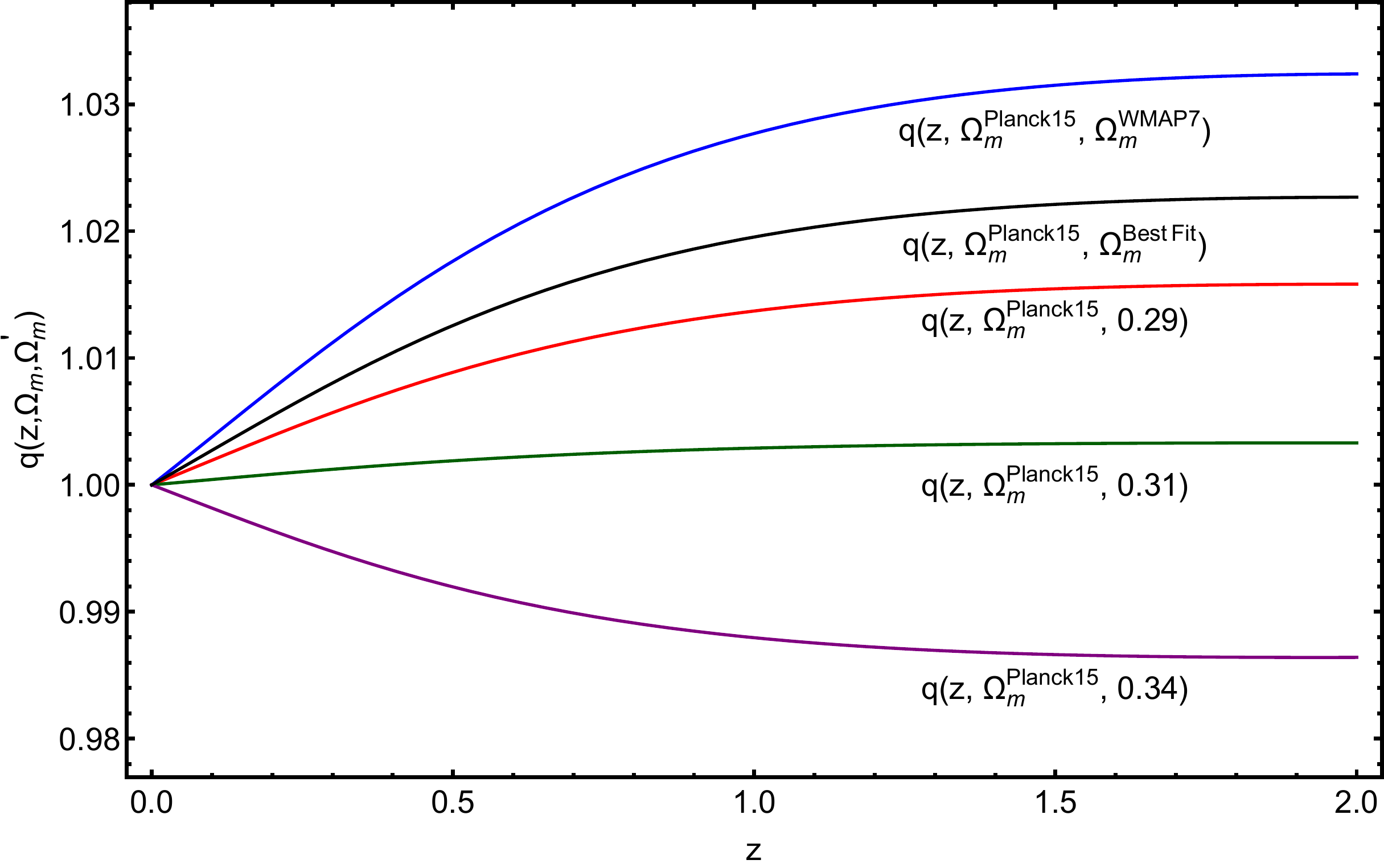}
\caption{Plot of the correction factor $q$ as a function of the redshift $z$ (from  Ref. \cite{Kazantzidis:2018rnb})}
\label{fig:qplot}
\end{figure}

This correction is small (at most it can be about $2-3\%$ at redshifts $z\simeq 1$ for reasonable values of $\Omega_{m}$) \cite{Kazantzidis:2018rnb}. The magnitude of this factor is demonstrated in Fig. \ref{fig:qplot} for typical values of fiducial and true cosmologies. 

\begin{figure}[!h]
\centering
\includegraphics[width = 0.41\textwidth]{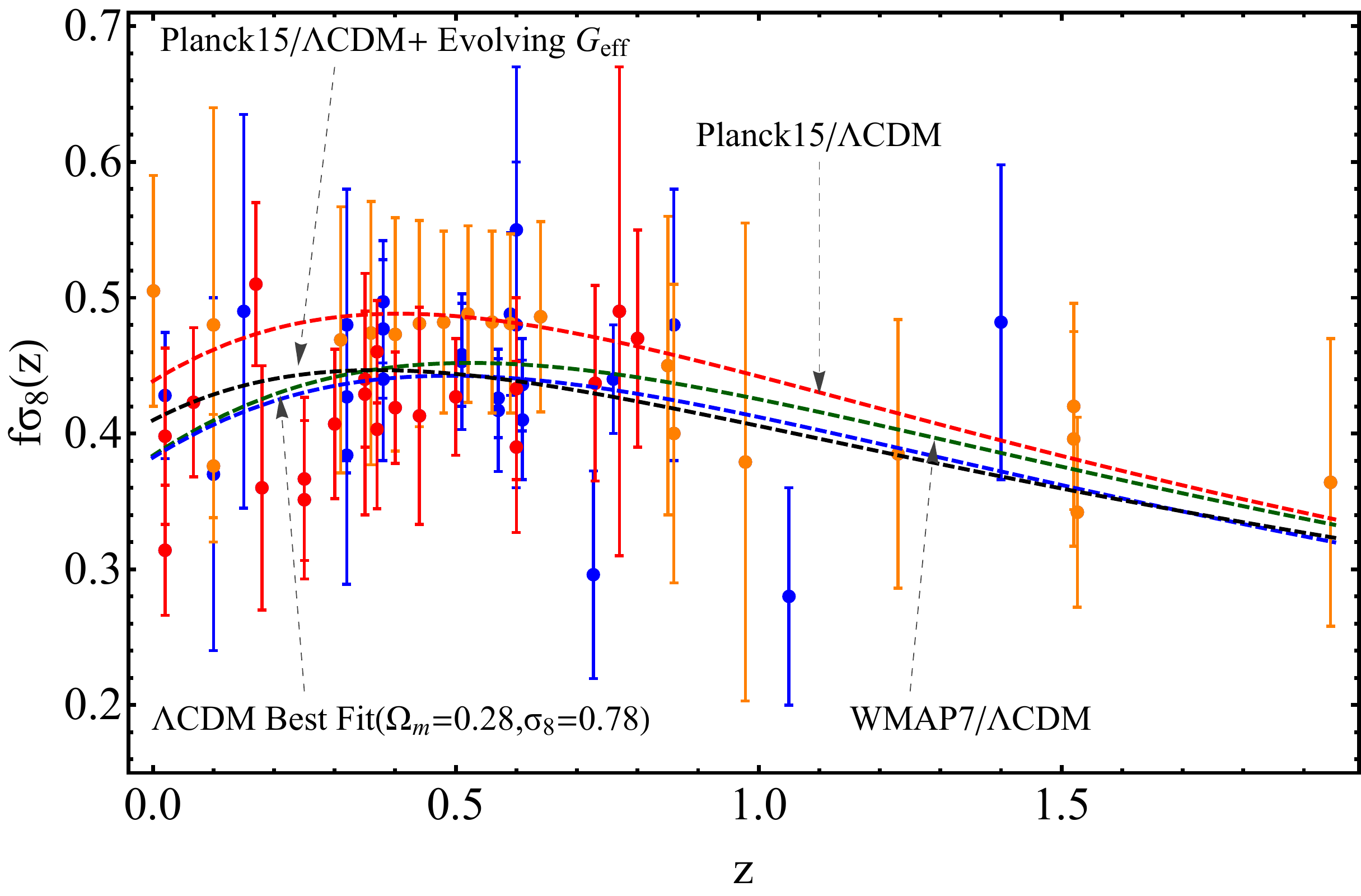}
\caption{Plot of $\fs$ as a function of the redshift $z$ for the full growth rate data set of Table \ref{tab:fs8-data-kazan} in Appendix \ref{sec:appA}. The green dashed line describes the best fits of \wlcdm while the red one the best fits of \plcdmnospace. The blue dashed line describes the best fit \lcdm ($\Omega_{m}=0.28\pm 0.02$) indicated by Table \ref{tab:fs8-data-kazan} while the black one corresponds to an evolving \geffz parametrization with a \plcdm background. The 20 earlier published data from the compilation are denoted as red points whereas the 20 latest published points are denoted as orange points (from Ref.  \cite{Kazantzidis:2018rnb})}
\label{fig:fs8zkaz}
\end{figure}

\begin{figure}[!h]
\centering
\includegraphics[width = 0.41\textwidth]{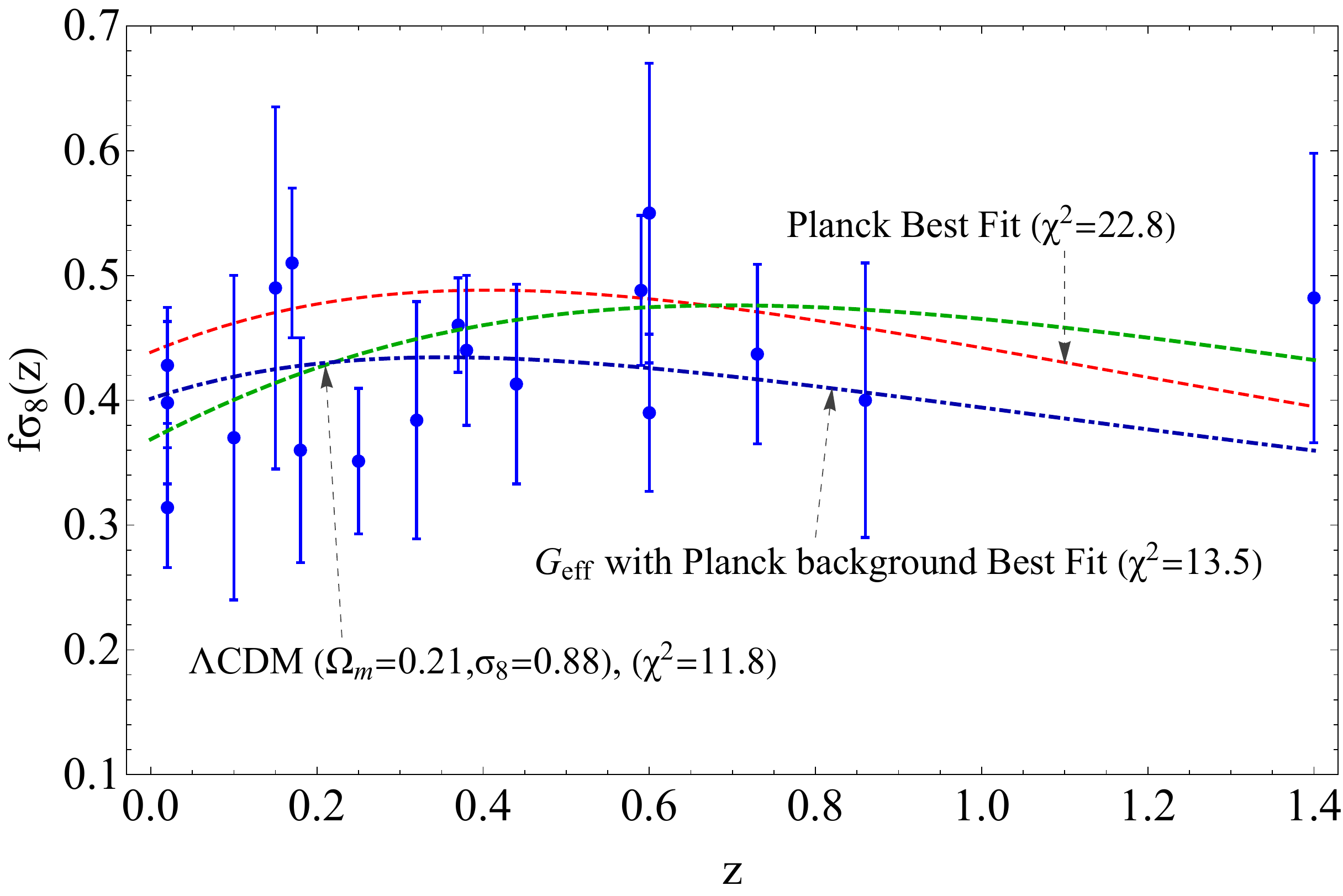}
\caption{Plot of $\fs$ as a function of the redshift $z$ for the 18 growth rate dataset of Ref.  \cite{Nesseris:2017vor}. The green dashed line describes the best fit of \lcdm ($\Omega_{m}=0.21$), the red one the best fit of \plcdmnospace. The blue dot-dashed one corresponds to an evolving \geffz parametrization with a \plcdm background.}
\label{fig:fs8z}
\end{figure}

The fiducial model corrected dataponts of Tables \ref{tab:fs8-data-kazan} and \ref{tab:fs8-data-gold} are shown in Figs. \ref{fig:fs8zkaz} and \ref{fig:fs8z}, respectively along with the predictions of specific models obtained by solving Eq. (\ref{odedeltaz}) with matter domination initial conditions and using Eq. (\ref{fs8}) for specific cosmological models: \plcdm with GR ($G_{\rm eff}=G_N$), the best fit \lcdm model to the $\fs$ data (with a reduced value of $\Omega_m$) and a modified gravity model where the background expansion is given by the \plcdm parameters while $G_{\rm eff}$ is allowed to vary with redshift with a specific

\begin{figure*}[ht!]
\centering
\includegraphics[width = 0.9\textwidth]{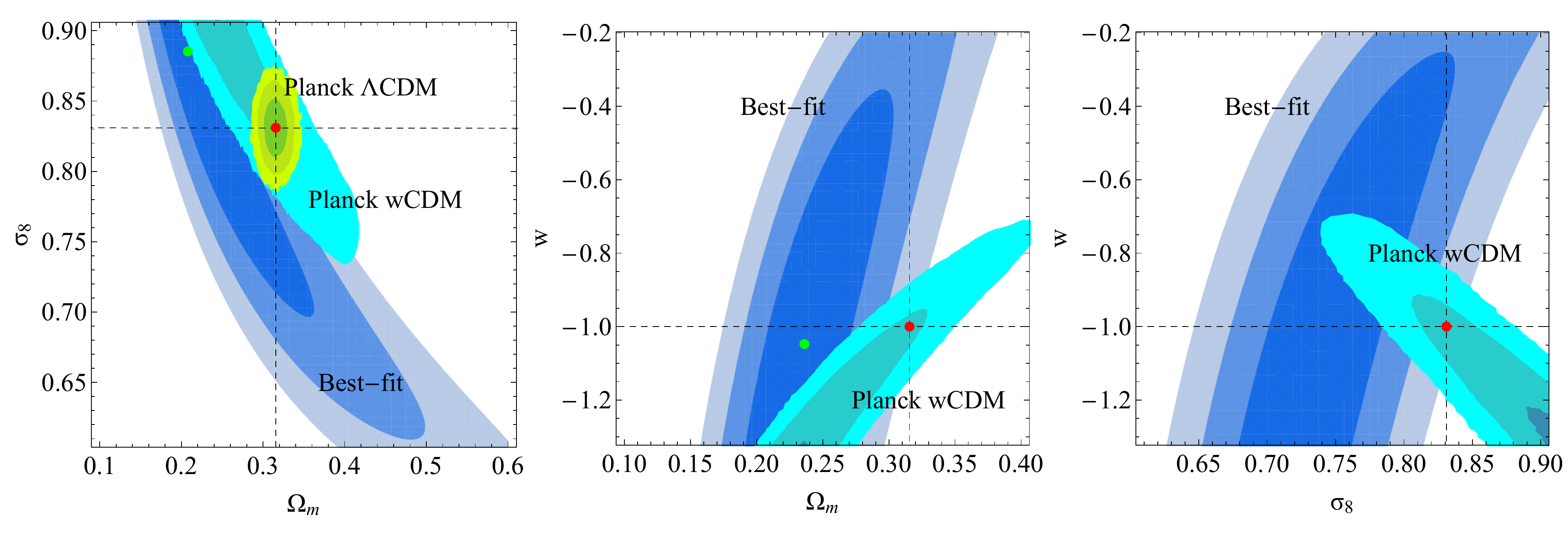}
\caption{The $1\sigma-3\sigma$ contours level in the parametric space $(w,\sigma_8,\Omega_{m})$ using the collection of the 18 points presented in  Ref. \cite{Nesseris:2017vor}. The blue contours describe the best fit of the data, the light green contours correspond to \plcdm while the light blue are constructed from the Planck data assuming a $wCDM$ background.}
\label{fig:contourness}
\end{figure*}
\begin{figure*}[ht!]
\centering
\includegraphics[width = 0.9\textwidth]{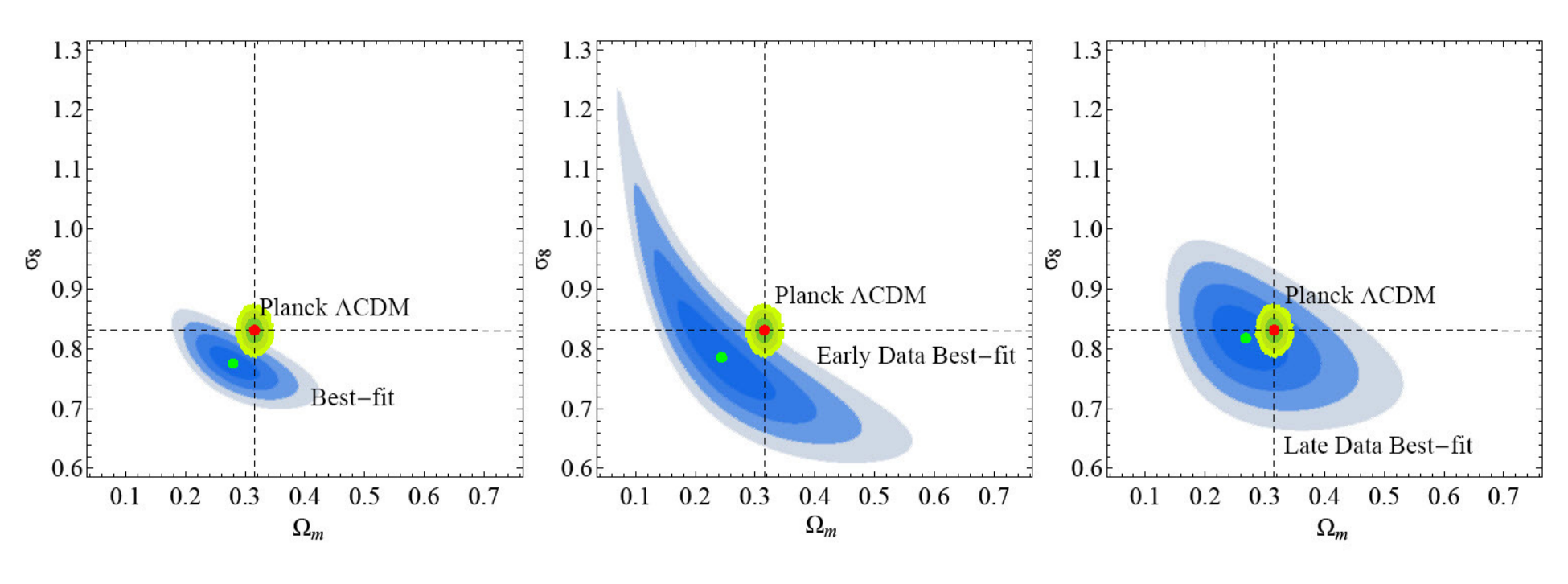}
\caption{The $1\sigma-4\sigma$ confidence contours in the parametric space $(\Omega_{m}-\sigma_8)$ using the full dataset of $\fs$ (Table \ref{tab:fs8-data-kazan}) from Ref. \cite{Kazantzidis:2018rnb}. The blue contours correspond to the best fit obtained using the full compilation of $\fs$  data from Table \ref{tab:fs8-data-kazan} (left panel), the 20 early data (middle panel) and the 20 late data (right panel). The light green contours describe the contours for the \plcdm model.}
\label{fig:fullcontour}
\end{figure*}

\noindent parametrization described below so that the best fit to the $\fs$ data is obtained. In both Figs. \ref{fig:fs8zkaz} and \ref{fig:fs8z} it is clear that the \plcdm prediction (red dashed line) is somewhat higher than the majority of the $\fs$ datapoints indicating that the growth rate is too large in this model. As shown in Figs. \ref{fig:fs8zkaz} and \ref{fig:fs8z}, this growth rate at low $z$ can be reduced (thus improving the fit to the data) by either decreasing $\Omega_m$  while maintaining GR and \lcdm (green line) or by allowing for a $G_{\rm eff}$ that evolves with redshift so that it is reduced at low $z$ (blue line). As a result, the growth data at low redshifts $(z<1)$ are more appropriate to detect possible deviations from GR than the points at high redshifts $(z>1)$ \cite{Kazantzidis:2018jtb}.

The tension between a \plcdm background (GR) and the growth data of Fig. \ref{fig:fs8z} is shown more clearly in Fig. \ref{fig:contourness} where we show the likelihood contours in two dimensional subspaces of the parameter space $\sigma_8,w,\Omega_m$. In each plot, the third parameter has a fixed value indicated by \plcdmnospace.

The blue $1\sigma-2\sigma$ parameter contours are obtained using the growth data of Fig. \ref{fig:fs8z} while the red dot corresponds to the \plcdmnospace. Clearly, there is a $2-3\sigma$ tension between the growth data contours and the best fit \plcdm parameter values. The Planck15 best fit $wCDM$ parameter contours are also shown indicating that if the equation of state parameter $w$ is allowed to vary, the tension level between the growth data parameter $1\sigma-2\sigma$ contours (blue contours) and the Planck15 \cite{Ade:2015xua} contours is significantly reduced.

\begin{figure*}[ht!]
\centering
\includegraphics[width = 0.8\textwidth]{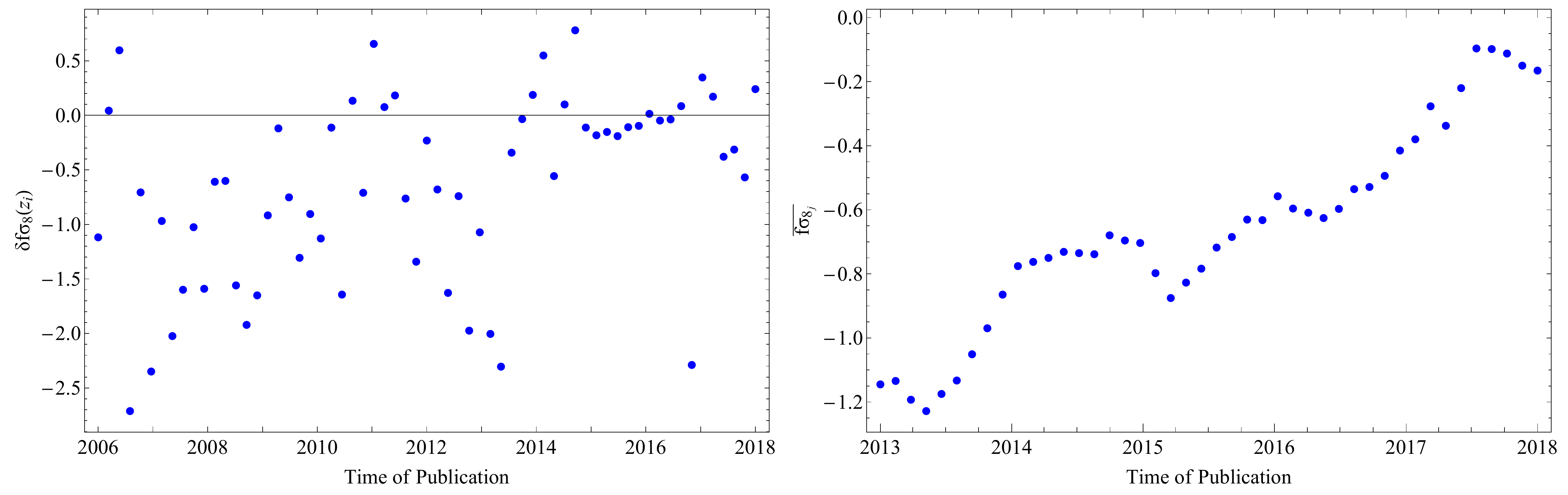}
\caption{Left panel: Residuals datapoints vs time of publication. Right panel: The 20 points moving average vs time of publication (from Ref. \cite{Kazantzidis:2018rnb}).}
\label{fig:Residual_mov_av}
\end{figure*}
\begin{figure*}[ht!]
\centering
\includegraphics[width = 0.8\textwidth]{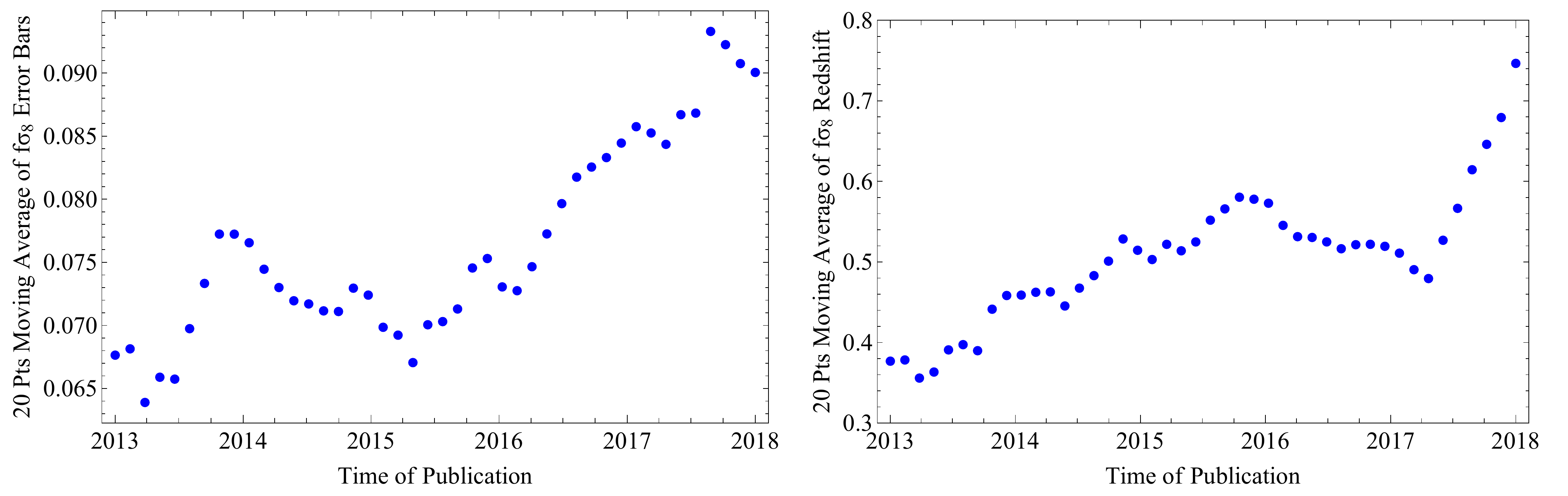}
\caption{Left panel: The 20 point moving average of the growth error bars with time of publication. Right panel: The 20 point moving average of the redshifts of $\fs$ redshifts with time of publication (from Ref. \cite{Kazantzidis:2018rnb})}
\label{fig:errorbar_redsh}
\end{figure*}

An interesting question to address is the following: ``How does the level of tension between the growth data and the \plcdm best fit parameter values evolve with time of publication?" or ``Are early growth data at the same tension level with \plcdm as more recently published data?" This question has been addressed in Ref. \cite{Kazantzidis:2018rnb} using the data of Table \ref{tab:fs8-data-kazan} and reviewed in what follows.

The evolution of the tension level is demonstrated in Fig. \ref{fig:fullcontour} where we show the ($\Omega_{m}-\sigma_8$) \lcdm ($w=-1$) best fit parameter contours obtained with the full dataset of Table \ref{tab:fs8-data-kazan} (left panel), with the earliest 20 datapoints (middle panel) and with the latest 20 datapoints of the same Table. 

Interestingly, the tension drops from a level more than $3\sigma$ for the early published data to less than $1\sigma$ for the $20$ most recent datapoints. The exact tension  level ($\sigma$ distance between the growth data best fit parameters and the \plcdm best fit parameters) is shown in Table \ref{tab:sigma}. 
\begin{table}[h]
\caption{The sigma differences of Fig. \ref{fig:fullcontour} Contours for the Full Dataset, Early and Late Data.}
\label{tab:sigma}
\begin{centering}
\begin{tabular}{|c|c|c|c|}
\hline
\rule{0pt}{3ex} 
& Full Dataset & Early Data & Late Data\\ 
\hline
\rule{0pt}{3ex} 
Fig. \ref{fig:fullcontour} Contours & $4.97\sigma$ & $3.89\sigma$ & $0.94\sigma$\\
\hline
\end{tabular}
\end{centering}
\end{table}

%%%%%%%%%%%%%%%%%%%%%%%%%%%%%%%%%%%%%%%%%
The evolving tension between \plcdm parameter values and $\fs$ data may also be described by defining the residuals
\be
\delta f\sigma_8(z_i)\equiv \frac{f\sigma_8(z_i)^{data}-f\sigma_8(z_i)^{Planck15/\Lambda CDM}}{\sigma_i}
\label{eq:resfs8}
\ee
Using these residuals, the 20 point moving average residual may be defined as
\be
\overline{{f\sigma_8}_j}\equiv \sum_{i=j-20}^j\frac{\delta f\sigma_8(z_i)}{20}
\label{eq:movavdef}
\ee 

These residual datapoints versus time of publication along with the corresponding 20 point moving average from  Eq. \eqref{eq:movavdef}, are shown in Fig. \ref{fig:Residual_mov_av} (from Ref. \cite{Kazantzidis:2018rnb}). There is a clear trend  for reduced tension with \plcdm in more recently published data.

More recent $\fs$ datapoints tend to probe higher redshifts and thus they also tend to have higher errors. This is demonstrated in Fig. \ref{fig:errorbar_redsh} where we show the 20 point moving average of the datapoint errorbars and redshifts  versus time of publication. Both of them show an increasing trend especially for more recent data. At higher redshifts the universe is matter dominated and GR is approximately restored in most models and thus there is degeneracy in the predictions of different models. Thus, more recent datapoints that tend to probe higher redshifts have less constraining power on cosmological models.

In fact, the increase of the average redshift is a possible explanation for the reduced tension of the recent data with \plcdm due to the degeneracy that exists between models at high z. This degeneracy can also be observed in Fig. \ref{fig:fs8zkaz}, where for high $z$ the four curves coincide.

As discussed above, the tension between $\fs$ data and \plcdm may be reduced by either reducing $\Omega_m$ or by extending GR and allowing for an evolving $G_{\rm eff}$. Such an evolving $G_{\rm eff}(z)$ may be described by a parametrization of the form
\ba
\frac{G_{\textrm{eff}}(a,g_a,n)}{G_{\textrm{N}}} &=& 1+g_a(1-a)^n - g_a(1-a)^{2n} \nn \\
&=&1+g_a\left(\frac{z}{1+z}\right)^n - g_a\left(\frac{z}{1+z}\right)^{2n}. \label{geffansatz}
\ea
where $g_a$ and $n$ are parameters to be fit. This parametrization for $n=2$ has been used for the construction of the Figs. \ref{fig:contourness} and \ref{fig:fullcontour}.

The parametrization \eqref{geffansatz} is  well motivated and consistent with solar system experiments. The solar system constraints entail for the first derivative that \cite{Nesseris:2006hp}
\be 
\lim_{z \to 0} G'_{\rm eff}(z) \simeq 0  \Rightarrow \Big\lvert \frac{1}{G_N} \frac{d G_{\rm eff}(z)}{dz} \Big \vert_{z=0} \Big\rvert < 10^{-3} h^{-1}
\label{geffconstr1}
\ee
This constraint implies that unless a Chameleon type mechanism \cite{Khoury:2003rn} is present we must have $n\geq 2$ in the parametrization \eqref{geffansatz}. The solar system experiments also leave the second derivative unconstrained since \cite{Nesseris:2006hp} 
\be 
\Big\vert \frac{1}{G_N}  \frac{d^2G_{\rm eff}(z)}{dz^2} \Big \vert_{z=0}\Big\vert < 10^{5} h^{-2}
\label{geffconstr2}
\ee

Finally, at high redshifts, the Big Bang Nucleosynthesis provides the following additional constraint at the $1\sigma$ level \cite{Copi:2003xd}
\be
\lvert G_{\rm eff}/G_{N} -1 \rvert \leq 0.2  
\label{geffnucconstr}
\ee 
which is also consistent with the parametrization \eqref{geffansatz}.

\begin{figure}[!t]
\centering
\includegraphics[width = 0.42\textwidth]{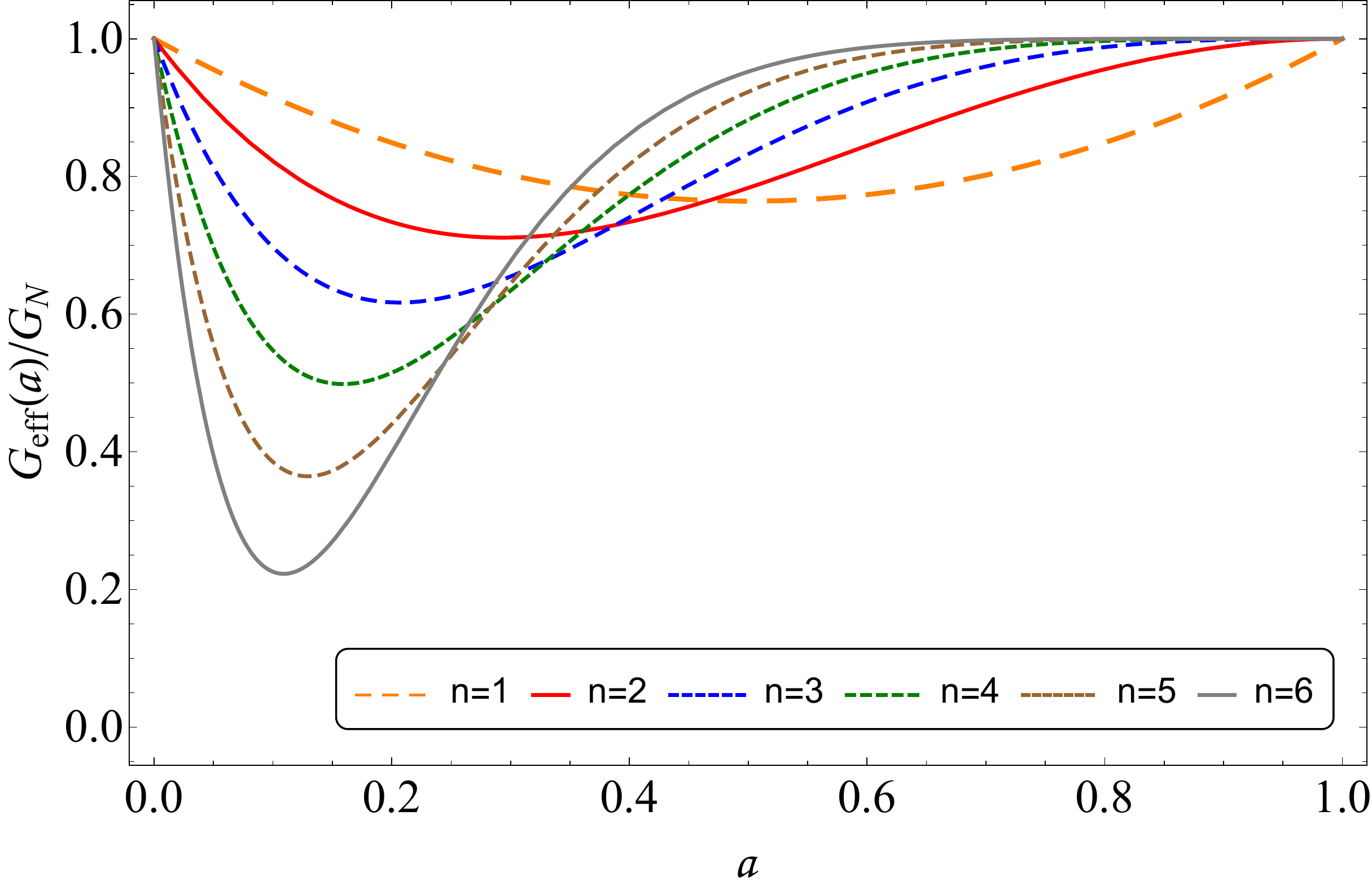}
\caption{Plot of $G_\text{eff}/G_\text{N}$ as a function of $a$ considering the values of $n$ and $g_a$ from Table \ref{tab:geff_ga} (from Ref. \cite{Nesseris:2017vor})}
\label{fig:geffa}
\end{figure}

In the context of this parametrization, which was presented in  Refs. \cite{Nesseris:2017vor,Kazantzidis:2018rnb}, two extra parameters have been inserted ($g_a$ and $n$) describing the possible deviation from GR. Setting $\Omega_{m}=\Omega_{m}^{Planck15}$ and $\sigma_8=\sigma_8^{Planck15}$, \ie eliminating the tension with respect to $\Omega_{m}$ and $\sigma_8$, we see that the value of $\chi^2$ gets reduced significantly for $g_a\neq 0$ (using the data of Table \ref{tab:fs8-data-gold} in the Appendix \ref{sec:appA} and setting $n=2$ it was found that  $g_a =-1.16 \pm 0.341$). This corresponds to the blue curve of Fig. \ref{fig:fs8z}. The best fit values of of $g_a$ for various values of $n$ are also shown in Table \ref{tab:geff_ga} and the corresponding forms of \geffz are shown in Fig. \ref{fig:geffa}. 

\begin{table}[h]
\caption{The best fit values of $g_a$ along with the $1 \sigma$ errors bars for various values of $n$.}
\label{tab:geff_ga}
\begin{centering}
\begin{tabular}{|c|c|}
   \hline
$n$ &  $g_a$ \\ 
\hline
$0.343$ & $-1.200 \pm 1.025$ \\
$1$ & $-0.944 \pm 0.253$ \\
$2$ & $-1.156 \pm 0.341$ \\
$3$ & $-1.534 \pm 0.453$ \\
$4$ & $-2.006 \pm 0.538$ \\
$5$ & $-2.542 \pm 0.689$ \\
$6$ & $-3.110 \pm 0.771$ 
\\
\hline
\end{tabular}
\end{centering}
\end{table}

Notice that a significant reduction of the gravitational constant is required at low $z$ to fit the $\fs$ data with $\Omega_{m}=\Omega_{m}^{Planck15}$. Such a large reduction is inconsistent with other cosmological observations (e.g. CMB large scale power spectrum where the ISW effect dominates \cite{Nesseris:2017vor} or the distance moduli of SnIa when their dependence on $G_{\rm eff}(z)$ is taken into account \cite{Gannouji:2018ncm}) and therefore it is unlikely that the tension implies only the existence of evolving   $G_{\rm eff}(z)$. It is more likely that the tension is also (or only) due to other factors like a reduced value of $\Omega_{m}$ or systematic/statistical errors of the $\fs$ data.

\begin{figure}[!h]
\centering
\includegraphics[width = 0.42\textwidth]{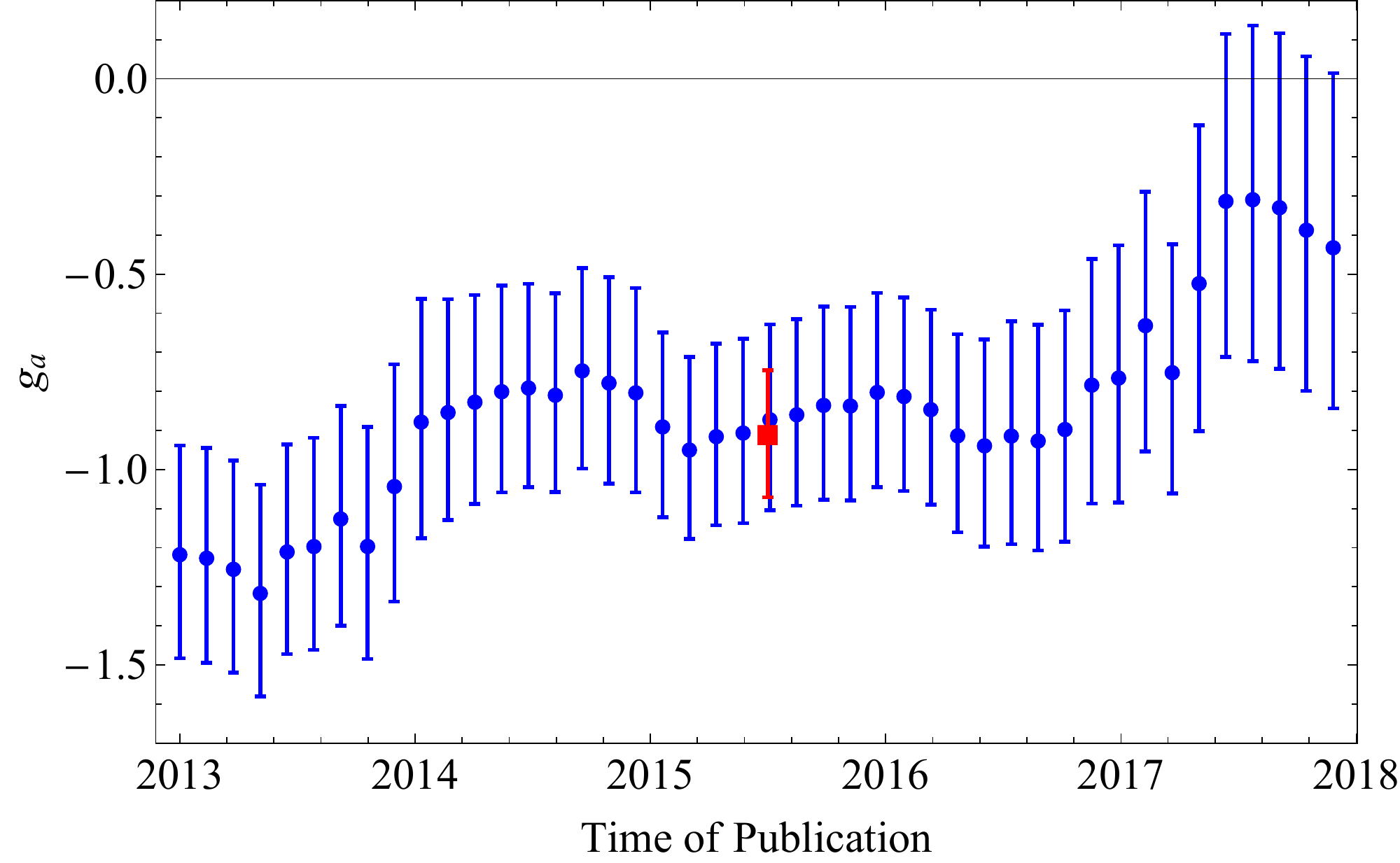}
\caption{The $1\sigma$ range of the parameter $g_a$ from the compilation of Table \ref{tab:fs8-data-kazan}. The red square point denotes the the best fit of $g_a$ obtained from the compilation of Table \ref{tab:fs8-data-kazan} along with its error bar (from Ref. \cite{Kazantzidis:2018rnb})}
\label{fig:gaplot}
\end{figure}

The evolution of the tension level with time of pubication of the $\fs$ data may also be seen by deriving the best fit value of the parameter $g_a$ assuming $n=2$ and a \plcdm background while using 20 datapoint subsamples from Table \ref{tab:fs8-data-kazan} starting from the earliest to the latest subsample (from left to right in Fig. \ref{fig:gaplot}).

\noindent Clearly the absolute value of $g_a$ required to eliminate the tension with \plcdm decreases significantly for more recent data indicating also the reduced level of the tension for more recent data. The best fit value of $g_a$ for the full dataset of Table \ref{tab:fs8-data-kazan}  ($g_a=-0.91\pm 0.17$) is also indicated in Fig. \ref{fig:gaplot} (red point).

\subsection{Consistency of Reduced $G_{\rm eff}(z)$  with Modified Gravity Theories}

The best fit form of $G_{\rm eff}(z)$ which appears to indicate reduced strength of gravity at low $z$ may lead to constraints on the fundamental parameters of modified theories of gravity. In fact, it may be shown that the simplest modified gravity theories including $f(R)$ and scalar-tensor theories tend to be inconsistent with a decreasing $G_{\rm eff}(z)$ especially in a \lcdm and in a phantom cosmological background \cite{Gannouji:2018ncm}.

In scalar-tensor gravity the action has the following form \cite{EspositoFarese:2000ij,Boisseau:2000pr}
\be
{\cal S}= \int d^4 x \sqrt{-g} \left[ \frac{1}{2}F(\phi)R - \frac{1}{2}Z(\phi) g^{\mu\nu} \partial_\mu \phi \partial_\nu \phi - U(\phi) \right] + S_m,
\label{action}
\ee
where we have set $8\pi G=1$. It is clear from Eq. \eqref{action} that the action depends on the scalar field $\phi$. Throughout this subsection we also consider $Z(\phi)=1$. The line element for a flat Friedmann-Robertson-Walker(FRW) metric is
\be
ds^2 = -dt^2 + a^2(t) \left[dr^2 + r^2 (d\theta^2 + \sin^2\theta \ d\phi^2) \right].
\ee
By varying the action \eqref{action} with respect to the inverse metric, considering that the scalar field is homogenous and that the background is that of a perfect fluid, the dynamical equations of motion are of the following form
\begin{eqnarray}
3F H^2 &=&  \rho +{\frac{1}{2}} \dot\phi^2 - 3 H \dot F + U \label{fe1}\\
-2F \dot H  &=& (\rho+p) + \dot \phi^2 +\ddot F - H \dot F \label{fe2} 
\end{eqnarray}

Usually it is convenient to express Eqs. \eqref{fe1} and \eqref{fe2} in terms of the redshift $z$. We define the squared rescaled Hubble parameter as
\be
q(z) \equiv E^2(z) = \frac{H^2(z)}{H_0^2}
\ee
After an additional rescaling of the potential ($U \rightarrow U \cdot H_0^2$) the equation of motion for $F(z)$ is \cite{EspositoFarese:2000ij,Boisseau:2000pr,Nesseris:2017vor}
\begin{widetext}
\be F^{\prime\prime}(z)+ \left[\frac{q^\prime(z)}{2q(z)}-\frac{2}{1+z}\right] F^{\prime}(z)  - \frac{1}{(1+z)}\frac{q^\prime(z)}{q(z)} F(z)+3 \frac{1+z}{q(z)} \Omega_{m}= -\phi^{\prime}(z)^2 \label{fe1a} \ee
\end{widetext}
where the prime denotes from now on differentiation with respect to $z$.

In scalar-tensor theories the effective Newton's constant may be expressed as \cite{Nesseris:2006jc}
\be
G_{\rm eff}(z)/G_{\textrm{N}}=\frac{1}{F(z)}\frac{F(z)+2F_{,\phi}^2}
{F(z)+\frac{3}{2}F_{,\phi}^2}
\label{Geff}
\ee
where $G_\text{N}$ is the usual Newton's constant in GR.

Using the best fit form of \geffz on the left hand side of Eq. \eqref{Geff}, we may obtain the corresponding form of $F(z)$ and then use Eq. \eqref{fe1a} with a $q(z)$ corresponding to \plcdm to find the corresponding form of $\phi'(z)^2$.  Therefore the question that we want to address is: ``Can the weakening effect of gravity indicated by the growth data be due to an underlying scalar-tensor theory?"

If this effect is due to an underlying scalar-tensor theory, then  the new reconstructed scalar field must obey $\phi^{\prime}(z)^2>0$ so that the theory is self consistent. However a decreasing $G_{\textrm{eff}}$ with redshift at low $z$ is inconsistent with $\phi^{\prime}(z)^2>0$ as it is demonstrated numerically in the following Fig. \ref{fig:phizplot}.

\begin{figure}[!h]
\centering
\includegraphics[width = 0.48\textwidth]{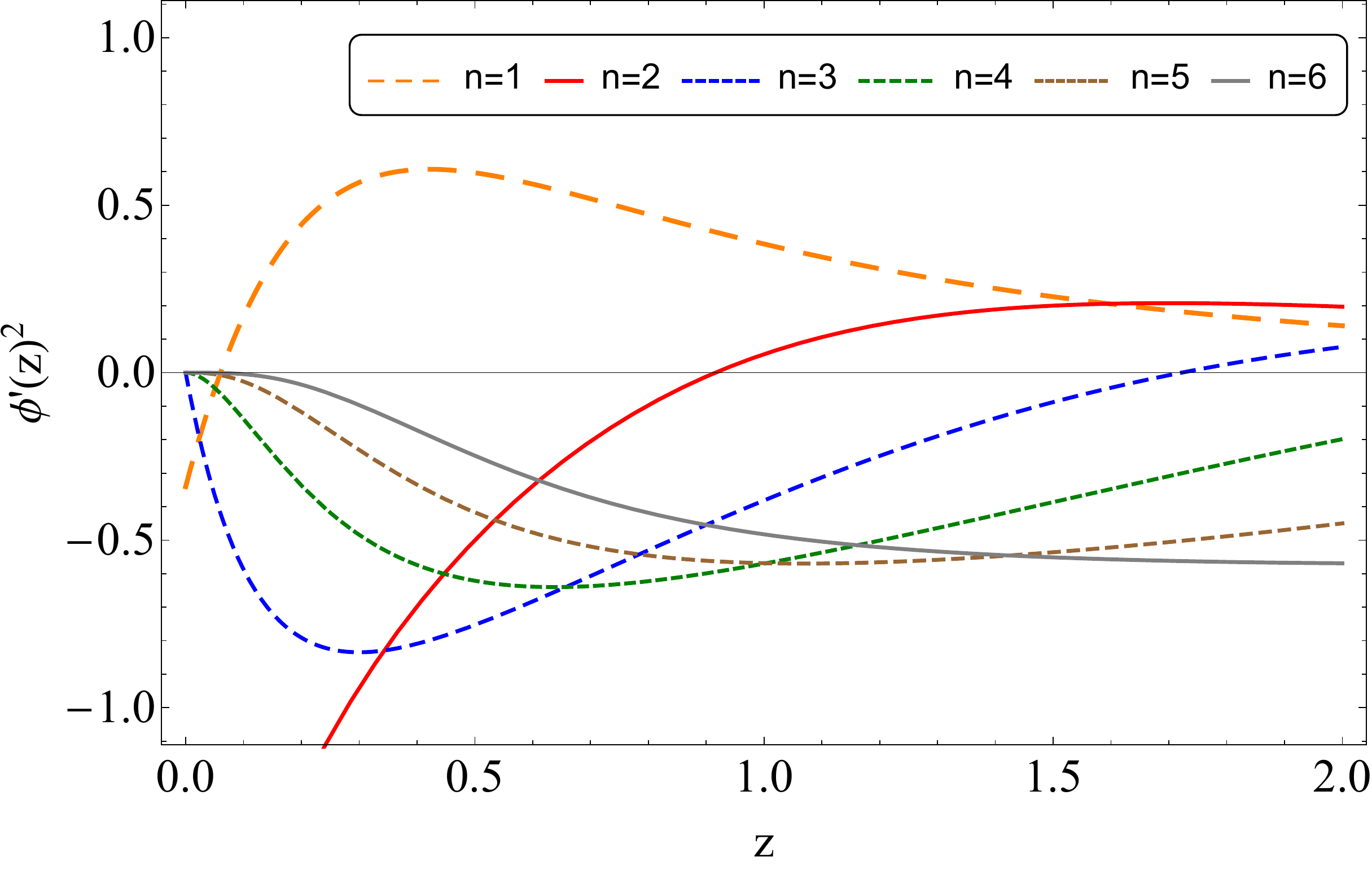}
\caption{Plot of $\phi^{\prime}(z)^2$ with redshift $z$ for various values of $n$ (from Ref. \cite{Nesseris:2017vor}). The best fit values of $g_a$ shown in Table \ref{tab:geff_ga} were assumed with a \plcdm background.}
\label{fig:phizplot}
\end{figure}

From Fig. \ref{fig:phizplot} it is clear that at low $z$, $\phi^{\prime}(z)^2$ is negative. As a result this behaviour can not be  supported by a self consistent scalar tensor theory.

This numerical result may also be demonstrated analytically.  For a \wcdm background we have
\be 
q(z)=\Omega_m(1+z)^3+\left(1-\Omega_m \right)(1+z)^{3(1+w)}\la{eq:wcdmq}
\ee
For low $z$, we can expand the dynamical Newton's constant \geffznospace, which up to the second order is of the form
\be 
G_{\rm eff}(z)=G_{\rm eff}(0)+G'_{\rm eff}(0)z+\frac{z^2}{2} G''_{\rm eff}(0)\la{eq:gefftaylor}
\ee
Applying the solar system constraints for the first derivative of \geffznospace, \ie Eq. \eqref{geffconstr1}, Eq. \eqref{eq:gefftaylor} is rewritten as
\be
G_{\rm eff}(z)=G_{\rm eff}(0)+\frac{z^2}{2} G''_{\rm eff}(0) \la{eq:gefftaylorconstr}
\ee

It is straightforward to show that the constraint of Eq. \eqref{geffconstr1} implies that $F'(0) \approx 0$. Therefore, setting $G_N=F(0)=1$ and differentiating \geffz with respect to $z$ we obtain
\be 
G_{\rm eff}''(0)=F''(0) \left(-1+\frac{F''(0)}{\phi'(0)^2} \right) \la{eq:geffzpp0}
\ee
Furthermore, using Eq. \eqref{eq:wcdmq} in  Eq. \eqref{eq:geffzpp0} and setting $z=0$, Eq. \eqref{eq:fppelim1} is derived
\be
3-3w(-1+\Omega_{0m})-3\Omega_{0m}-\phi'(0)^2-F''(0)=0 \la{eq:fppelim1}
\ee
\noindent Substituting it to Eq. \eqref{eq:geffzpp0}, the second derivative of \geffz takes the following form \cite{Gannouji:2018ncm}
\be 
G_{\rm eff}''(0)=9(1+w)(-1+\Omega_{m})+\frac{9(1+w)^2 (-1+\Omega_{m})^2}{\phi'(0)^2} +2\phi'(0)^2  \label{geff0sct}
\ee

\begin{figure}[!h]
\centering
\includegraphics[width = 0.45\textwidth]{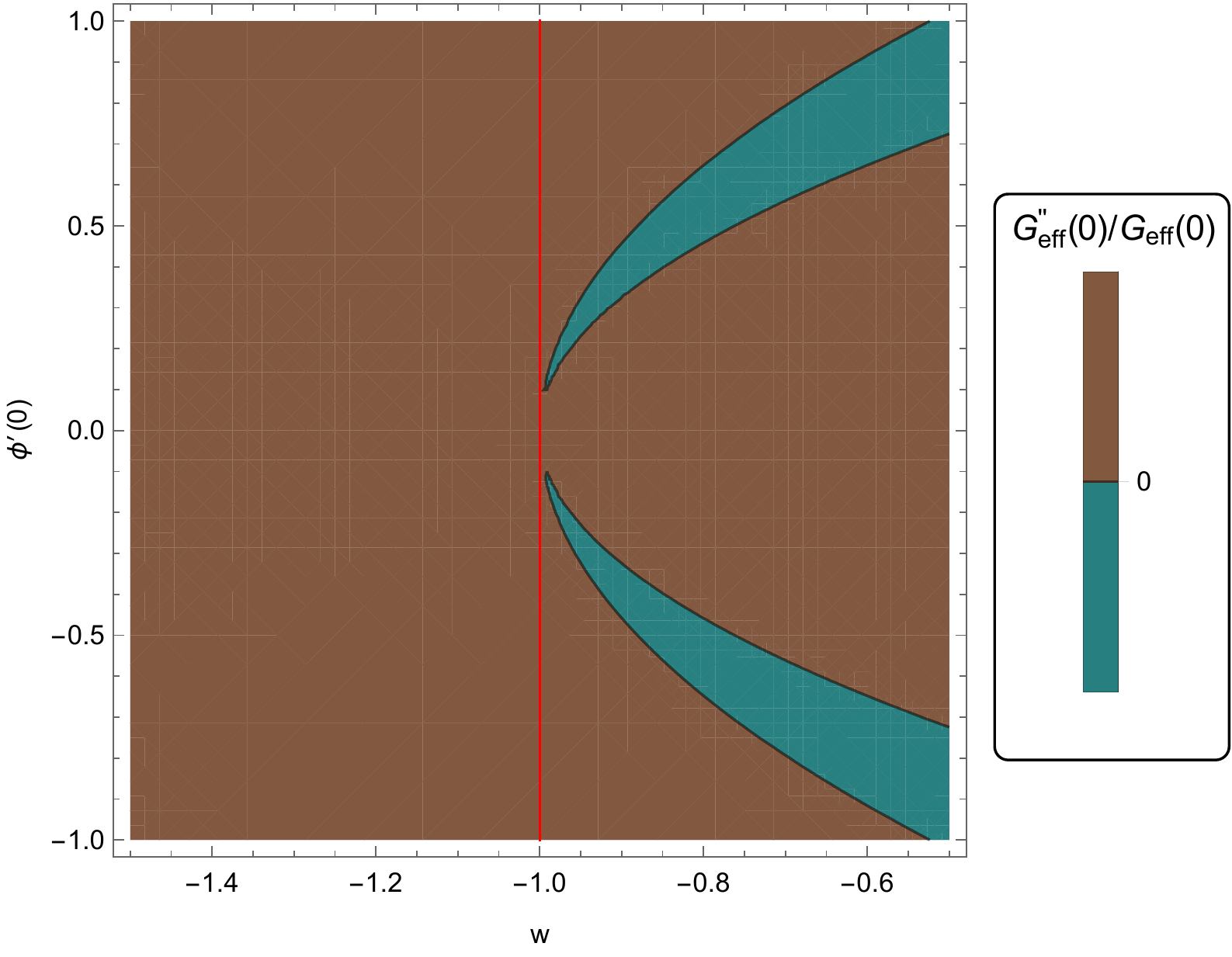}
\caption{The second derivative of $G_{\rm eff}$ in the parametric space  $(\phi'(0)-w)$, setting $\Omega_m=0.3$. With blue we denote the parameter values in which $G_{\rm eff}''(0)<0$, while the brown regions describes $G_{\rm eff}''(0)>0$ which is achieved only for $w>-1$.}
\label{fig:Geffdoubpr}
\end{figure}

Fixing a \lcdm background, \ie setting $w=-1$, Eq. \eqref{eq:gefftaylor} takes the form 
\be 
G_{\rm eff}(z) \approx  G_{\rm eff}(0)+ \frac{1}{2} G_{\rm eff}''(0)z^2=G_{\rm eff}(0)+\phi'^2(0)z^2\label{eq:geffform}
\ee
which  is always an increasing function of $z$ if we assume that the kinetic term of $\phi^{\prime}(z)$ is always positive, an assumption which is crucial if we want to have a self-consistent theory. This is demonstrated in Fig. \ref{fig:Geffdoubpr}.

From the above analysis the following result is extracted: If a \lcdm background is assumed, any \geffz initially decreasing with $z$ leads to a reconstructed scalar-tensor negative kinetic term for some range of low $z$ \cite{Gannouji:2018ncm}.

\begin{figure}[h]
\centering
\includegraphics[width = 0.49\textwidth]{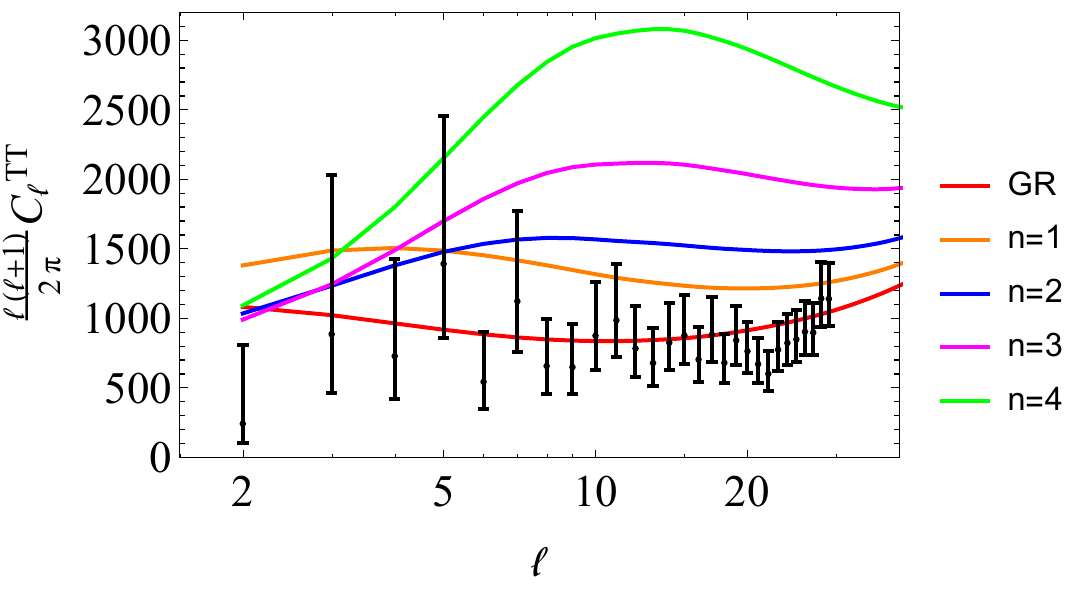}
\includegraphics[width = 0.51\textwidth]{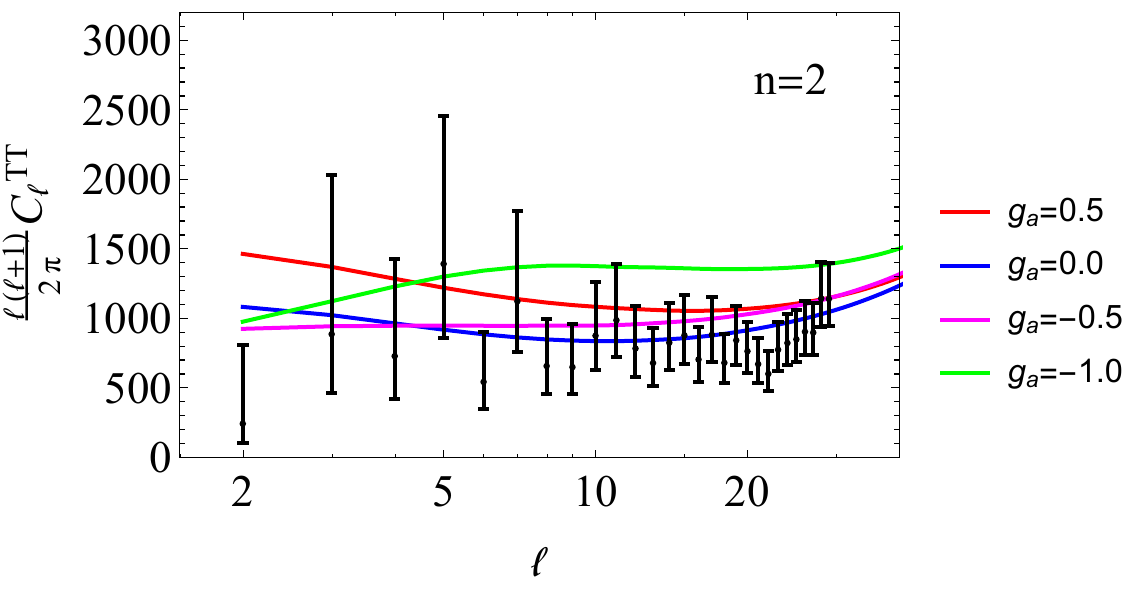}
\caption{Top panel:  The theoretically predicted CMB power spectra for the best fit parameter values $g_a$ for various values of $n$ from Table \ref{tab:geff_ga}. Clearly these best fit forms of $G_{\rm eff}(z)$ are not consistent with the observed values of $C_l's$. Therefore, the tension between the \plcdm best fit parameter values and those indicated by the $\fs$ data can not be attributed solely to an evolving $G_{\rm eff}(z)$. Bottom panel: The theoretically predicted low $l$ CMB power spectra for $n=2$ and various values of $g_a$. Only values $|g_a|\lsim 0.5$ are consistent with the observed CMB power spectrum (from Ref. \cite{Nesseris:2017vor}).}
\label{fig:ISW}
\end{figure}

Since the magnitude of the best fit parameter $g_a$ is relatively large, it is important to test its consistency with other observational probes and in particular with the low $l$ angular power spectrum of the CMB which is affected by the ISW effects and therefore can probe the strength of gravity. Using MGCAMB \cite{Hojjati:2011ix} the predicted CMB angular power spectrum may be derived assuming a \plcdm background cosmology and a $G_{\rm eff}(z)$ parametrized by the ansatz (\ref{geffansatz}). Such an analysis \cite{Nesseris:2017vor} indicates that for $l \gsim 80$ the CMB angular power spectrum remains practically unaffected by the evolving form of $G_{\rm eff}(z)$ and thus the \plcdm best fit parameter value for $\Omega_m$ remains also practically unaffected. On the other hand, the low $l$ CMB spectrum is affected significantly due to the ISW effect. This is demonstrated in Fig. \ref{fig:ISW} \cite{Nesseris:2017vor} where the measured low $l$ values of the $C_l$ components are superposed with the theoretical prediction obtained for an evolving $G_{\rm eff}(z)$ for various values of the parameters $g_a$ and $n$.

\noindent Clearly the low $l$ CMB power spectra impose strong constraints on the allowed values of $g_a$ and only the range $|g_a|\lsim 0.5$ appears to be consistent with the observed values of low $l$ $C_l's$.

\section{Hints of Modified Gravity on Sub millimetre Scales}
\label{sec:submmscales}
\subsection{Review of the Washington Experiment}
\label{sec:WashRev}

As discussed in the Introduction, the small scale frontier of the gravitational physics research is on sub-mm scales. This scale, however, is also connected with macrophysics and with dark energy. In fact, the dark energy scale may be written as
\be 
\lambda_{de} \equiv\sqrt[4]{\hbar c/\rho_{ de}}\approx  0.085 mm
\ee
where it is assumed that $\Omega_m=0.3$ and $H_0=70 km sec^{-1} Mpc^{-1}$. Hence, if the accelerating expansion is connected with modified gravity, it is natural to expect signatures of  modified theories of gravity on scales $\lambda \approx 0.1 mm$.

In the last decade a large number of experiments \cite{Murata:2014nra-good-review-ofexperiments,Kapner:2006si-washington3,Hoyle:2004cw-washington2,Hoyle:2000cv-washington1} have imposed constraints on parametrizations which are extensions of Newton's gravitational potential.

\begin{figure}[!h]
\centering
\includegraphics[width = 0.3\textwidth]{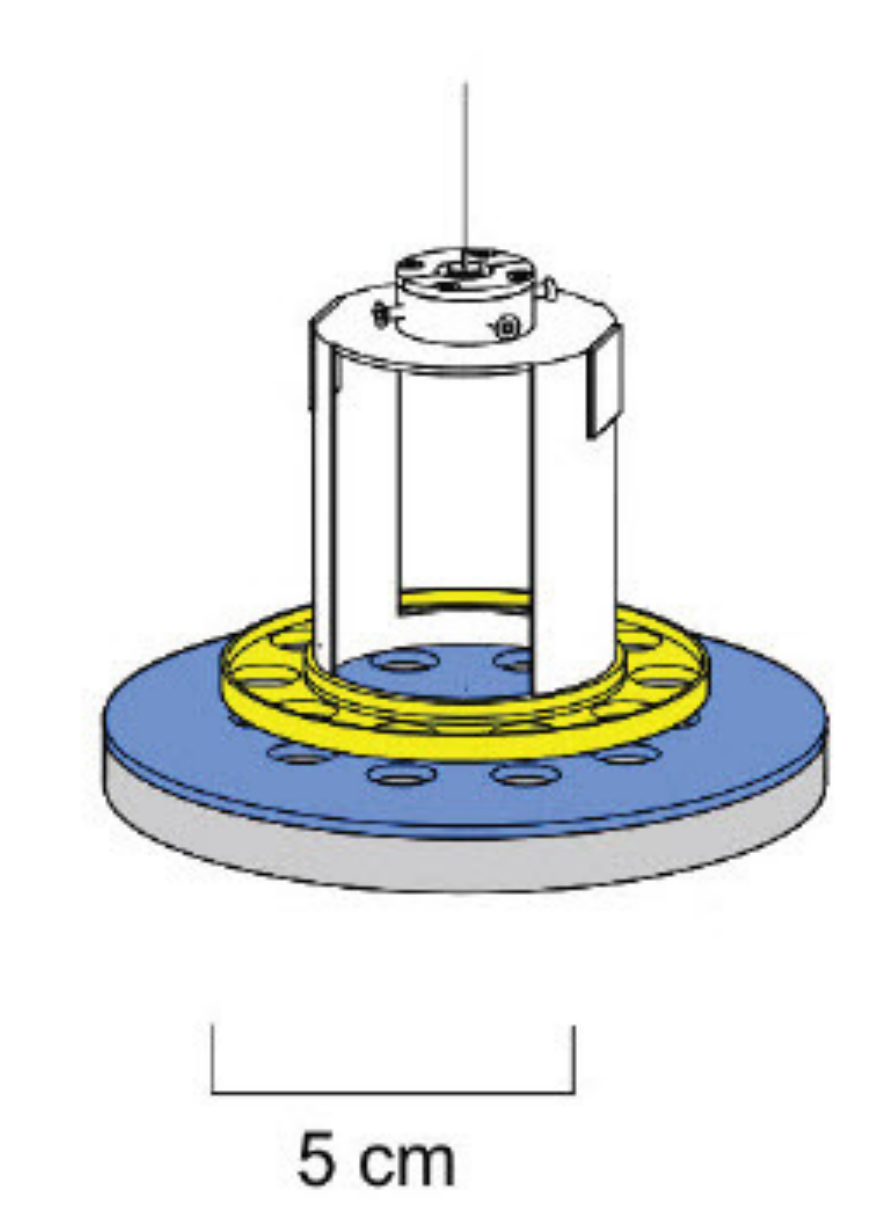}
\caption{The Washington Experiment set-up (from Ref. \cite{Hoyle:2004cw})}
\label{fig:washexp}
\end{figure}

One of the most sensitive such experiments which have imposed the best constraints so far on the Yukawa parametrization discussed in the Introduction is the Washington experiment \cite{Hoyle:2004cw} which consists of three similar setups (Experiments I, II, III). It is based on a  torsion-balance set-up shown in Fig. \ref{fig:washexp}.

It consists of  a fiber pendulum, 82 cm long, attached to a thin plate ring (yellow in Fig. \ref{fig:washexp}) placed above a rotating plate with holes (blue in Fig. \ref{fig:washexp}). The blue ring was an attractor which, like the pendulum ring, contained ten equally spaced holes with diameters about 9.5mm. The test-bodies used to measure the gravitational interaction were  the holes excerting  a torque of the form
\be 
N(\phi)= -\frac{\partial V(\phi)}{\partial \phi }\label{eq:torwash}
\ee
where $V(\phi)$ is the potential energy of the attractor ring-pendulum system when the holes of the ring twisted and formed an angle $\phi$ with respect to the pendulum.

\begin{figure}[!h]
\centering
\includegraphics[width = 0.48\textwidth]{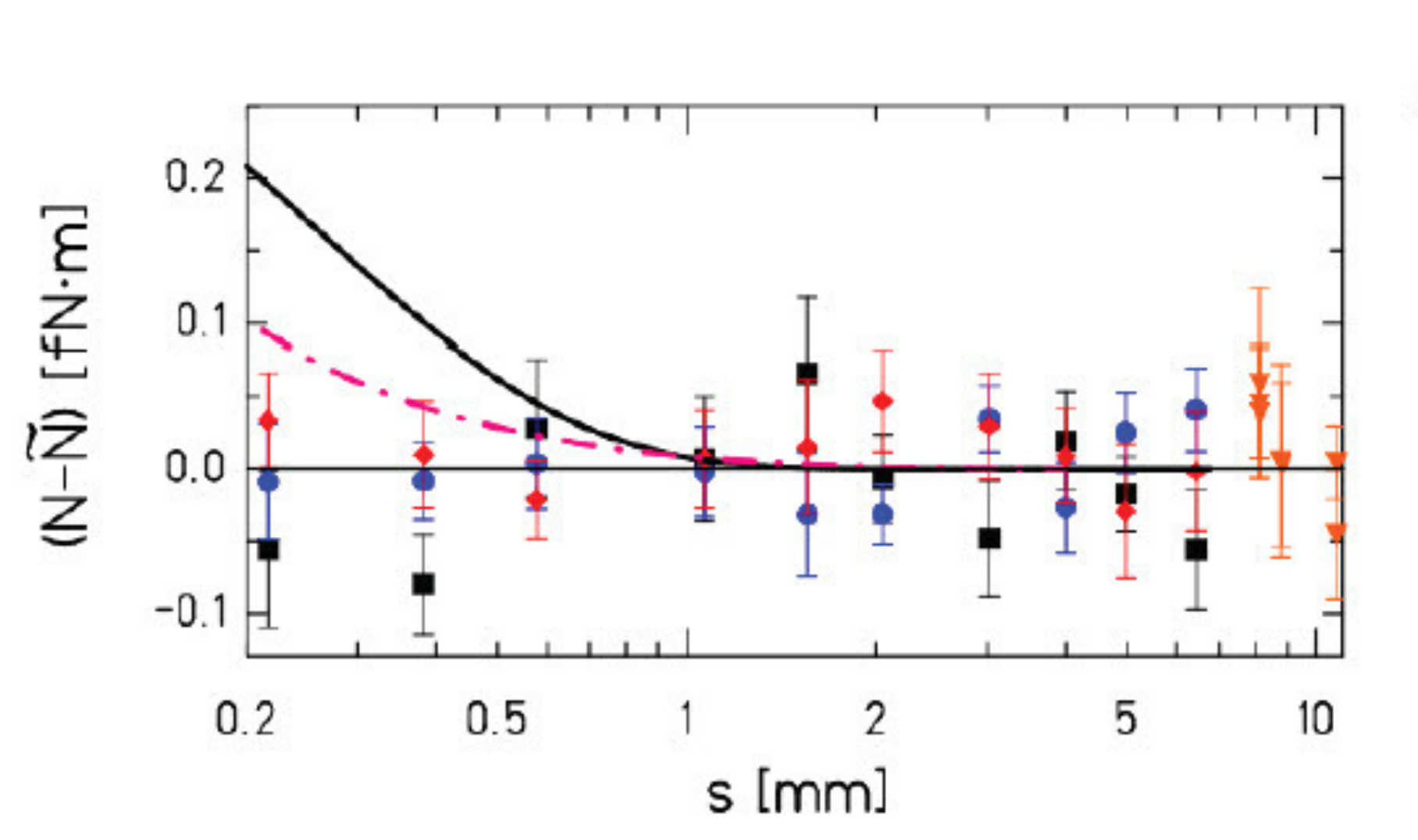}
\caption{The residuals of the datapoints used in the analysis of Ref. \cite{Hoyle:2004cw} (datapoints of Experiment I). The solid curve corresponds to residuals of the Yukawa parametrization of Eq. \eqref{ntpot-wash} for $a=1$, $m^{-1}=\lambda=250 \mu m$, whereas the dot-dashed describes the power law of  Eq. \eqref{powlaw-wash}, where $k=5$ and $\beta_k=0.005$ (from Ref. \cite{Hoyle:2004cw}). }
\label{fig:reswashexp}
\end{figure}

The torque residuals that were measured in this experiment were fit assuming two different forms of a gravitational potential: A Yukawa parametrization of the form
\be 
V_{\rm eff}= -G \frac{M}{r}(1+\alpha e^{- m r}) \label{ntpot-wash}
\ee
and a power law parametrization of the form \cite{Kapner:2006si}
\be 
V_{\rm eff}= -G \frac{M}{r} \left(1+\beta_k \left[\frac{1 mm}{r} \right]^{k-1} \right) \label{powlaw-wash}
\ee
This power law ansatz emerges naturally from some  brane world models \cite{Donini:2016kgu,Benichou:2011dx,Bronnikov:2006jy,Nojiri:2002wn-newtonptl-brane}. The torque residuals from the Newtonian torques are shown Fig. \ref{fig:reswashexp} along with the predicted residuals in the context of the above generalized gravitational potentials for specific parameter values.

\subsection{Yukawa and Oscillating Newtonian Potenial in $f(R)$ Theories}

The simplest form of $f(R)$ theories is $f(R)=R+\frac{1}{6m^2}R^2 + ...$. In the weak field limit for   $m^2 >0$ the theory is self consistent and stable \cite{Berry:2011pb-weak-field-fr-incl-osc,Capozziello:2007ms-fR-newtonian-limit-with-osc-good-disc} leading to Yukawa type correction to the gravitational potential. For $m^2 <0$ the Yukawa type gravitational potential gets modified and the exponentially suppressed correction transforms to an oscillating correction of the form 
\be 
V_{\rm eff}= -G \frac{M}{r}(1+\alpha \cos(m r+\theta))
\label{oscillanz}
\ee
where $\theta$ is a parameter.

In order to demonstrate the validity of the modified Newtonian potentials \eqref{ntpot-wash} and \eqref{oscillanz} in the context of $f(R)$ theories, consider the generalized  Einstein-Hilbert action \cite{Perivolaropoulos:2016ucs,Starobinsky:2007hu,DeFelice:2010aj}
\be 
S_R=\frac{1}{16 \pi G}\int d^4 x \sqrt{-g} f(R)+S_{matter}
\label{fraction1}
\ee
where $R$ is the Ricci scalar. It is easy to show that this action can be rewritten in the equivalent form \cite{Chiba:2003ir-gamma05-weak-field,Faraoni:2006hx-ill-defined-equivalency-st-fR}
\be 
S_{BD}=\frac{1}{16 \pi G}\int d^4 x \sqrt{-g} \left[ f(\phi)+f_\phi(\phi)(R-\phi)\right]+ S_{matter}
\label{bdequiv}
\ee

By varying action (\ref{bdequiv}) with respect to the scalar field $\phi$ and assuming  $f_{\phi\phi}\neq 0$ we obtain
\be 
\phi=R
\ee
Setting $f(R)=R+\frac{1}{6m^2}R^2$, and defining $\Phi \equiv 1 + \frac{1}{3m^2}\phi$, then it is straightforward to rewrite Eq. \eqref{bdequiv} as
\be
S_{BD}=\frac{1}{16 \pi G}\int d^4 x \sqrt{-g} \left[\Phi R -\frac{3}{2}m^2 (\Phi-1)^2\right]+ S_{matter}
\label{bdequiv3}
\ee
which is the action of a massive BD scalar field with $\omega=0$.

Furthermore, varying Eq. \eqref{bdequiv3} with respect to the inverse metric and the scalar field $\Phi$, we obtain the dynamical equations
\begin{widetext}
\be
\Phi \left(R_{\mu\nu}-{\frac{1}{2}}g_{\mu\nu}R\right)
= 8\pi G T_{\mu\nu}
+\nabla_\mu\partial_\nu \Phi - g_{\mu\nu}\Box \Phi
- g_{\mu\nu} \frac{3}{4}  m^2 (\Phi-1)^2 \label{metdyneq} \ee
\be
\Box\Phi =
\frac{8\pi G}{3} \,T + m^2 \left((\Phi-1)^2 + (\Phi-1)\Phi\right)
\label{phidyneq}
\ee
\end{widetext}
respectively. Considering the weak gravitational field for a point mass of the form
\be
T_{\mu\nu}=diag(M\delta(\vec r),0,0,0)
\label{tmnpointmass}
\ee
the quantities $\Phi$ and $g_{\mu \nu}$ can be expanded as 
\ba 
\Phi&=&1 +\varphi \label{expf}  \\ 
g_{\mu\nu}&=&\eta_{\mu\nu}+h_{\mu\nu} \label{expg} 
\ea
Substituting Eqs. \eqref{expf} and \eqref{expg} in the dynamical equations and keeping terms up to linear order we obtain the perturbative dynamical equations around the vacuum solution as
\be
\left(\Box -  m^2 \right)\varphi = -\frac{8\pi G}{3} M \delta(\vec r) \label{linphieq} \ee
\be -\frac{1}{2} \left[\Box \left(h_{\mu\nu}-\eta_{\mu\nu}\frac{h}{2}\right)\right]= 8\pi G T_{\mu\nu} +\partial_\mu\partial_\nu \varphi-\eta_{\mu\nu}\Box\varphi \label{linmeteq} \ee
where $h=h^\mu_\mu$. For static configurations, these equations convert to 
\ba \nabla^2\varphi -  m^2 \varphi &=& -\frac{8\pi G}{3}M \delta(\vec r) \label{linphieq1} \\
\nabla^2 h_{00} - \nabla^2 \varphi &=& -8\pi G M \delta(\vec r) \label{linmeteq1a} \\
\nabla^2 h_{ij} - \delta_{ij} \nabla^2 \varphi &=& -8\pi GM \delta(\vec r)  \delta_{ij} \label{linmeteq1b} 
\ea
which lead to the following weak field solution for $\varphi$ and $h_{\mu \nu}$:
\ba \varphi &=&\frac{2 G M}{3r}e^{-m r} \label{phisol} \\
h_{00}&=&\frac{2GM}{r}\left(1+\frac{1}{3}e^{-m r}\right)\label{h00sol} \\
h_{ij}&=&\frac{2GM}{r}\delta_{ij}\left(1-\frac{1}{3}e^{-m r}\right)\label{hijsol} 
\ea 
Thus, the Yukawa generalization for the gravitational potential of a point mass is obtained under the assumption $m^2>0$ while for $m^2<0$ an oscillating solution is obtained
\be
\varphi =\frac{2 G M}{r}\frac{1}{3}Cos(\vert m \vert r+\theta)
\label{phisolnegm2} 
\ee
where $\theta$ is a parameter that describes an arbitrary phase.
This solution leads to an oscillating gravitational potential of the form 
\be 
V_{\rm eff}=-\frac{h_{00}}{2}=-\frac{G M}{r}\left(1+\frac{1}{3}Cos(\vert m \vert r+\theta)\right) \label{ntpot2} 
\ee 
In order to study the stability of these solutions, 
we allow for a time-dependent perturbation $\delta \varphi$, ($\varphi=\varphi_0(r) + \delta\varphi(r,t)$), where $\varphi_0$ is the unperturbed part of the solution. The perturbed part $\delta\varphi(r,t)$ satisfies the following equation 
\be
-\ddot \delta\varphi + \nabla^2 \delta\varphi - m^2 \delta\varphi =0
\label{perteq1}
\ee
which for positive $m^2$ is the usual Klein-Gordon equation and leads to a wavelike stable solution (even if we consider higher order terms in the initial Lagrangian \cite{Dolan:2007mj}). However if a negative $m^2$ is considered, Eq. \eqref{linphieq1} leads to instabilities and exponentially increasing perturbations \cite{Frolov:2015usa-newton-potential-nonlocal,Kehagias:2014sda-nonlocal-oscillations,Perivolaropoulos:2016ucs}. 

Thus, $f(R)$ theories are unable to predict oscillatory behavior of the Newtonian potential without the presence of tachyonic instabilities unless higher-order terms in the action or a nontrivial background energy momentum tensor are considered. As discussed in the following, however, such oscillatory behavior is more natural in the context of nonlocal gravity theories.

\subsection{Fit of Oscillating Parametrization on the Washington Experiment Data}
In the context of the Washington experiment \cite{Hoyle:2004cw} the data were reported as differences between the measured torques for a Yukawa type potential and the expected torques from a Newtonian potential. These differences (residuals) have been reported in three different experiments denoted as Experiment I, II and III respectively in Ref. \cite{Kapner:2006si}. Each experiment involved variations of the attractor and detector thickness in such a way that the systematic errors were minimized.

Therefore a total of $N=87$ residuals points \cite{Perivolaropoulos:2016ucs} were shown \cite{Kapner:2006si-washington3,Hoyle:2000cv-washington1,Hoyle:2004cw-washington2,Kapner:2005qy-thesis,Perivolaropoulos:2016ucs} along with the residual curves. These $87$ residual points could be either statistical fluctuations around a Newtonian gravitational potential or could emerge from generalized gravitational potentials, e.g. Eq. \eqref{ntpot-wash}, deviating from the Newtonian potential. 

In Ref. \cite{Perivolaropoulos:2016ucs} the 87 residual datapoints $\delta \tau \equiv \tau-\tau_N$ from the three experiments were fit to the following parametrizations: 
\ba  
\delta \tau_1(\alpha',m',r) &=& \alpha' \label{constpar} \\
\delta \tau_2(\alpha',m',r) &=& \alpha' e^{-m' r}
\label{exppar} \\
\delta \tau_3(\alpha',m',r) &=& \alpha' \cos(m' r + \frac{3\pi}{4})
\label{oscilpar}
\ea
\ie an offset Newtonian, a Yukawa and an oscillating ansatz
where $\alpha'$ and $m'$ are the parameters that were fitted. The parameter $\theta'$ was fixed in $\theta'=\frac{3\pi}{4}$, since it provided the best fit compared to other selected phases. The primes were used in order to avoid confusion with the fundamental parameters of Eq. \eqref{ntpot-wash}. It is important to note that the connection between the dotted and undotted parameters $\alpha, \, m \text{ and } \theta$ is not obvious unless specific details of the apparatus of the experiment are known.  This is discussed in detail in Ref. \cite{Perivolaropoulos:2016ucs}.

The parametrizations \eqref{constpar}-\eqref{oscilpar} were used in Ref. \cite{Perivolaropoulos:2016ucs} to minimize $\chi^2(\alpha',m')$ which was defined the usual way as
\be 
\chi^2(\alpha',m')=\sum_{j=1}^N\frac{\left(\delta \tau(j)-\delta \tau_i(\alpha',m',r_j)\right)^2}{\sigma_j^2}
\label{defchi2}
\ee
where $j$ refered to the $j^{th}$ residual of the experiment, $i$ to the selected parametrization ($i$ runs from 1 to 3) and $N=87$. In the following Table \ref{tab:name} the best fit values of $\chi^2$ for each parametrization are shown.

\begin{table}[h]
\caption{The best fit value of $\chi^2$ for each parametrization using the $87$ residual datapoints. Note the improved quality of fit for the oscillating parametrization.}
\label{tab:name}
\begin{centering}
\begin{tabular}{|c|c|}
   \hline
   \rule{0pt}{3ex} 
\textbf {Parametrization} & $\chi^2$\\
\hline
$\delta \tau=\alpha'$ & $ 85.5 $\\
$\delta \tau=\alpha'e^{-m' r}$ & $85.4$ \\
$\delta \tau=\alpha'\cos(m' r +\frac{3\pi}{4})$ & $70.7$ 
\\
\hline
\end{tabular}
\end{centering}
\end{table}

\begin{figure}[!h]
\centering
\includegraphics[width = 0.48\textwidth]{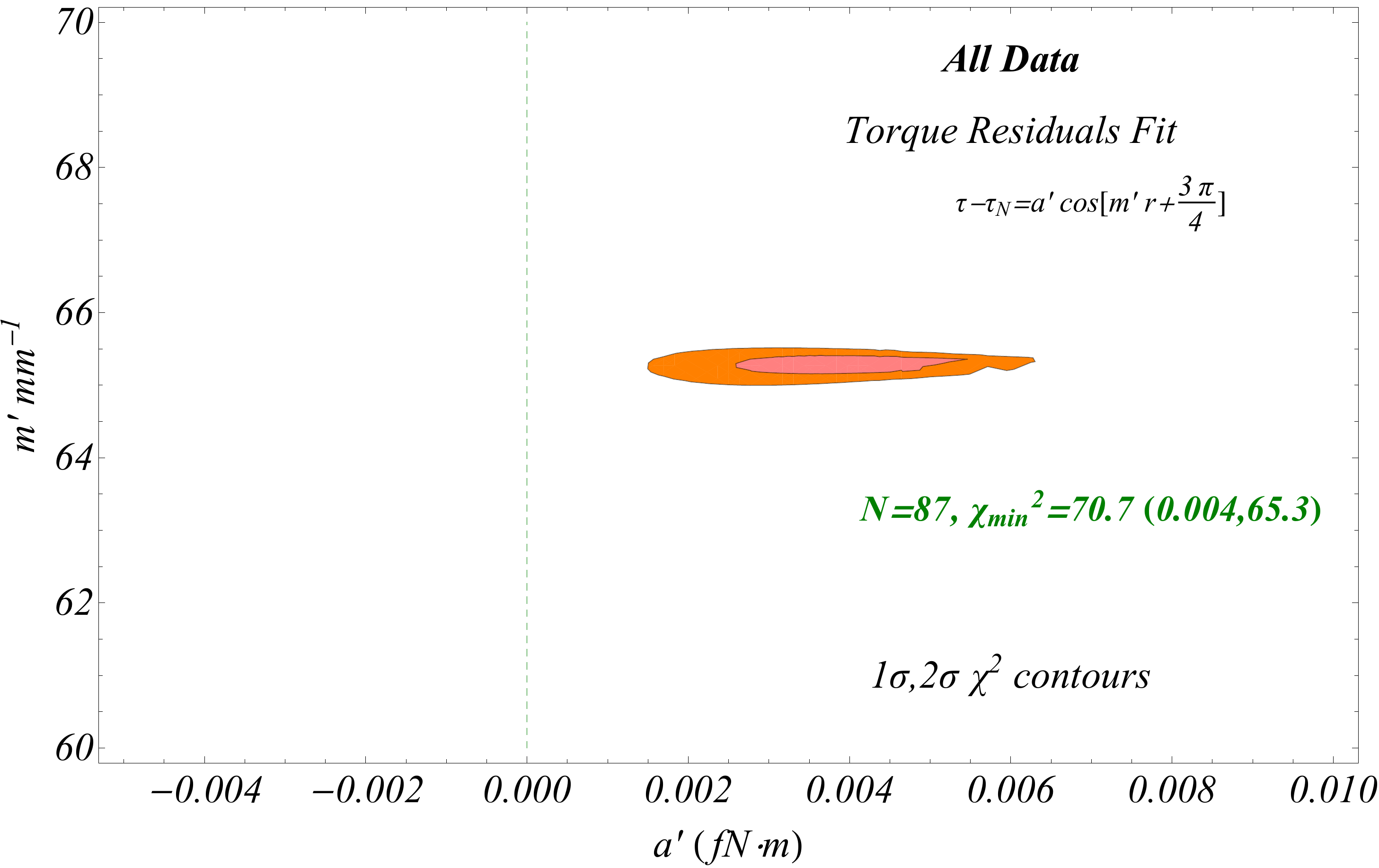}
\caption{The $1\sigma$ and $2\sigma$ contours in the parametric space $(\alpha',m')$  for the oscillating parametrization \eqref{oscilpar}. The quality of fit is significantly improved compared to the Newtonian ansatz. (from Ref. \cite{Perivolaropoulos:2016ucs}) }
\label{fig:contoursoscil}
\end{figure}

\begin{figure}[!h]
\centering
\includegraphics[width = 0.48\textwidth]{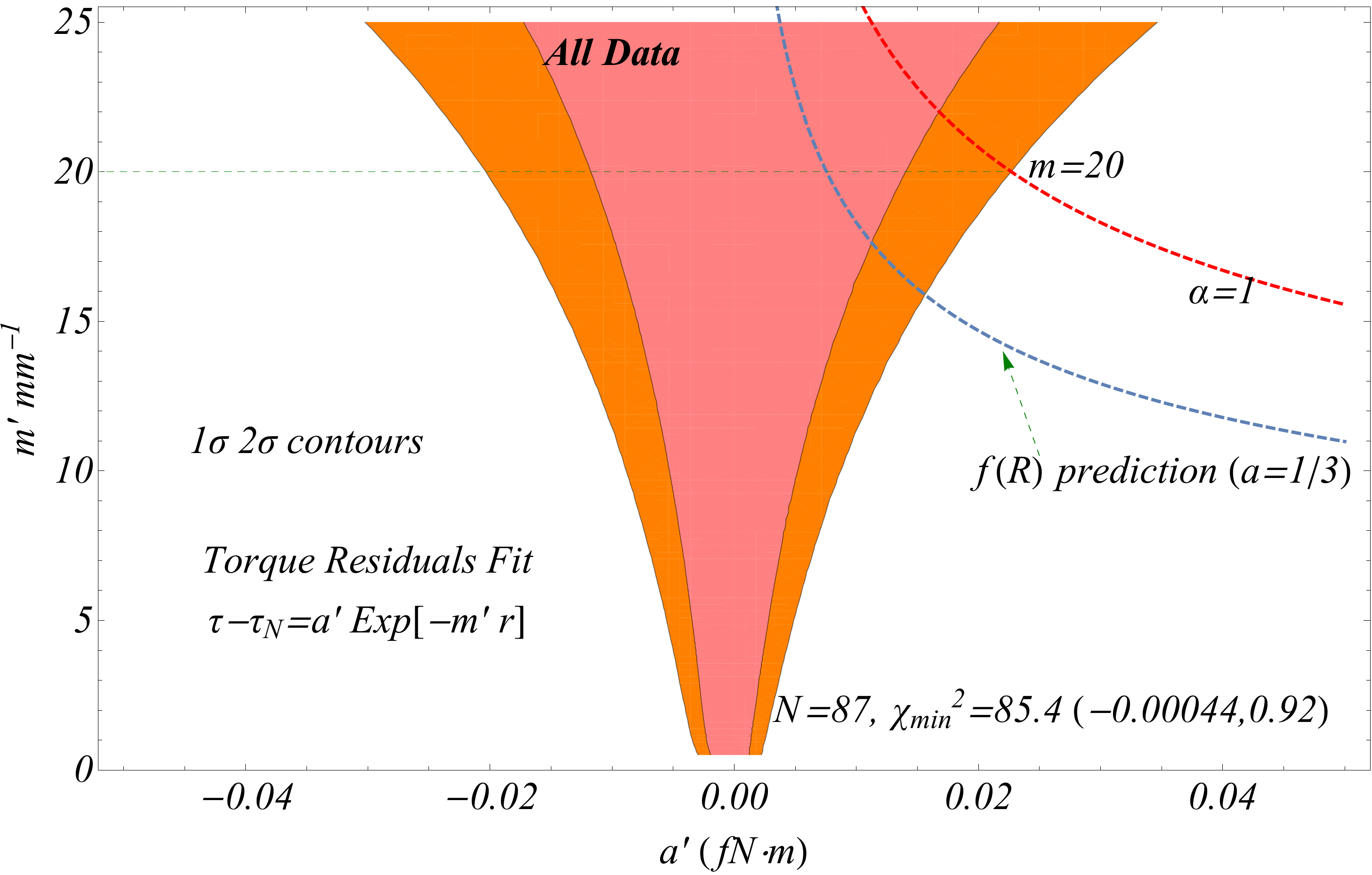}
\caption{The $1\sigma$ and $2\sigma$ contours in the parametric space $(\alpha',m')$  for the Yukawa parametrization, \eqref{exppar}. Despite the additional parameters the quality of fit is practically not improved compared to the Newtonian ansatz. (from Ref. \cite{Perivolaropoulos:2016ucs}) }
\label{fig:contoursexp}
\end{figure}

Table \ref{tab:name} indicates that the value of $\chi^2$  for the oscillating parametrization, \ie Eq. \eqref{oscilpar}, is significantly smaller ($\delta \chi^2 \simeq -15$) compared with the other two parametrizations. For the oscillating parametrization, the best fit value of the spatial frequency $m$ was obtained as  $m\simeq 65 mm^{-1}$ corresponding to a wavelength $\lambda = \frac{2\pi}{m}\simeq 0.1 mm$. The corresponding $1\sigma$ and $2\sigma$ contours in the parametric space $(\alpha',m')$ are shown in Fig. \ref{fig:contoursoscil} for the oscillating parametrization and in Fig. \ref{fig:contoursexp} for the Yukawa parametrization which does not provide a better fit than the offset Newtonian potential. 

\begin{figure}[!h]
\centering
\includegraphics[width = 0.48\textwidth]{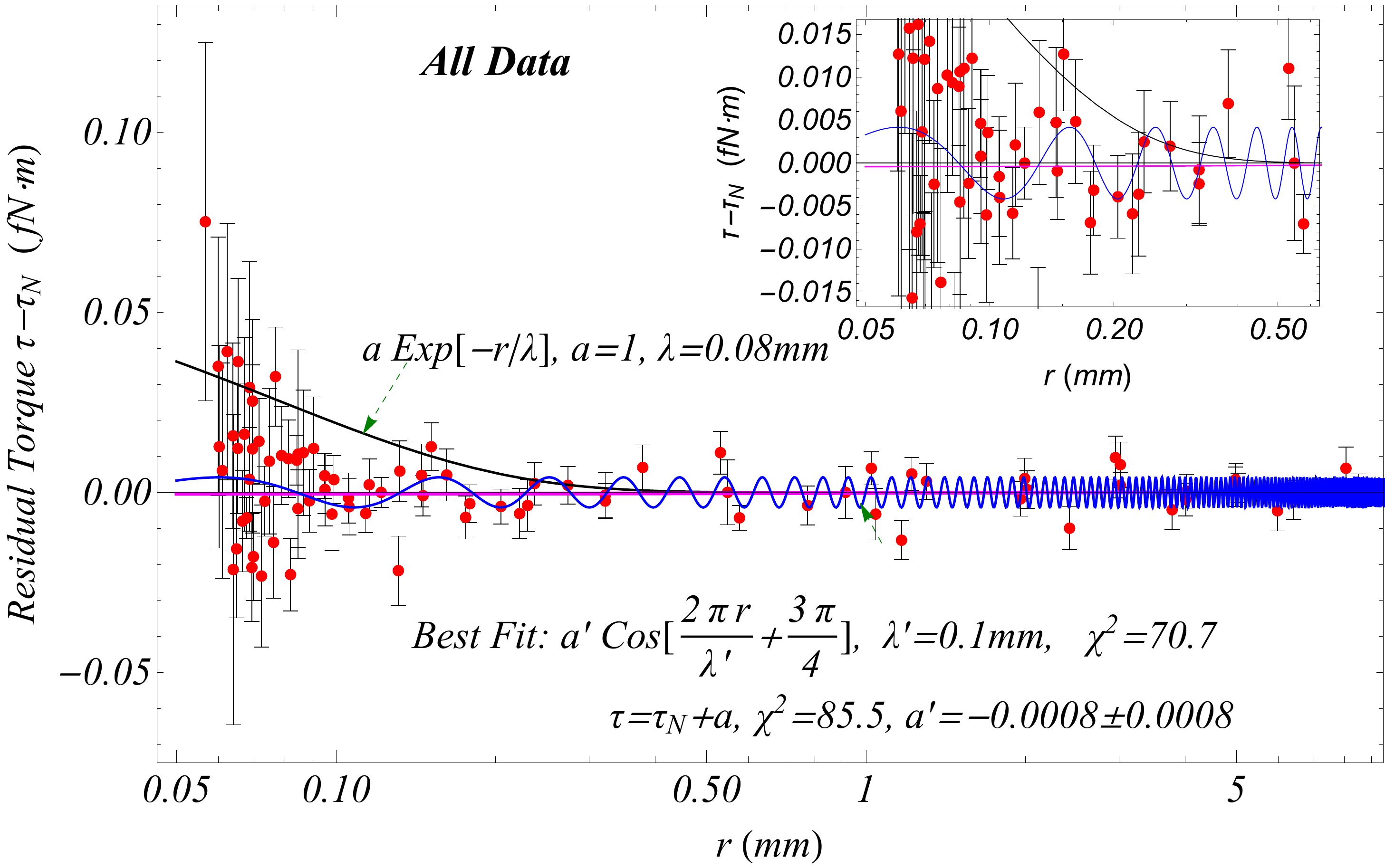}
\caption{The residual data (torques) that it was considered  in Ref. \cite{Perivolaropoulos:2016ucs}. The pink line denotes the best fit Yukawa parametrization, while the thin blue line describes the oscillating parametrization.}
\label{fig:dataplotd}
\end{figure}

The eighty seven residual datapoints superposed with the best fit Yukawa  and the oscillating parametrizations  are illustrated n Fig. \ref{fig:dataplotd} along with the best fit values of $\chi^2$.

\begin{figure}[!h]
\centering
\includegraphics[width = 0.48\textwidth]{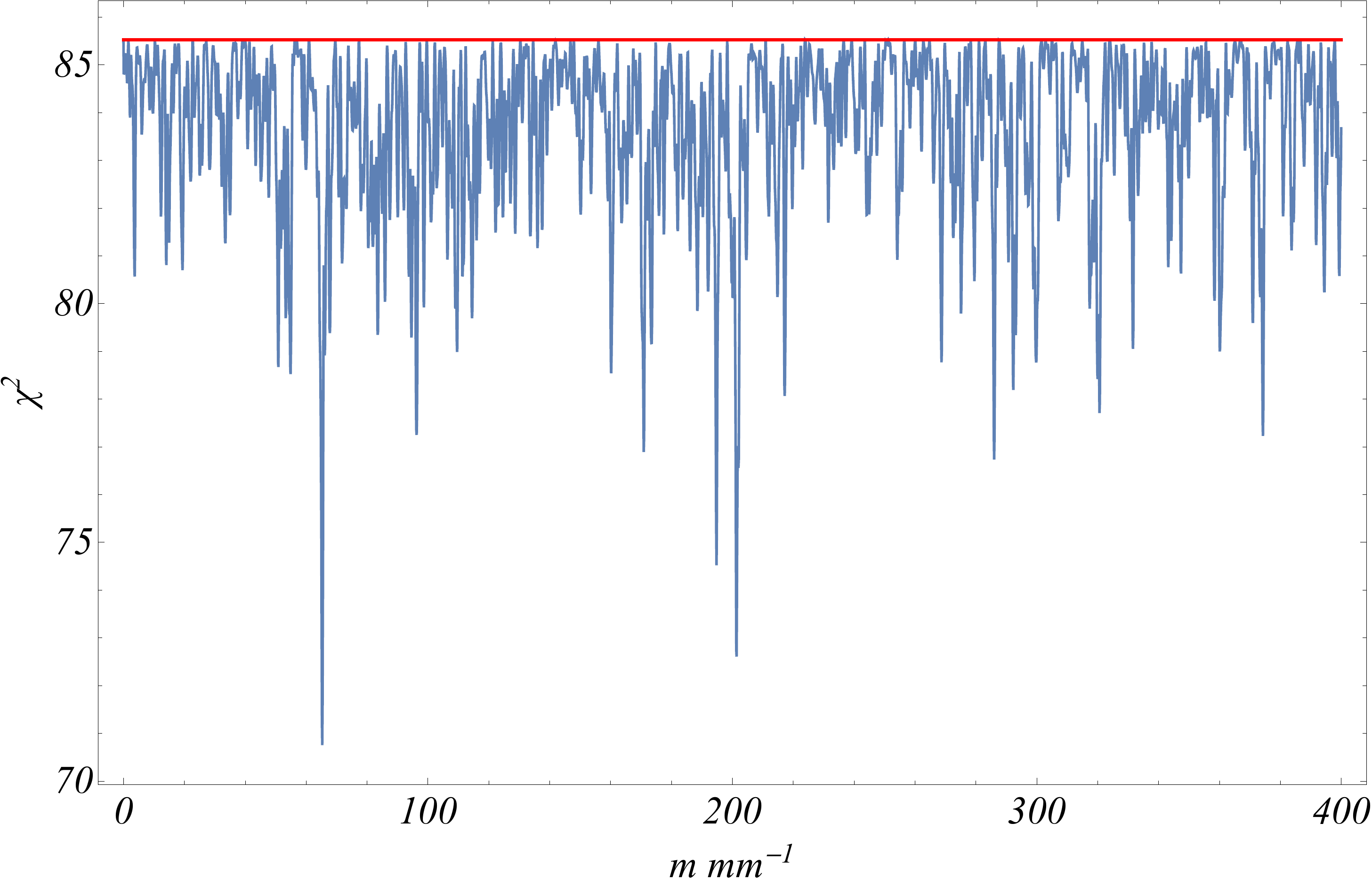}
\caption{The value of the minimized $\chi^2$ as a function of the spatial frequency $m$. The red straight line is for $\delta \tau=0$ (from  Ref. \cite{Perivolaropoulos:2016ucs}) }
\label{figchi2vsm}
\end{figure}

The statistical significance of the $\chi^2$ minimum ($\delta \chi^2 \simeq -15$) corresponding to the best fit parameters $(\alpha',m')=(0.004 fN\cdot m,65mm^{-1})$ of the oscillating parametrization is more than $3\sigma$ for a two-parameter parametrization. However, the existence of multiple minima in the parameter space $(\alpha',m')$ with similar depths reduces the statistical significance of this signal. The existence of such additional minima is demonstrated in Fig. \ref{figchi2vsm} where we show $\chi^2(m')$ where for each $m'$ minimization with respect to $\alpha'$ has been performed. For example the minima corresponding to $m'\simeq 195 mm^{-1}$ and $m'\simeq 202mm^{-1}$ have comparable depths with the main minimum at $m'\simeq 65 mm^{-1}$ but they are effectively higher harmonics of this deepest minimum.

\begin{figure}[!h]
\centering
\vspace{-0.5 cm}\includegraphics[width = 0.48\textwidth]{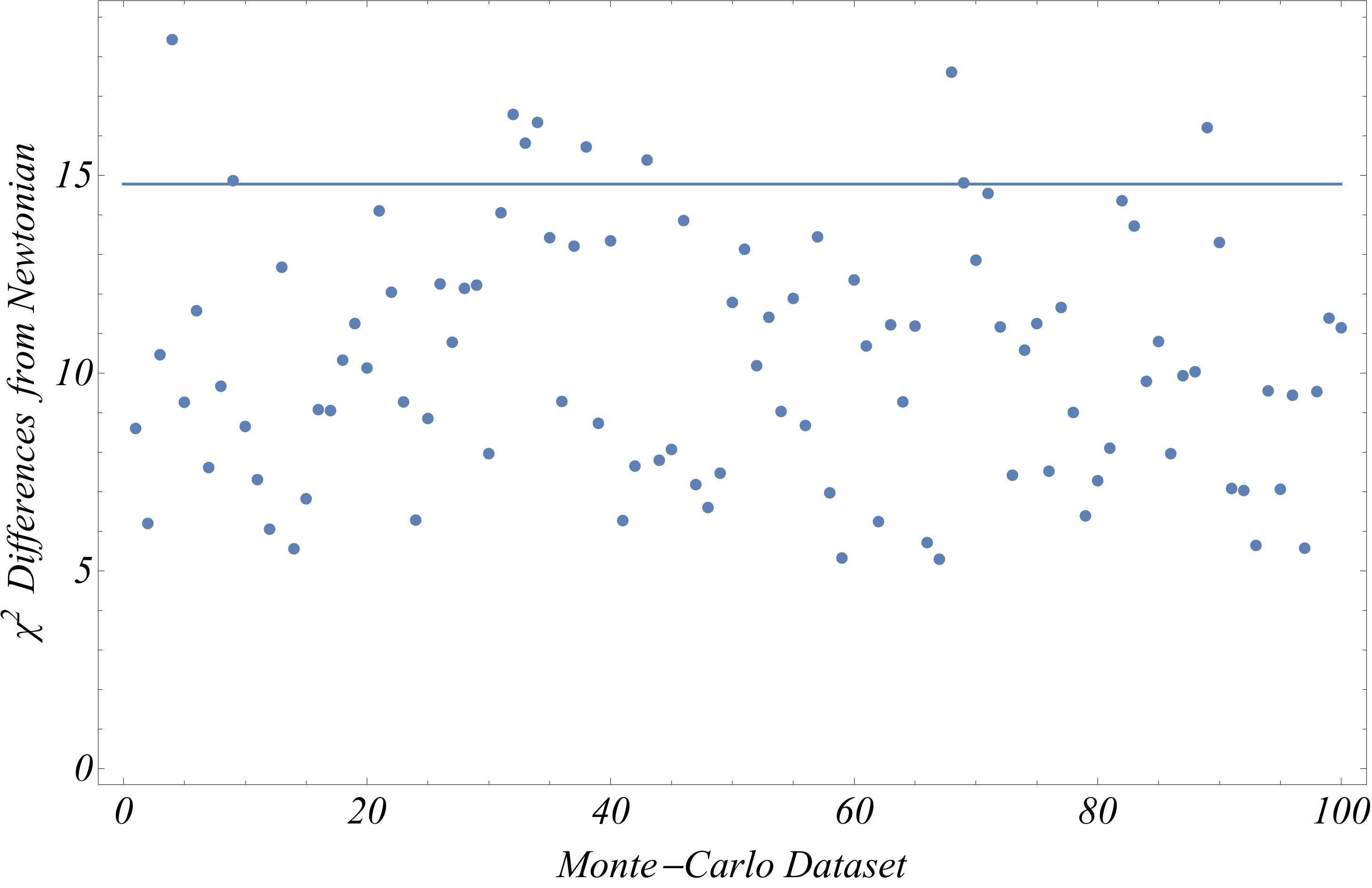}
\caption{About $10\%$ of the  Monte Carlo datasets created under the assumption of zero residuals (Newtonian model) have a $\delta \chi^2$ larger than the $\delta \chi^2$ of the real data when fitting the oscillating parametrization \eqref{oscilpar} for the torque residuals.}
\label{montecarlofits}
\end{figure}

In order to estimate the significance of the deepest $\chi^2$ minimum at $m\simeq 65 mm^{-1}$ Monte Carlo simulations were performed \cite{Perivolaropoulos:2016ucs} (100 realizations) of the 87 data from the Washington experiment assuming a Gaussian distribution of the residuals around a Newtonian potential ($\delta \tau=0$) and standard deviation equal to the errorbars of the residuals. Using each Monte-Carlo realization $\chi^2(\alpha',m',\theta)$ was minimized for $m'$ in the range $0-100mm^{-1}$ using the oscillating parametrization (\ref{oscilpar}). The depth $\delta \chi^2\equiv \chi^2(\alpha'=0)-\chi_{min}^2$  corresponding to each simulated Monte-Carlo dataset was then compared to the corresponding depth $\delta \chi^2\simeq 15$ of the real data. About $10\%$ of the simulated Newtonian data had a larger depth $\delta \chi^2$ than the real data (Fig. \ref{montecarlofits}). Thus, the probability that the oscillating signal in the Washington experiment data is a statistical fluctuation is about $10\%$.

Therefore, there is an oscillation signal in the data whose origin could be either statistical, systematic or physical. In the later case, it is important to identify physical theories that are consistent with such an oscillating signal since as discussed above such a signal is not consistent with the simplest modified gravity theories as it is associated with instabilities. As shown in the next section however a class of theories involving infinite derivatives in the Lagrangian (nonlocal theories of gravity) naturally predict the existence of such oscillations on sub-mm scales.

\subsection{Oscillating Newtonian Potential from Non-Local Gravity Theories}
The Lagrangian of non-local gravity theories may be written as \cite{Biswas:2011ar}
\begin{widetext}
\be
L_{IDG}=\frac{1}{8 \pi G} \sqrt{-g} \left[R+\alpha \left(R F_1 (\square) R+R^{\mu \nu}F_2 (\square) R_{\mu \nu}+R^{\mu \nu \rho \sigma} F_3 (\square) R_{\mu \nu \rho \sigma} \right) \right]
\ee
\end{widetext}
where
\be 
F_i(\square)= \sum_{n=0}^{\infty} f_{i,n} \left( \frac{\square}{m^2}\right)^n \quad \square=g^{\mu \nu} \nabla_\mu \nabla_\nu
\ee

Such a Lagrangian involving infinite higher derivative terms may offer the solution to some basic problems of GR, such as the behaviour of GR at small scales (GR predicts singularities at small scales). A related issue is the existence of unrenormalisable UV divergences in GR \cite{Goroff:1985th}. These divergences can be alleviated if an Einstein-Hilbert action with higher derivative terms is considered \cite{Stelle:1977ry}. These higher terms, however, are related with instabilities at the quantum level since the gravitational propagator imposed from these theories has a spin 2 component, which leads to an unhealthy classical vacuum theory (unstable). These extra problems can be resolved if we take infinite number of higher derivatives in the action, \ie making the theory nonlocal, which modifies appropriately the gravitational propagator \cite{Biswas:2011ar}. These infinite derivatives are usually condensed in an exponential term for the avoidance of introducing new poles. \cite{Tomboulis:1997gg,Siegel:2003vt,Deser:1986xr,Modesto:2011kw}

Thus nonlocal gravity theories provide the following advantages:
\begin{itemize}
\item They  soften the  UV divergences present at the quantum level along with the singularities of Big Bang and Black Holes. \cite{Biswas:2011ar,Frolov:2008uf-singularity-problem-forfR}.
\item They modify the Newtonian potential at the scale of nonlocality $m$, removing the divergences of the Newtonian potential at $r=0$, while in many cases they predict the existence decaying spatial oscillations of the gravitational potential on scales smaller than the nonlocality scale $m$. \cite{Edholm:2016hbt-nonlocal-potential-stable-spatial-oscillations,Kehagias:2014sda-nonlocal-oscillations,Frolov:2015usa-newton-potential-nonlocal,Maggiore:2014sia-nonlocal-gravity-oscil}
\item They are consistent with cosmological observations as they can provide a mechanism producing the observed accelerating expansion of the universe. \cite{Park:2012cp,Calcagni:2010ab,Barvinsky:2011hd,Dirian:2016puz}
\end{itemize}

\begin{figure}[!h]
\centering
\includegraphics[width = 0.48\textwidth]{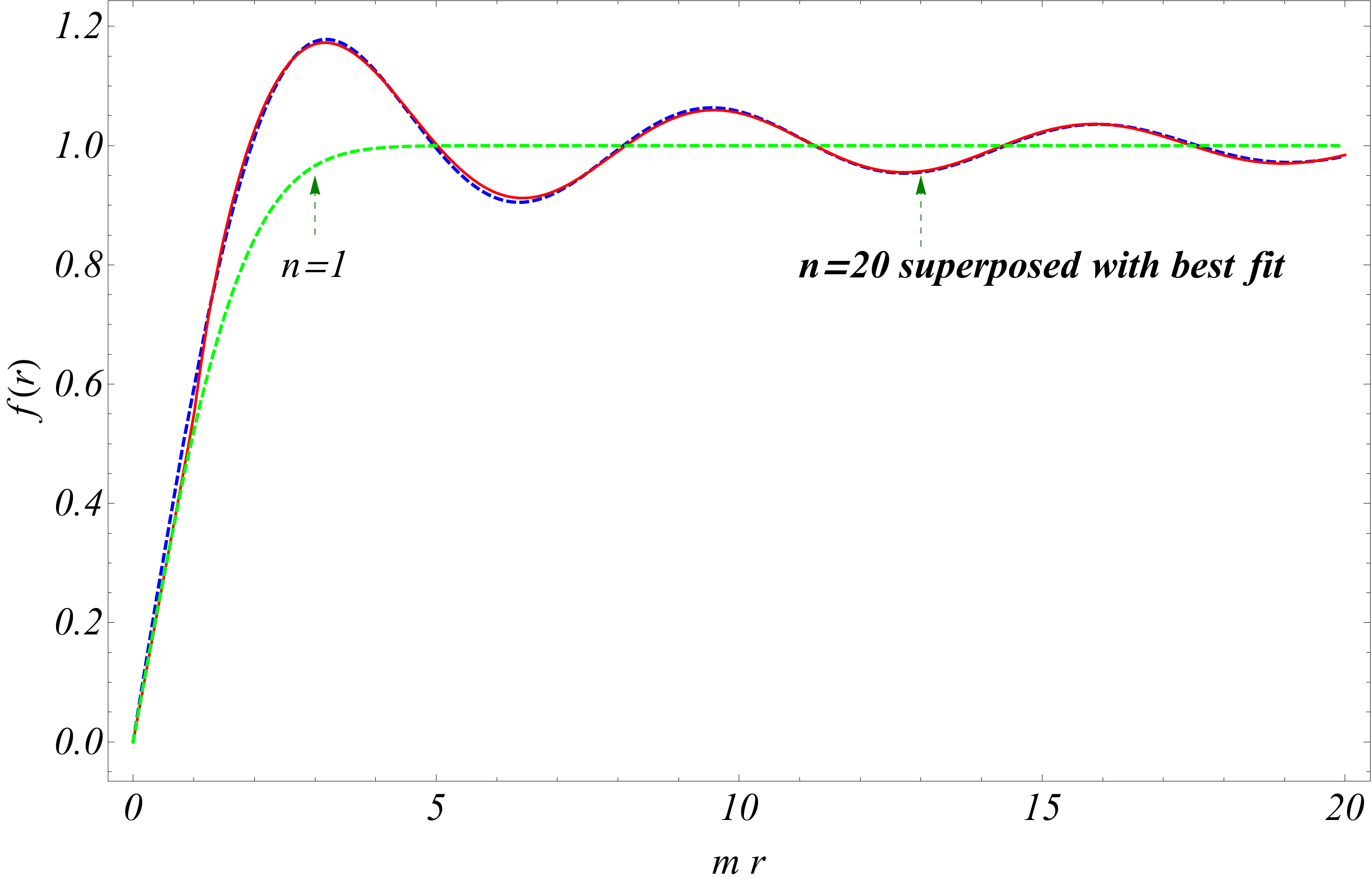}
\caption{The plot of $f(r)$ for $n=1$ and $n=20$ denoted with dashed green and red line respectively along with the fit of (\ref{fitfr1}),(\ref{fitfr2}) (from Ref. \cite{Perivolaropoulos:2016ucs})}
\label{fig:nonlocalfr}
\end{figure}

In the nonlocal theories the modified Newtonian potential is \cite{Edholm:2016hbt-nonlocal-potential-stable-spatial-oscillations}
\be 
V_{\rm eff}(r)= -\frac{G M}{r} f(r,m)
\label{newtpotnonlocal}
\ee
where
\be 
f(r,m)=\frac{1}{\pi} \int_{-\infty}^{+\infty} dk \frac{sin(k r) e^{-\tau(k,m)}}{k}
\label{frm-lonloc}
\ee
Setting $\tau$  as
\be
\tau=\frac{k^{2n}}{m^{2n}}
\label{tauform}
\ee
which is a typical form for $\tau$ and $n=1$, then Eq.\eqref{frm-lonloc} is rewritten as
\be
f(r)=Erf(m\frac{r}{2})
\ee
which for $\bar{r}\equiv m\; r<<1$ takes a linear form and for $\bar{r}\gg 1$ it approaches unity.  $f(r)$ is shown in Fig. \ref{fig:nonlocalfr}, for two different values of $n$

For large $n$, $f(r)$ can be approximated very well by the following functions \cite{Perivolaropoulos:2016ucs}
\ba  
f(r)&=&\alpha_1 \bar r \hspace{0.2cm} 0<\bar r<1 
\label{fitfr1}\\
f(r)&=&1 + \alpha_2 \frac{\cos(\bar r+\theta)}{\bar r} \hspace{0.2cm} 1<\bar r
\label{fitfr2}
\ea
where $\alpha_1=0.544$, $\alpha_2 = 0.572$, $\theta=0.885 \pi$ (see also Ref. \cite{Conroy:2017nkc} for a similar parametrization).

Such models are interesting since not only they are free from UV divergences and singularities but the have a well-defined Newtonian limit. Therefore, this type of oscillating behavior may have been the origin of the oscillating signal in the data of the Washington experiment discussed in the previous sub-section.

\section{Conclusions}
\label{sec:concl}

In the present brief review, we discussed experimental and observational data on the smallest and the largest scales where gravity can be directly probed with current technology. We demonstrated that current data indicate the presence of hints of modified gravity in both the cosmological and the sub-millimeter scales. Concerning the cosmological scales, we showed that the best fit \plcdm $\sigma_8-\Omega_m$ parameter values are more than $3 \sigma$ away from the corresponding best fit parameter values obtained using the latest RSD growth rate data $\fs$, assuming a \plcdm background cosmology \cite{Nesseris:2017vor}. This tension has also been observed from various weak gravitational lensing analyses \cite{Joudaki:2017zdt, Abbott:2017wau}.

The tension can be reduced either by reducing the value of $\Omega_m$ in the context of a \lcdm cosmology or by allowing for an evolving Newton's constant  \geffz leading to weaker gravity at $z\simeq 1$. In particular we showed that a \plcdm $H(z)$ cosmological background with a well motivated form of $G_{\rm eff}(z)=1+g_a\left(\frac{z}{1+z}\right)^n - g_a\left(\frac{z}{1+z}\right)^{2n}$, which is a decreasing function of $z$ (for $g_a<0$), can be significantly more consistent with the full dataset of $\fs$ from Ref. \cite{Kazantzidis:2018rnb}. This type of evolution cannot be reproduced in scalar-tensor theories with a \lcdm background, since it leads to negative kinetic term of the scalar field $\phi(z)$. One possible way to reproduce a decreasing \geffz in scalar-tensor theories would be to assume  for a \wcdm expansion background with $w>-1$. For example we demonstrated that for a  \wcdm background, \geffz can be a decreasing function of $z$ in the context of scalar-tensor theories.

Finally on sub-mm scales higher derivative gravity models generically predict sub-mm spatial oscillations of the gravitational potential. Hints for such oscillations have been demonstrated to exist in the Washington torsion-balance experiment. 

Thus we have demonstrated the existence of  hints for deviations from GR on both the largest scales where a $G_{\rm eff}(z<1)<G_N$ is favored and on the smallest probed scales (sub-mm) where an oscillating $G_{\rm eff}(r)$ is favored. It is therefore important to clarify if these hints are due to systematic or statistical effects or they constitute early manifestation for new physics. This clarification may be achieved by considering new larger datasets  focusing on the redshifts/scales where these hints appear $(z\simeq 0.3 \text{ and } r\simeq 80\mu m). $

\section*{Acknowledgements}
This research is co-financed by Greece and the European Union (European Social Fund- ESF) through the Operational Programme ``Human Resources Development, Education and Lifelong Learning" in the context of the project ``Strengthening Human Resources Research Potential via Doctorate Research" (MIS-5000432), implemented by the State Scholarships Foundation (IKY)

\appendix

\section{Data Used in the Analysis}
\label{sec:appA}
\begin{widetext}
\begin{longtable}{ | c | c | c | c | c | c | c | }
\caption{The $\fs$ data compilation from Ref. \cite{Kazantzidis:2018rnb}} 
\label{tab:fs8-data-kazan}\\
% header and footer information
\hline
    Index & Dataset & $z$ & $f\sigma_8(z)$ & Refs. & Year & Fiducial Cosmology \\
\hline
1 & SDSS-LRG & $0.35$ & $0.440\pm 0.050$ & \cite{Song:2008qt} &  30 October 2006 &$(\Omega_{m},\Omega_K,\sigma_8$)$=(0.25,0,0.756)$\cite{Tegmark:2006az} \\

2 & VVDS & $0.77$ & $0.490\pm 0.18$ & \cite{Song:2008qt}  & 6 October 2009 & $(\Omega_{m},\Omega_K,\sigma_8)=(0.25,0,0.78)$ \\

3 & 2dFGRS & $0.17$ & $0.510\pm 0.060$ & \cite{Song:2008qt}  &  6 October 2009 & $(\Omega_{m},\Omega_K)=(0.3,0,0.9)$ \\

4 & 2MRS &0.02& $0.314 \pm 0.048$ &  \cite{Davis:2010sw}, \cite{Hudson:2012gt}& 13 Novemver 2010 & $(\Omega_{m},\Omega_K,\sigma_8)=(0.266,0,0.65)$ \\

5 & SnIa+IRAS &0.02& $0.398 \pm 0.065$ & \cite{Turnbull:2011ty}, \cite{Hudson:2012gt}& 20 October 2011 & $(\Omega_{m},\Omega_K,\sigma_8)=(0.3,0,0.814)$\\

6 & SDSS-LRG-200 & $0.25$ & $0.3512\pm 0.0583$ & \cite{Samushia:2011cs} & 9 December 2011 & $(\Omega_{m},\Omega_K,\sigma_8)=(0.276,0,0.8)$  \\

7 & SDSS-LRG-200 & $0.37$ & $0.4602\pm 0.0378$ & \cite{Samushia:2011cs} & 9 December 2011 & \\

8 & SDSS-LRG-60 & $0.25$ & $0.3665\pm0.0601$ & \cite{Samushia:2011cs} & 9 December 2011 & $(\Omega_{m},\Omega_K,\sigma_8)=(0.276,0,0.8)$ \\

9 & SDSS-LRG-60 & $0.37$ & $0.4031\pm0.0586$ & \cite{Samushia:2011cs} & 9 December 2011 &\\

10 & WiggleZ & $0.44$ & $0.413\pm 0.080$ & \cite{Blake:2012pj} & 12 June 2012  & $(\Omega_{m},h,\sigma_8)=(0.27,0.71,0.8)$ \\

11 & WiggleZ & $0.60$ & $0.390\pm 0.063$ & \cite{Blake:2012pj} & 12 June 2012 &  \\

12 & WiggleZ & $0.73$ & $0.437\pm 0.072$ & \cite{Blake:2012pj} & 12 June 2012 &\\

13 & 6dFGS& $0.067$ & $0.423\pm 0.055$ & \cite{Beutler:2012px} & 4 July 2012 & $(\Omega_{m},\Omega_K,\sigma_8)=(0.27,0,0.76)$ \\

14 & SDSS-BOSS& $0.30$ & $0.407\pm 0.055$ & \cite{Tojeiro:2012rp} & 11 August 2012 & $(\Omega_{m},\Omega_K,\sigma_8)=(0.25,0,0.804)$ \\

15 & SDSS-BOSS& $0.40$ & $0.419\pm 0.041$ & \cite{Tojeiro:2012rp} & 11 August 2012 & \\

16 & SDSS-BOSS& $0.50$ & $0.427\pm 0.043$ & \cite{Tojeiro:2012rp} & 11 August 2012 & \\

17 & SDSS-BOSS& $0.60$ & $0.433\pm 0.067$ & \cite{Tojeiro:2012rp} & 11 August 2012 & \\

18 & Vipers& $0.80$ & $0.470\pm 0.080$ & \cite{delaTorre:2013rpa} & 9 July 2013 & $(\Omega_{m},\Omega_K,\sigma_8)=(0.25,0,0.82)$  \\

19 & SDSS-DR7-LRG & $0.35$ & $0.429\pm 0.089$ & \cite{Chuang:2012qt}  & 8 August 2013 & $(\Omega_{m},\Omega_K,\sigma_8$)$=(0.25,0,0.809)$\cite{Komatsu:2010fb}\\

20 & GAMA & $0.18$ & $0.360\pm 0.090$ & \cite{Blake:2013nif}  & 22 September 2013 & $(\Omega_{m},\Omega_K,\sigma_8)=(0.27,0,0.8)$ \\

21& GAMA & $0.38$ & $0.440\pm 0.060$ & \cite{Blake:2013nif}  & 22 September 2013 & \\

22 & BOSS-LOWZ& $0.32$ & $0.384\pm 0.095$ & \cite{Sanchez:2013tga}  & 17 December 2013  & $(\Omega_{m},\Omega_K,\sigma_8)=(0.274,0,0.8)$ \\

23 & SDSS DR10 and DR11 & $0.32$ & $0.48 \pm 0.10$ & \cite{Sanchez:2013tga} &   17 December 2013 & $(\Omega_{m},\Omega_K,\sigma_8$)$=(0.274,0,0.8)$\cite{Anderson:2013zyy}\\

24 & SDSS DR10 and DR11 & $0.57$ & $0.417 \pm 0.045$ & \cite{Sanchez:2013tga} &  17 December 2013 &  \\

25 & SDSS-MGS & $0.15$ & $0.490\pm0.145$ & \cite{Howlett:2014opa} & 30 January 2015 & $(\Omega_{m},h,\sigma_8)=(0.31,0.67,0.83)$ \\

26 & SDSS-veloc & $0.10$ & $0.370\pm 0.130$ & \cite{Feix:2015dla}  & 16 June 2015 & $(\Omega_{m},\Omega_K,\sigma_8$)$=(0.3,0,0.89)$\cite{Tegmark:2003uf} \\

27 & FastSound& $1.40$ & $0.482\pm 0.116$ & \cite{Okumura:2015lvp}  & 25 November 2015 & $(\Omega_{m},\Omega_K,\sigma_8$)$=(0.27,0,0.82)$\cite{Hinshaw:2012aka} \\

28 & SDSS-CMASS & $0.59$ & $0.488\pm 0.060$ & \cite{Chuang:2013wga} & 8 July 2016 & $\ \ (\Omega_{m},h,\sigma_8)=(0.307115,0.6777,0.8288)$ \\

29 & BOSS DR12 & $0.38$ & $0.497\pm 0.045$ & \cite{Alam:2016hwk} & 11 July 2016 & $(\Omega_{m},\Omega_K,\sigma_8)=(0.31,0,0.8)$ \\

30 & BOSS DR12 & $0.51$ & $0.458\pm 0.038$ & \cite{Alam:2016hwk} & 11 July 2016 & \\

31 & BOSS DR12 & $0.61$ & $0.436\pm 0.034$ & \cite{Alam:2016hwk} & 11 July 2016 & \\

32 & BOSS DR12 & $0.38$ & $0.477 \pm 0.051$ & \cite{Beutler:2016arn} & 11 July 2016 & $(\Omega_{m},h,\sigma_8)=(0.31,0.676,0.8)$ \\

33 & BOSS DR12 & $0.51$ & $0.453 \pm 0.050$ & \cite{Beutler:2016arn} & 11 July 2016 & \\

34 & BOSS DR12 & $0.61$ & $0.410 \pm 0.044$ & \cite{Beutler:2016arn} & 11 July 2016 &  \\

35 &Vipers v7& $0.76$ & $0.440\pm 0.040$ & \cite{Wilson:2016ggz} & 26 October 2016  & $(\Omega_{m},\sigma_8)=(0.308,0.8149)$ \\

36 &Vipers v7 & $1.05$ & $0.280\pm 0.080$ & \cite{Wilson:2016ggz} & 26 October 2016 &\\

37 &  BOSS LOWZ & $0.32$ & $0.427\pm 0.056$ & \cite{Gil-Marin:2016wya} & 26 October 2016 & $(\Omega_{m},\Omega_K,\sigma_8)=(0.31,0,0.8475)$\\

38 & BOSS CMASS & $0.57$ & $0.426\pm 0.029$ & \cite{Gil-Marin:2016wya} & 26 October 2016 & \\

39 & Vipers  & $0.727$ & $0.296 \pm 0.0765$ & \cite{Hawken:2016qcy} &  21 November 2016 & $(\Omega_{m},\Omega_K,\sigma_8)=(0.31,0,0.7)$\\

40 & 6dFGS+SnIa & $0.02$ & $0.428\pm 0.0465$ & \cite{Huterer:2016uyq} & 29 November 2016 & $(\Omega_{m},h,\sigma_8)=(0.3,0.683,0.8)$ \\

41 & Vipers  & $0.6$ & $0.48 \pm 0.12$ & \cite{delaTorre:2016rxm} & 16 December 2016 & $(\Omega_{m},\Omega_b,n_s,\sigma_8$)= $(0.3, 0.045, 0.96,0.831)$\cite{Ade:2015xua} \\

42 & Vipers  & $0.86$ & $0.48 \pm 0.10$ & \cite{delaTorre:2016rxm} & 16 December 2016  & \\

43 &Vipers PDR-2& $0.60$ & $0.550\pm 0.120$ & \cite{Pezzotta:2016gbo} & 16 December 2016 & $(\Omega_{m},\Omega_b,\sigma_8)=(0.3,0.045,0.823)$ \\

44 & Vipers PDR-2& $0.86$ & $0.400\pm 0.110$ & \cite{Pezzotta:2016gbo} & 16 December 2016 &\\

45 & SDSS DR13  & $0.1$ & $0.48 \pm 0.16$ & \cite{Feix:2016qhh} & 22 December 2016 & $(\Omega_{m},\sigma_8$)$=(0.25,0.89)$\cite{Tegmark:2003uf} \\

46 & 2MTF & 0.001 & $0.505 \pm 0.085$ &  \cite{Howlett:2017asq} & 16 June 2017 & $(\Omega_{m},\sigma_8)=(0.3121,0.815)$\\

47 & Vipers PDR-2 & $0.85$ & $0.45 \pm 0.11$ & \cite{Mohammad:2017lzz} & 31 July 2017  &  $(\Omega_b,\Omega_{m},h)=(0.045,0.30,0.8)$ \\

48 & BOSS DR12 & $0.31$ & $0.469 \pm 0.098$ &  \cite{Wang:2017wia} & 15 September 2017 & $(\Omega_{m},h,\sigma_8)=(0.307,0.6777,0.8288)$\\

49 & BOSS DR12 & $0.36$ & $0.474 \pm 0.097$ &  \cite{Wang:2017wia} & 15 September 2017 & \\

50 & BOSS DR12 & $0.40$ & $0.473 \pm 0.086$ &  \cite{Wang:2017wia} & 15 September 2017 & \\

51 & BOSS DR12 & $0.44$ & $0.481 \pm 0.076$ &  \cite{Wang:2017wia} & 15 September 2017 & \\

52 & BOSS DR12 & $0.48$ & $0.482 \pm 0.067$ &  \cite{Wang:2017wia} & 15 September 2017 & \\

53 & BOSS DR12 & $0.52$ & $0.488 \pm 0.065$ &  \cite{Wang:2017wia} & 15 September 2017 & \\

54 & BOSS DR12 & $0.56$ & $0.482 \pm 0.067$ &  \cite{Wang:2017wia} & 15 September 2017 & \\

55 & BOSS DR12 & $0.59$ & $0.481 \pm 0.066$ &  \cite{Wang:2017wia} & 15 September 2017 & \\

56 & BOSS DR12 & $0.64$ & $0.486 \pm 0.070$ &  \cite{Wang:2017wia} & 15 September 2017 & \\

57 & SDSS DR7 & $0.1$ & $0.376\pm 0.038$ & \cite{Shi:2017qpr} & 12 December 2017 & $(\Omega_{m},\Omega_b,\sigma_8)=(0.282,0.046,0.817)$ \\

58 & SDSS-IV & $1.52$ & $0.420 \pm 0.076$ &  \cite{Gil-Marin:2018cgo} & 8 January 2018  & $(\Omega_{m},\Omega_b h^2,\sigma_8)=(0.26479, 0.02258,0.8)$ \\ 

59 & SDSS-IV & $1.52$ & $0.396 \pm 0.079$ & \cite{Hou:2018yny} & 8 January 2018 & $(\Omega_{m},\Omega_b h^2,\sigma_8)=(0.31,0.022,0.8225)$ \\ 

60 & SDSS-IV & $0.978$ & $0.379 \pm 0.176$ &  \cite{Zhao:2018jxv} & 9 January 2018 &$(\Omega_{m},\sigma_8)=(0.31,0.8)$\\

61 & SDSS-IV & $1.23$ & $0.385 \pm 0.099$ &  \cite{Zhao:2018jxv} & 9 January 2018 & \\

62 & SDSS-IV & $1.526$ & $0.342 \pm 0.070$ &  \cite{Zhao:2018jxv} & 9 January 2018 & \\

63 & SDSS-IV & $1.944$ & $0.364 \pm 0.106$ &  \cite{Zhao:2018jxv} & 9 January 2018 & \\
\hline
\end{longtable}

\begin{longtable}{ | c | c | c | c | c | c | c | }
\caption{A compilation of robust and independent $f\sigma_8(z)$ measurements from different surveys. In the columns we show in ascending order with respect to redshift, the name and year of the survey that made the measurement, the redshift and value of $f\sigma_8(z)$ and the corresponding reference and fiducial cosmology} 
\label{tab:fs8-data-gold}\\
% header and footer information
\hline
    Index & Dataset & $z$ & $f\sigma_8(z)$ & Refs. & Year & Fiducial Cosmology \\
\hline
1 & 6dFGS+SnIa & $0.02$ & $0.428\pm 0.0465$ & \cite{Huterer:2016uyq} & 2016 & $(\Omega_m,h,\sigma_8)=(0.3,0.683,0.8)$ \\

2 & SnIa+IRAS &0.02& $0.398 \pm 0.065$ &  \cite{Turnbull:2011ty},\cite{Hudson:2012gt} & 2011& $(\Omega_m,\Omega_K)=(0.3,0)$\\

3 & 2MASS &0.02& $0.314 \pm 0.048$ &  \cite{Davis:2010sw},\cite{Hudson:2012gt} & 2010& $(\Omega_m,\Omega_K)=(0.266,0)$ \\

4 & SDSS-veloc & $0.10$ & $0.370\pm 0.130$ & \cite{Feix:2015dla}  &2015 &$(\Omega_m,\Omega_K)=(0.3,0)$ \\

5 & SDSS-MGS & $0.15$ & $0.490\pm0.145$ & \cite{Howlett:2014opa} & 2014& $(\Omega_m,h,\sigma_8)=(0.31,0.67,0.83)$ \\

6 & 2dFGRS & $0.17$ & $0.510\pm 0.060$ & \cite{Song:2008qt}  & 2009& $(\Omega_m,\Omega_K)=(0.3,0)$ \\

7 & GAMA & $0.18$ & $0.360\pm 0.090$ & \cite{Blake:2013nif}  & 2013& $(\Omega_m,\Omega_K)=(0.27,0)$ \\

8 & GAMA & $0.38$ & $0.440\pm 0.060$ & \cite{Blake:2013nif}  & 2013& \\

9 &SDSS-LRG-200 & $0.25$ & $0.3512\pm 0.0583$ & \cite{Samushia:2011cs} & 2011& $(\Omega_m,\Omega_K)=(0.25,0)$  \\

10 &SDSS-LRG-200 & $0.37$ & $0.4602\pm 0.0378$ & \cite{Samushia:2011cs} & 2011& \\

11 &BOSS-LOWZ& $0.32$ & $0.384\pm 0.095$ & \cite{Sanchez:2013tga}  &2013 & $(\Omega_m,\Omega_K)=(0.274,0)$ \\

12 & SDSS-CMASS & $0.59$ & $0.488\pm 0.060$ & \cite{Chuang:2013wga} &2013& $\ \ (\Omega_m,h,\sigma_8)=(0.307115,0.6777,0.8288)$ \\

13 &WiggleZ & $0.44$ & $0.413\pm 0.080$ & \cite{Blake:2012pj} & 2012&$(\Omega_m,h)=(0.27,0.71)$ \\

14 &WiggleZ & $0.60$ & $0.390\pm 0.063$ & \cite{Blake:2012pj} & 2012& \\

15 &WiggleZ & $0.73$ & $0.437\pm 0.072$ & \cite{Blake:2012pj} & 2012 &\\

16 &Vipers PDR-2& $0.60$ & $0.550\pm 0.120$ & \cite{Pezzotta:2016gbo} & 2016& $(\Omega_m,\Omega_b)=(0.3,0.045)$ \\

17 &Vipers PDR-2& $0.86$ & $0.400\pm 0.110$ & \cite{Pezzotta:2016gbo} & 2016&\\

18 &FastSound& $1.40$ & $0.482\pm 0.116$ & \cite{Okumura:2015lvp}  & 2015& $(\Omega_m,\Omega_K)=(0.270,0)$\\
\hline
\end{longtable}
\end{widetext}

\raggedleft
\bibliography{Bibliography}

\begin{thebibliography}{149}%
\makeatletter
\providecommand \@ifxundefined [1]{%
 \@ifx{#1\undefined}
}%
\providecommand \@ifnum [1]{%
 \ifnum #1\expandafter \@firstoftwo
 \else \expandafter \@secondoftwo
 \fi
}%
\providecommand \@ifx [1]{%
 \ifx #1\expandafter \@firstoftwo
 \else \expandafter \@secondoftwo
 \fi
}%
\providecommand \natexlab [1]{#1}%
\providecommand \enquote  [1]{``#1''}%
\providecommand \bibnamefont  [1]{#1}%
\providecommand \bibfnamefont [1]{#1}%
\providecommand \citenamefont [1]{#1}%
\providecommand \href@noop [0]{\@secondoftwo}%
\providecommand \href [0]{\begingroup \@sanitize@url \@href}%
\providecommand \@href[1]{\@@startlink{#1}\@@href}%
\providecommand \@@href[1]{\endgroup#1\@@endlink}%
\providecommand \@sanitize@url [0]{\catcode `\\12\catcode `\$12\catcode
  `\&12\catcode `\#12\catcode `\^12\catcode `\_12\catcode `\%12\relax}%
\providecommand \@@startlink[1]{}%
\providecommand \@@endlink[0]{}%
\providecommand \url  [0]{\begingroup\@sanitize@url \@url }%
\providecommand \@url [1]{\endgroup\@href {#1}{\urlprefix }}%
\providecommand \urlprefix  [0]{URL }%
\providecommand \Eprint [0]{\href }%
\providecommand \doibase [0]{http://dx.doi.org/}%
\providecommand \selectlanguage [0]{\@gobble}%
\providecommand \bibinfo  [0]{\@secondoftwo}%
\providecommand \bibfield  [0]{\@secondoftwo}%
\providecommand \translation [1]{[#1]}%
\providecommand \BibitemOpen [0]{}%
\providecommand \bibitemStop [0]{}%
\providecommand \bibitemNoStop [0]{.\EOS\space}%
\providecommand \EOS [0]{\spacefactor3000\relax}%
\providecommand \BibitemShut  [1]{\csname bibitem#1\endcsname}%
\let\auto@bib@innerbib\@empty
%</preamble>
\bibitem [{\citenamefont {Kapner}\ \emph
  {et~al.}(2007{\natexlab{a}})\citenamefont {Kapner}, \citenamefont {Cook},
  \citenamefont {Adelberger}, \citenamefont {Gundlach}, \citenamefont {Heckel},
  \citenamefont {Hoyle},\ and\ \citenamefont
  {Swanson}}]{Kapner:2006si-washington3}%
  \BibitemOpen
  \bibfield  {author} {\bibinfo {author} {\bibfnamefont {D.~J.}\ \bibnamefont
  {Kapner}}, \bibinfo {author} {\bibfnamefont {T.~S.}\ \bibnamefont {Cook}},
  \bibinfo {author} {\bibfnamefont {E.~G.}\ \bibnamefont {Adelberger}},
  \bibinfo {author} {\bibfnamefont {J.~H.}\ \bibnamefont {Gundlach}}, \bibinfo
  {author} {\bibfnamefont {Blayne~R.}\ \bibnamefont {Heckel}}, \bibinfo
  {author} {\bibfnamefont {C.~D.}\ \bibnamefont {Hoyle}}, \ and\ \bibinfo
  {author} {\bibfnamefont {H.~E.}\ \bibnamefont {Swanson}},\ }\bibfield
  {title} {\enquote {\bibinfo {title} {{Tests of the gravitational
  inverse-square law below the dark-energy length scale}},}\ }\href {\doibase
  10.1103/PhysRevLett.98.021101} {\bibfield  {journal} {\bibinfo  {journal}
  {Phys. Rev. Lett.}\ }\textbf {\bibinfo {volume} {98}},\ \bibinfo {pages}
  {021101} (\bibinfo {year} {2007}{\natexlab{a}})},\ \Eprint
  {http://arxiv.org/abs/hep-ph/0611184} {arXiv:hep-ph/0611184 [hep-ph]}
  \BibitemShut {NoStop}%
%%CITATION = HEP-PH/0611184;%%
\bibitem [{\citenamefont {Will}(2014)}]{Will:2014kxa}%
  \BibitemOpen
  \bibfield  {author} {\bibinfo {author} {\bibfnamefont {Clifford~M.}\
  \bibnamefont {Will}},\ }\bibfield  {title} {\enquote {\bibinfo {title} {{The
  Confrontation between General Relativity and Experiment}},}\ }\href {\doibase
  10.12942/lrr-2014-4} {\bibfield  {journal} {\bibinfo  {journal} {Living Rev.
  Rel.}\ }\textbf {\bibinfo {volume} {17}},\ \bibinfo {pages} {4} (\bibinfo
  {year} {2014})},\ \Eprint {http://arxiv.org/abs/1403.7377} {arXiv:1403.7377
  [gr-qc]} \BibitemShut {NoStop}%
%%CITATION = ARXIV:1403.7377;%%
\bibitem [{\citenamefont {Anderson}\ \emph {et~al.}(2002)\citenamefont
  {Anderson}, \citenamefont {Laing}, \citenamefont {Lau}, \citenamefont
  {Nieto},\ and\ \citenamefont {Turyshev}}]{Anderson:2001ks}%
  \BibitemOpen
  \bibfield  {author} {\bibinfo {author} {\bibfnamefont {John~D.}\ \bibnamefont
  {Anderson}}, \bibinfo {author} {\bibfnamefont {Philip~A.}\ \bibnamefont
  {Laing}}, \bibinfo {author} {\bibfnamefont {Eunice~L.}\ \bibnamefont {Lau}},
  \bibinfo {author} {\bibfnamefont {Michael~Martin}\ \bibnamefont {Nieto}}, \
  and\ \bibinfo {author} {\bibfnamefont {Slava~G.}\ \bibnamefont {Turyshev}},\
  }\bibfield  {title} {\enquote {\bibinfo {title} {{The search for a standard
  explanation of the Pioneer anomaly}},}\ }\href {\doibase
  10.1142/S0217732302007107} {\bibfield  {journal} {\bibinfo  {journal} {Mod.
  Phys. Lett.}\ }\textbf {\bibinfo {volume} {A17}},\ \bibinfo {pages}
  {875--886} (\bibinfo {year} {2002})},\ \Eprint
  {http://arxiv.org/abs/gr-qc/0107022} {arXiv:gr-qc/0107022 [gr-qc]}
  \BibitemShut {NoStop}%
%%CITATION = GR-QC/0107022;%%
\bibitem [{\citenamefont {{Sokolov}}(2016)}]{Sokolov}%
  \BibitemOpen
  \bibfield  {author} {\bibinfo {author} {\bibfnamefont {V.~V.}\ \bibnamefont
  {{Sokolov}}},\ }\bibfield  {title} {\enquote {\bibinfo {title} {{On the
  observed mass distribution of compact stellar remnants in close binary
  systems and possible interpretations proposed for the time being}},}\ }in\
  \href@noop {} {\emph {\bibinfo {booktitle} {Quark Phase Transition in Compact
  Objects and Multimessenger Astronomy: Neutrino Signals, Supernovae and
  Gamma-Ray Bursts}}},\ \bibinfo {editor} {edited by\ \bibinfo {editor}
  {\bibfnamefont {V.~V.}\ \bibnamefont {{Sokolov}}}, \bibinfo {editor}
  {\bibfnamefont {V.~V.}\ \bibnamefont {{Vlasyuk}}}, \ and\ \bibinfo {editor}
  {\bibfnamefont {V.~B.}\ \bibnamefont {{Petkov}}}}\ (\bibinfo {year} {2016})\
  pp.\ \bibinfo {pages} {121--132}\BibitemShut {NoStop}%
\bibitem [{\citenamefont {Jungman}\ \emph {et~al.}(1996)\citenamefont
  {Jungman}, \citenamefont {Kamionkowski},\ and\ \citenamefont
  {Griest}}]{Jungman:1995df}%
  \BibitemOpen
  \bibfield  {author} {\bibinfo {author} {\bibfnamefont {Gerard}\ \bibnamefont
  {Jungman}}, \bibinfo {author} {\bibfnamefont {Marc}\ \bibnamefont
  {Kamionkowski}}, \ and\ \bibinfo {author} {\bibfnamefont {Kim}\ \bibnamefont
  {Griest}},\ }\bibfield  {title} {\enquote {\bibinfo {title} {{Supersymmetric
  dark matter}},}\ }\href {\doibase 10.1016/0370-1573(95)00058-5} {\bibfield
  {journal} {\bibinfo  {journal} {Phys. Rept.}\ }\textbf {\bibinfo {volume}
  {267}},\ \bibinfo {pages} {195--373} (\bibinfo {year} {1996})},\ \Eprint
  {http://arxiv.org/abs/hep-ph/9506380} {arXiv:hep-ph/9506380 [hep-ph]}
  \BibitemShut {NoStop}%
%%CITATION = HEP-PH/9506380;%%
\bibitem [{\citenamefont {Bekenstein}(2004)}]{Bekenstein:2004ne}%
  \BibitemOpen
  \bibfield  {author} {\bibinfo {author} {\bibfnamefont {Jacob~D.}\
  \bibnamefont {Bekenstein}},\ }\bibfield  {title} {\enquote {\bibinfo {title}
  {{Relativistic gravitation theory for the MOND paradigm}},}\ }\href {\doibase
  10.1103/PhysRevD.70.083509, 10.1103/PhysRevD.71.069901} {\bibfield  {journal}
  {\bibinfo  {journal} {Phys. Rev.}\ }\textbf {\bibinfo {volume} {D70}},\
  \bibinfo {pages} {083509} (\bibinfo {year} {2004})},\ \bibinfo {note}
  {[Erratum: Phys. Rev.D71,069901(2005)]},\ \Eprint
  {http://arxiv.org/abs/astro-ph/0403694} {arXiv:astro-ph/0403694 [astro-ph]}
  \BibitemShut {NoStop}%
%%CITATION = ASTRO-PH/0403694;%%
\bibitem [{\citenamefont {Abbott}\ \emph
  {et~al.}(2017{\natexlab{a}})\citenamefont {Abbott} \emph
  {et~al.}}]{TheLIGOScientific:2017qsa}%
  \BibitemOpen
  \bibfield  {author} {\bibinfo {author} {\bibfnamefont {B.~P.}\ \bibnamefont
  {Abbott}} \emph {et~al.} (\bibinfo {collaboration} {LIGO Scientific,
  Virgo}),\ }\bibfield  {title} {\enquote {\bibinfo {title} {{GW170817:
  Observation of Gravitational Waves from a Binary Neutron Star Inspiral}},}\
  }\href {\doibase 10.1103/PhysRevLett.119.161101} {\bibfield  {journal}
  {\bibinfo  {journal} {Phys. Rev. Lett.}\ }\textbf {\bibinfo {volume} {119}},\
  \bibinfo {pages} {161101} (\bibinfo {year} {2017}{\natexlab{a}})},\ \Eprint
  {http://arxiv.org/abs/1710.05832} {arXiv:1710.05832 [gr-qc]} \BibitemShut
  {NoStop}%
%%CITATION = ARXIV:1710.05832;%%
\bibitem [{\citenamefont {Boran}\ \emph {et~al.}(2018)\citenamefont {Boran},
  \citenamefont {Desai}, \citenamefont {Kahya},\ and\ \citenamefont
  {Woodard}}]{Boran:2017rdn}%
  \BibitemOpen
  \bibfield  {author} {\bibinfo {author} {\bibfnamefont {S.}~\bibnamefont
  {Boran}}, \bibinfo {author} {\bibfnamefont {S.}~\bibnamefont {Desai}},
  \bibinfo {author} {\bibfnamefont {E.~O.}\ \bibnamefont {Kahya}}, \ and\
  \bibinfo {author} {\bibfnamefont {R.~P.}\ \bibnamefont {Woodard}},\
  }\bibfield  {title} {\enquote {\bibinfo {title} {{GW170817 Falsifies Dark
  Matter Emulators}},}\ }\href {\doibase 10.1103/PhysRevD.97.041501} {\bibfield
   {journal} {\bibinfo  {journal} {Phys. Rev.}\ }\textbf {\bibinfo {volume}
  {D97}},\ \bibinfo {pages} {041501} (\bibinfo {year} {2018})},\ \Eprint
  {http://arxiv.org/abs/1710.06168} {arXiv:1710.06168 [astro-ph.HE]}
  \BibitemShut {NoStop}%
%%CITATION = ARXIV:1710.06168;%%
\bibitem [{\citenamefont {Green}\ \emph {et~al.}(2018)\citenamefont {Green},
  \citenamefont {Moffat},\ and\ \citenamefont {Toth}}]{Green:2017qcv}%
  \BibitemOpen
  \bibfield  {author} {\bibinfo {author} {\bibfnamefont {M.~A.}\ \bibnamefont
  {Green}}, \bibinfo {author} {\bibfnamefont {J.~W.}\ \bibnamefont {Moffat}}, \
  and\ \bibinfo {author} {\bibfnamefont {V.~T.}\ \bibnamefont {Toth}},\
  }\bibfield  {title} {\enquote {\bibinfo {title} {{Modified Gravity (MOG), the
  speed of gravitational radiation and the event GW170817/GRB170817A}},}\
  }\href {\doibase 10.1016/j.physletb.2018.03.015} {\bibfield  {journal}
  {\bibinfo  {journal} {Phys. Lett.}\ }\textbf {\bibinfo {volume} {B780}},\
  \bibinfo {pages} {300--302} (\bibinfo {year} {2018})},\ \Eprint
  {http://arxiv.org/abs/1710.11177} {arXiv:1710.11177 [gr-qc]} \BibitemShut
  {NoStop}%
%%CITATION = ARXIV:1710.11177;%%
\bibitem [{\citenamefont {Moffat}(2006)}]{Moffat:2005si}%
  \BibitemOpen
  \bibfield  {author} {\bibinfo {author} {\bibfnamefont {J.~W.}\ \bibnamefont
  {Moffat}},\ }\bibfield  {title} {\enquote {\bibinfo {title}
  {{Scalar-tensor-vector gravity theory}},}\ }\href {\doibase
  10.1088/1475-7516/2006/03/004} {\bibfield  {journal} {\bibinfo  {journal}
  {JCAP}\ }\textbf {\bibinfo {volume} {0603}},\ \bibinfo {pages} {004}
  (\bibinfo {year} {2006})},\ \Eprint {http://arxiv.org/abs/gr-qc/0506021}
  {arXiv:gr-qc/0506021 [gr-qc]} \BibitemShut {NoStop}%
%%CITATION = GR-QC/0506021;%%
\bibitem [{\citenamefont {Milgrom}(1983{\natexlab{a}})}]{Milgrom:1983ca}%
  \BibitemOpen
  \bibfield  {author} {\bibinfo {author} {\bibfnamefont {M.}~\bibnamefont
  {Milgrom}},\ }\bibfield  {title} {\enquote {\bibinfo {title} {{A Modification
  of the Newtonian dynamics as a possible alternative to the hidden mass
  hypothesis}},}\ }\href {\doibase 10.1086/161130} {\bibfield  {journal}
  {\bibinfo  {journal} {Astrophys. J.}\ }\textbf {\bibinfo {volume} {270}},\
  \bibinfo {pages} {365--370} (\bibinfo {year}
  {1983}{\natexlab{a}})}\BibitemShut {NoStop}%
%%CITATION = ASJOA,270,365;%%
\bibitem [{\citenamefont {Milgrom}(1983{\natexlab{b}})}]{Milgrom:1983pn}%
  \BibitemOpen
  \bibfield  {author} {\bibinfo {author} {\bibfnamefont {M.}~\bibnamefont
  {Milgrom}},\ }\bibfield  {title} {\enquote {\bibinfo {title} {{A Modification
  of the Newtonian dynamics: Implications for galaxies}},}\ }\href {\doibase
  10.1086/161131} {\bibfield  {journal} {\bibinfo  {journal} {Astrophys. J.}\
  }\textbf {\bibinfo {volume} {270}},\ \bibinfo {pages} {371--383} (\bibinfo
  {year} {1983}{\natexlab{b}})}\BibitemShut {NoStop}%
%%CITATION = ASJOA,270,371;%%
\bibitem [{\citenamefont {Milgrom}(1983{\natexlab{c}})}]{Milgrom:1983zz}%
  \BibitemOpen
  \bibfield  {author} {\bibinfo {author} {\bibfnamefont {M.}~\bibnamefont
  {Milgrom}},\ }\bibfield  {title} {\enquote {\bibinfo {title} {{A modification
  of the Newtonian dynamics: implications for galaxy systems}},}\ }\href
  {\doibase 10.1086/161132} {\bibfield  {journal} {\bibinfo  {journal}
  {Astrophys. J.}\ }\textbf {\bibinfo {volume} {270}},\ \bibinfo {pages}
  {384--389} (\bibinfo {year} {1983}{\natexlab{c}})}\BibitemShut {NoStop}%
%%CITATION = ASJOA,270,384;%%
\bibitem [{\citenamefont {Christodoulou}\ and\ \citenamefont
  {Kazanas}(2018)}]{Christodoulou:2018xxw}%
  \BibitemOpen
  \bibfield  {author} {\bibinfo {author} {\bibfnamefont {Dimitris~M.}\
  \bibnamefont {Christodoulou}}\ and\ \bibinfo {author} {\bibfnamefont
  {Demosthenes}\ \bibnamefont {Kazanas}},\ }\bibfield  {title} {\enquote
  {\bibinfo {title} {{Interposing a Varying Gravitational Constant Between
  Modified Newtonian Dynamics and Weak Weyl Gravity}},}\ }\href {\doibase
  10.1093/mnrasl/sly118} {\bibfield  {journal} {\bibinfo  {journal} {Mon. Not.
  Roy. Astron. Soc.}\ }\textbf {\bibinfo {volume} {479}},\ \bibinfo {pages}
  {L143--L147} (\bibinfo {year} {2018})},\ \Eprint
  {http://arxiv.org/abs/1806.09778} {arXiv:1806.09778 [gr-qc]} \BibitemShut
  {NoStop}%
%%CITATION = ARXIV:1806.09778;%%
\bibitem [{\citenamefont {Christodoulou}\ and\ \citenamefont
  {Kazanas}(2019{\natexlab{a}})}]{Christodoulou:2018xzg}%
  \BibitemOpen
  \bibfield  {author} {\bibinfo {author} {\bibfnamefont {Dimitris~M.}\
  \bibnamefont {Christodoulou}}\ and\ \bibinfo {author} {\bibfnamefont
  {Demosthenes}\ \bibnamefont {Kazanas}},\ }\bibfield  {title} {\enquote
  {\bibinfo {title} {{Gravitational potential and non-relativistic Lagrangian
  in modified gravity with varying G}},}\ }\href {\doibase
  10.1093/mnrasl/sly222} {\bibfield  {journal} {\bibinfo  {journal} {Mon. Not.
  Roy. Astron. Soc.}\ }\textbf {\bibinfo {volume} {483}},\ \bibinfo {pages}
  {L85--L87} (\bibinfo {year} {2019}{\natexlab{a}})},\ \Eprint
  {http://arxiv.org/abs/1811.08920} {arXiv:1811.08920 [astro-ph.GA]}
  \BibitemShut {NoStop}%
%%CITATION = ARXIV:1811.08920;%%
\bibitem [{\citenamefont {Christodoulou}\ and\ \citenamefont
  {Kazanas}(2019{\natexlab{b}})}]{Christodoulou:2019ixr}%
  \BibitemOpen
  \bibfield  {author} {\bibinfo {author} {\bibfnamefont {Dimitris~M.}\
  \bibnamefont {Christodoulou}}\ and\ \bibinfo {author} {\bibfnamefont
  {Demosthenes}\ \bibnamefont {Kazanas}},\ }\bibfield  {title} {\enquote
  {\bibinfo {title} {{Gauss’s law and the source for Poisson’s equation in
  modified gravity with Varying G}},}\ }\href {\doibase 10.1093/mnras/stz120}
  {\bibfield  {journal} {\bibinfo  {journal} {Mon. Not. Roy. Astron. Soc.}\
  }\textbf {\bibinfo {volume} {484}},\ \bibinfo {pages} {1421--1425} (\bibinfo
  {year} {2019}{\natexlab{b}})},\ \Eprint {http://arxiv.org/abs/1901.02589}
  {arXiv:1901.02589 [astro-ph.GA]} \BibitemShut {NoStop}%
%%CITATION = ARXIV:1901.02589;%%
\bibitem [{\citenamefont {Christodoulou}\ and\ \citenamefont
  {Kazanas}(2019{\natexlab{c}})}]{Christodoulou:2019atk}%
  \BibitemOpen
  \bibfield  {author} {\bibinfo {author} {\bibfnamefont {Dimitris~M.}\
  \bibnamefont {Christodoulou}}\ and\ \bibinfo {author} {\bibfnamefont
  {Demosthenes}\ \bibnamefont {Kazanas}},\ }\bibfield  {title} {\enquote
  {\bibinfo {title} {{Universal expansion with spatially varying $G$}},}\
  }\href {\doibase 10.1093/mnrasl/slz074} {\  (\bibinfo {year}
  {2019}{\natexlab{c}}),\ 10.1093/mnrasl/slz074},\ \Eprint
  {http://arxiv.org/abs/1905.04296} {arXiv:1905.04296 [gr-qc]} \BibitemShut
  {NoStop}%
%%CITATION = ARXIV:1905.04296;%%
\bibitem [{\citenamefont {Murata}\ and\ \citenamefont
  {Tanaka}(2015)}]{Murata:2014nra-good-review-ofexperiments}%
  \BibitemOpen
  \bibfield  {author} {\bibinfo {author} {\bibfnamefont {Jiro}\ \bibnamefont
  {Murata}}\ and\ \bibinfo {author} {\bibfnamefont {Saki}\ \bibnamefont
  {Tanaka}},\ }\bibfield  {title} {\enquote {\bibinfo {title} {{A review of
  short-range gravity experiments in the LHC era}},}\ }\href {\doibase
  10.1088/0264-9381/32/3/033001} {\bibfield  {journal} {\bibinfo  {journal}
  {Class. Quant. Grav.}\ }\textbf {\bibinfo {volume} {32}},\ \bibinfo {pages}
  {033001} (\bibinfo {year} {2015})},\ \Eprint {http://arxiv.org/abs/1408.3588}
  {arXiv:1408.3588 [hep-ex]} \BibitemShut {NoStop}%
%%CITATION = ARXIV:1408.3588;%%
\bibitem [{\citenamefont {Nesseris}\ \emph {et~al.}(2017)\citenamefont
  {Nesseris}, \citenamefont {Pantazis},\ and\ \citenamefont
  {Perivolaropoulos}}]{Nesseris:2017vor}%
  \BibitemOpen
  \bibfield  {author} {\bibinfo {author} {\bibfnamefont {Savvas}\ \bibnamefont
  {Nesseris}}, \bibinfo {author} {\bibfnamefont {George}\ \bibnamefont
  {Pantazis}}, \ and\ \bibinfo {author} {\bibfnamefont {Leandros}\ \bibnamefont
  {Perivolaropoulos}},\ }\bibfield  {title} {\enquote {\bibinfo {title}
  {{Tension and constraints on modified gravity parametrizations of
  $G_{\textrm{eff}}(z)$ from growth rate and Planck data}},}\ }\href {\doibase
  10.1103/PhysRevD.96.023542} {\bibfield  {journal} {\bibinfo  {journal} {Phys.
  Rev.}\ }\textbf {\bibinfo {volume} {D96}},\ \bibinfo {pages} {023542}
  (\bibinfo {year} {2017})},\ \Eprint {http://arxiv.org/abs/1703.10538}
  {arXiv:1703.10538 [astro-ph.CO]} \BibitemShut {NoStop}%
%%CITATION = ARXIV:1703.10538;%%
\bibitem [{\citenamefont {Macaulay}\ \emph {et~al.}(2013)\citenamefont
  {Macaulay}, \citenamefont {Wehus},\ and\ \citenamefont
  {Eriksen}}]{Macaulay:2013swa}%
  \BibitemOpen
  \bibfield  {author} {\bibinfo {author} {\bibfnamefont {Edward}\ \bibnamefont
  {Macaulay}}, \bibinfo {author} {\bibfnamefont {Ingunn~Kathrine}\ \bibnamefont
  {Wehus}}, \ and\ \bibinfo {author} {\bibfnamefont {Hans~Kristian}\
  \bibnamefont {Eriksen}},\ }\bibfield  {title} {\enquote {\bibinfo {title}
  {{Lower Growth Rate from Recent Redshift Space Distortion Measurements than
  Expected from Planck}},}\ }\href {\doibase 10.1103/PhysRevLett.111.161301}
  {\bibfield  {journal} {\bibinfo  {journal} {Phys. Rev. Lett.}\ }\textbf
  {\bibinfo {volume} {111}},\ \bibinfo {pages} {161301} (\bibinfo {year}
  {2013})},\ \Eprint {http://arxiv.org/abs/1303.6583} {arXiv:1303.6583
  [astro-ph.CO]} \BibitemShut {NoStop}%
%%CITATION = ARXIV:1303.6583;%%
\bibitem [{\citenamefont {Tsujikawa}(2015)}]{Tsujikawa:2015mga}%
  \BibitemOpen
  \bibfield  {author} {\bibinfo {author} {\bibfnamefont {Shinji}\ \bibnamefont
  {Tsujikawa}},\ }\bibfield  {title} {\enquote {\bibinfo {title} {{Possibility
  of realizing weak gravity in redshift space distortion measurements}},}\
  }\href {\doibase 10.1103/PhysRevD.92.044029} {\bibfield  {journal} {\bibinfo
  {journal} {Phys. Rev.}\ }\textbf {\bibinfo {volume} {D92}},\ \bibinfo {pages}
  {044029} (\bibinfo {year} {2015})},\ \Eprint
  {http://arxiv.org/abs/1505.02459} {arXiv:1505.02459 [astro-ph.CO]}
  \BibitemShut {NoStop}%
%%CITATION = ARXIV:1505.02459;%%
\bibitem [{\citenamefont {Johnson}\ \emph {et~al.}(2016)\citenamefont
  {Johnson}, \citenamefont {Blake}, \citenamefont {Dossett}, \citenamefont
  {Koda}, \citenamefont {Parkinson},\ and\ \citenamefont
  {Joudaki}}]{Johnson:2015aaa}%
  \BibitemOpen
  \bibfield  {author} {\bibinfo {author} {\bibfnamefont {Andrew}\ \bibnamefont
  {Johnson}}, \bibinfo {author} {\bibfnamefont {Chris}\ \bibnamefont {Blake}},
  \bibinfo {author} {\bibfnamefont {Jason}\ \bibnamefont {Dossett}}, \bibinfo
  {author} {\bibfnamefont {Jun}\ \bibnamefont {Koda}}, \bibinfo {author}
  {\bibfnamefont {David}\ \bibnamefont {Parkinson}}, \ and\ \bibinfo {author}
  {\bibfnamefont {Shahab}\ \bibnamefont {Joudaki}},\ }\bibfield  {title}
  {\enquote {\bibinfo {title} {{Searching for Modified Gravity: Scale and
  Redshift Dependent Constraints from Galaxy Peculiar Velocities}},}\ }\href
  {\doibase 10.1093/mnras/stw447} {\bibfield  {journal} {\bibinfo  {journal}
  {Mon. Not. Roy. Astron. Soc.}\ }\textbf {\bibinfo {volume} {458}},\ \bibinfo
  {pages} {2725--2744} (\bibinfo {year} {2016})},\ \Eprint
  {http://arxiv.org/abs/1504.06885} {arXiv:1504.06885 [astro-ph.CO]}
  \BibitemShut {NoStop}%
%%CITATION = ARXIV:1504.06885;%%
\bibitem [{\citenamefont {Basilakos}\ and\ \citenamefont
  {Nesseris}(2017)}]{Basilakos:2017rgc}%
  \BibitemOpen
  \bibfield  {author} {\bibinfo {author} {\bibfnamefont {Spyros}\ \bibnamefont
  {Basilakos}}\ and\ \bibinfo {author} {\bibfnamefont {Savvas}\ \bibnamefont
  {Nesseris}},\ }\bibfield  {title} {\enquote {\bibinfo {title} {{Conjoined
  constraints on modified gravity from the expansion history and cosmic
  growth}},}\ }\href {\doibase 10.1103/PhysRevD.96.063517} {\bibfield
  {journal} {\bibinfo  {journal} {Phys. Rev.}\ }\textbf {\bibinfo {volume}
  {D96}},\ \bibinfo {pages} {063517} (\bibinfo {year} {2017})},\ \Eprint
  {http://arxiv.org/abs/1705.08797} {arXiv:1705.08797 [astro-ph.CO]}
  \BibitemShut {NoStop}%
%%CITATION = ARXIV:1705.08797;%%
\bibitem [{\citenamefont {Kazantzidis}\ and\ \citenamefont
  {Perivolaropoulos}(2018)}]{Kazantzidis:2018rnb}%
  \BibitemOpen
  \bibfield  {author} {\bibinfo {author} {\bibfnamefont {Lavrentios}\
  \bibnamefont {Kazantzidis}}\ and\ \bibinfo {author} {\bibfnamefont
  {Leandros}\ \bibnamefont {Perivolaropoulos}},\ }\bibfield  {title} {\enquote
  {\bibinfo {title} {{Evolution of the $f\sigma_8$ tension with the
  Planck15/$\Lambda$CDM determination and implications for modified gravity
  theories}},}\ }\href {\doibase 10.1103/PhysRevD.97.103503} {\bibfield
  {journal} {\bibinfo  {journal} {Phys. Rev.}\ }\textbf {\bibinfo {volume}
  {D97}},\ \bibinfo {pages} {103503} (\bibinfo {year} {2018})},\ \Eprint
  {http://arxiv.org/abs/1803.01337} {arXiv:1803.01337 [astro-ph.CO]}
  \BibitemShut {NoStop}%
%%CITATION = ARXIV:1803.01337;%%
\bibitem [{\citenamefont {Joudaki}\ \emph {et~al.}(2018)\citenamefont {Joudaki}
  \emph {et~al.}}]{Joudaki:2017zdt}%
  \BibitemOpen
  \bibfield  {author} {\bibinfo {author} {\bibfnamefont {Shahab}\ \bibnamefont
  {Joudaki}} \emph {et~al.},\ }\bibfield  {title} {\enquote {\bibinfo {title}
  {{KiDS-450 + 2dFLenS: Cosmological parameter constraints from weak
  gravitational lensing tomography and overlapping redshift-space galaxy
  clustering}},}\ }\href {\doibase 10.1093/mnras/stx2820} {\bibfield  {journal}
  {\bibinfo  {journal} {Mon. Not. Roy. Astron. Soc.}\ }\textbf {\bibinfo
  {volume} {474}},\ \bibinfo {pages} {4894--4924} (\bibinfo {year} {2018})},\
  \Eprint {http://arxiv.org/abs/1707.06627} {arXiv:1707.06627 [astro-ph.CO]}
  \BibitemShut {NoStop}%
%%CITATION = ARXIV:1707.06627;%%
\bibitem [{\citenamefont {Hildebrandt}\ \emph {et~al.}(2017)\citenamefont
  {Hildebrandt} \emph {et~al.}}]{Hildebrandt:2016iqg}%
  \BibitemOpen
  \bibfield  {author} {\bibinfo {author} {\bibfnamefont {H.}~\bibnamefont
  {Hildebrandt}} \emph {et~al.},\ }\bibfield  {title} {\enquote {\bibinfo
  {title} {{KiDS-450: Cosmological parameter constraints from tomographic weak
  gravitational lensing}},}\ }\href {\doibase 10.1093/mnras/stw2805} {\bibfield
   {journal} {\bibinfo  {journal} {Mon. Not. Roy. Astron. Soc.}\ }\textbf
  {\bibinfo {volume} {465}},\ \bibinfo {pages} {1454} (\bibinfo {year}
  {2017})},\ \Eprint {http://arxiv.org/abs/1606.05338} {arXiv:1606.05338
  [astro-ph.CO]} \BibitemShut {NoStop}%
%%CITATION = ARXIV:1606.05338;%%
\bibitem [{\citenamefont {Troxel}\ \emph {et~al.}(2018)\citenamefont {Troxel}
  \emph {et~al.}}]{Troxel:2017xyo}%
  \BibitemOpen
  \bibfield  {author} {\bibinfo {author} {\bibfnamefont {M.~A.}\ \bibnamefont
  {Troxel}} \emph {et~al.} (\bibinfo {collaboration} {DES}),\ }\bibfield
  {title} {\enquote {\bibinfo {title} {{Dark Energy Survey Year 1 results:
  Cosmological constraints from cosmic shear}},}\ }\href {\doibase
  10.1103/PhysRevD.98.043528} {\bibfield  {journal} {\bibinfo  {journal} {Phys.
  Rev.}\ }\textbf {\bibinfo {volume} {D98}},\ \bibinfo {pages} {043528}
  (\bibinfo {year} {2018})},\ \Eprint {http://arxiv.org/abs/1708.01538}
  {arXiv:1708.01538 [astro-ph.CO]} \BibitemShut {NoStop}%
%%CITATION = ARXIV:1708.01538;%%
\bibitem [{\citenamefont {Köhlinger}\ \emph {et~al.}(2017)\citenamefont
  {Köhlinger} \emph {et~al.}}]{Kohlinger:2017sxk}%
  \BibitemOpen
  \bibfield  {author} {\bibinfo {author} {\bibfnamefont {F.}~\bibnamefont
  {Köhlinger}} \emph {et~al.},\ }\bibfield  {title} {\enquote {\bibinfo
  {title} {{KiDS-450: The tomographic weak lensing power spectrum and
  constraints on cosmological parameters}},}\ }\href {\doibase
  10.1093/mnras/stx1820} {\bibfield  {journal} {\bibinfo  {journal} {Mon. Not.
  Roy. Astron. Soc.}\ }\textbf {\bibinfo {volume} {471}},\ \bibinfo {pages}
  {4412--4435} (\bibinfo {year} {2017})},\ \Eprint
  {http://arxiv.org/abs/1706.02892} {arXiv:1706.02892 [astro-ph.CO]}
  \BibitemShut {NoStop}%
%%CITATION = ARXIV:1706.02892;%%
\bibitem [{\citenamefont {Lamoreaux}(1997)}]{Lamoreaux:1996wh}%
  \BibitemOpen
  \bibfield  {author} {\bibinfo {author} {\bibfnamefont {S.~K.}\ \bibnamefont
  {Lamoreaux}},\ }\bibfield  {title} {\enquote {\bibinfo {title}
  {{Demonstration of the Casimir force in the 0.6 to 6 micrometers range}},}\
  }\href {\doibase 10.1103/PhysRevLett.81.5475, 10.1103/PhysRevLett.78.5}
  {\bibfield  {journal} {\bibinfo  {journal} {Phys. Rev. Lett.}\ }\textbf
  {\bibinfo {volume} {78}},\ \bibinfo {pages} {5--8} (\bibinfo {year}
  {1997})},\ \bibinfo {note} {[Erratum: Phys. Rev.
  Lett.81,5475(1998)]}\BibitemShut {NoStop}%
%%CITATION = PRLTA,78,5;%%
\bibitem [{\citenamefont {Chiaverini}\ \emph {et~al.}(2003)\citenamefont
  {Chiaverini}, \citenamefont {Smullin}, \citenamefont {Geraci}, \citenamefont
  {Weld},\ and\ \citenamefont {Kapitulnik}}]{Chiaverini:2002cb}%
  \BibitemOpen
  \bibfield  {author} {\bibinfo {author} {\bibfnamefont {J.}~\bibnamefont
  {Chiaverini}}, \bibinfo {author} {\bibfnamefont {S.~J.}\ \bibnamefont
  {Smullin}}, \bibinfo {author} {\bibfnamefont {A.~A.}\ \bibnamefont {Geraci}},
  \bibinfo {author} {\bibfnamefont {D.~M.}\ \bibnamefont {Weld}}, \ and\
  \bibinfo {author} {\bibfnamefont {A.}~\bibnamefont {Kapitulnik}},\ }\bibfield
   {title} {\enquote {\bibinfo {title} {{New experimental constraints on
  nonNewtonian forces below 100 microns}},}\ }\href {\doibase
  10.1103/PhysRevLett.90.151101} {\bibfield  {journal} {\bibinfo  {journal}
  {Phys. Rev. Lett.}\ }\textbf {\bibinfo {volume} {90}},\ \bibinfo {pages}
  {151101} (\bibinfo {year} {2003})},\ \Eprint
  {http://arxiv.org/abs/hep-ph/0209325} {arXiv:hep-ph/0209325 [hep-ph]}
  \BibitemShut {NoStop}%
%%CITATION = HEP-PH/0209325;%%
\bibitem [{\citenamefont {Smullin}\ \emph {et~al.}(2005)\citenamefont
  {Smullin}, \citenamefont {Geraci}, \citenamefont {Weld}, \citenamefont
  {Chiaverini}, \citenamefont {Holmes},\ and\ \citenamefont
  {Kapitulnik}}]{Smullin:2005iv}%
  \BibitemOpen
  \bibfield  {author} {\bibinfo {author} {\bibfnamefont {S.~J.}\ \bibnamefont
  {Smullin}}, \bibinfo {author} {\bibfnamefont {A.~A.}\ \bibnamefont {Geraci}},
  \bibinfo {author} {\bibfnamefont {D.~M.}\ \bibnamefont {Weld}}, \bibinfo
  {author} {\bibfnamefont {J.}~\bibnamefont {Chiaverini}}, \bibinfo {author}
  {\bibfnamefont {Susan~P.}\ \bibnamefont {Holmes}}, \ and\ \bibinfo {author}
  {\bibfnamefont {A.}~\bibnamefont {Kapitulnik}},\ }\bibfield  {title}
  {\enquote {\bibinfo {title} {{New constraints on Yukawa-type deviations from
  Newtonian gravity at 20 microns}},}\ }\href {\doibase
  10.1103/PhysRevD.72.122001, 10.1103/PhysRevD.72.129901} {\bibfield  {journal}
  {\bibinfo  {journal} {Phys. Rev.}\ }\textbf {\bibinfo {volume} {D72}},\
  \bibinfo {pages} {122001} (\bibinfo {year} {2005})},\ \bibinfo {note}
  {[Erratum: Phys. Rev.D72,129901(2005)]},\ \Eprint
  {http://arxiv.org/abs/hep-ph/0508204} {arXiv:hep-ph/0508204 [hep-ph]}
  \BibitemShut {NoStop}%
%%CITATION = HEP-PH/0508204;%%
\bibitem [{\citenamefont {Geraci}\ \emph {et~al.}(2008)\citenamefont {Geraci},
  \citenamefont {Smullin}, \citenamefont {Weld}, \citenamefont {Chiaverini},\
  and\ \citenamefont {Kapitulnik}}]{Geraci:2008hb}%
  \BibitemOpen
  \bibfield  {author} {\bibinfo {author} {\bibfnamefont {Andrew~A.}\
  \bibnamefont {Geraci}}, \bibinfo {author} {\bibfnamefont {Sylvia~J.}\
  \bibnamefont {Smullin}}, \bibinfo {author} {\bibfnamefont {David~M.}\
  \bibnamefont {Weld}}, \bibinfo {author} {\bibfnamefont {John}\ \bibnamefont
  {Chiaverini}}, \ and\ \bibinfo {author} {\bibfnamefont {Aharon}\ \bibnamefont
  {Kapitulnik}},\ }\bibfield  {title} {\enquote {\bibinfo {title} {{Improved
  constraints on non-Newtonian forces at 10 microns}},}\ }\href {\doibase
  10.1103/PhysRevD.78.022002} {\bibfield  {journal} {\bibinfo  {journal} {Phys.
  Rev.}\ }\textbf {\bibinfo {volume} {D78}},\ \bibinfo {pages} {022002}
  (\bibinfo {year} {2008})},\ \Eprint {http://arxiv.org/abs/0802.2350}
  {arXiv:0802.2350 [hep-ex]} \BibitemShut {NoStop}%
%%CITATION = ARXIV:0802.2350;%%
\bibitem [{\citenamefont {Long}\ \emph {et~al.}(2002)\citenamefont {Long},
  \citenamefont {Chan}, \citenamefont {Churnside}, \citenamefont {Gulbis},
  \citenamefont {Varney},\ and\ \citenamefont {Price}}]{Long:2002wn}%
  \BibitemOpen
  \bibfield  {author} {\bibinfo {author} {\bibfnamefont {Joshua~C.}\
  \bibnamefont {Long}}, \bibinfo {author} {\bibfnamefont {Hilton~W.}\
  \bibnamefont {Chan}}, \bibinfo {author} {\bibfnamefont {Allison~B.}\
  \bibnamefont {Churnside}}, \bibinfo {author} {\bibfnamefont {Eric~A.}\
  \bibnamefont {Gulbis}}, \bibinfo {author} {\bibfnamefont {Michael C.~M.}\
  \bibnamefont {Varney}}, \ and\ \bibinfo {author} {\bibfnamefont {John~C.}\
  \bibnamefont {Price}},\ }\bibfield  {title} {\enquote {\bibinfo {title}
  {{Upper limits to submillimeter-range forces from extra space-time
  dimensions}},}\ }\href {\doibase 10.1038/nature01432} {\  (\bibinfo {year}
  {2002}),\ 10.1038/nature01432},\ \bibinfo {note} {[Nature421,922(2003)]},\
  \Eprint {http://arxiv.org/abs/hep-ph/0210004} {arXiv:hep-ph/0210004 [hep-ph]}
  \BibitemShut {NoStop}%
%%CITATION = HEP-PH/0210004;%%
\bibitem [{\citenamefont {Mitrofanov}\ and\ \citenamefont
  {Ponomareva}(1988)}]{Mitrofanov1}%
  \BibitemOpen
  \bibfield  {author} {\bibinfo {author} {\bibfnamefont {V.~P.}\ \bibnamefont
  {Mitrofanov}}\ and\ \bibinfo {author} {\bibfnamefont {O.~I.}\ \bibnamefont
  {Ponomareva}},\ }\href@noop {} {\bibfield  {journal} {\bibinfo  {journal}
  {Sov. Phys. JETP}\ }\textbf {\bibinfo {volume} {87}},\ \bibinfo {pages}
  {1963--1966} (\bibinfo {year} {1988})}\BibitemShut {NoStop}%
\bibitem [{\citenamefont {Hoyle}\ \emph
  {et~al.}(2004{\natexlab{a}})\citenamefont {Hoyle}, \citenamefont {Kapner},
  \citenamefont {Heckel}, \citenamefont {Adelberger}, \citenamefont {Gundlach},
  \citenamefont {Schmidt},\ and\ \citenamefont {Swanson}}]{Hoyle:2004cw}%
  \BibitemOpen
  \bibfield  {author} {\bibinfo {author} {\bibfnamefont {C.~D.}\ \bibnamefont
  {Hoyle}}, \bibinfo {author} {\bibfnamefont {D.~J.}\ \bibnamefont {Kapner}},
  \bibinfo {author} {\bibfnamefont {Blayne~R.}\ \bibnamefont {Heckel}},
  \bibinfo {author} {\bibfnamefont {E.~G.}\ \bibnamefont {Adelberger}},
  \bibinfo {author} {\bibfnamefont {J.~H.}\ \bibnamefont {Gundlach}}, \bibinfo
  {author} {\bibfnamefont {U.}~\bibnamefont {Schmidt}}, \ and\ \bibinfo
  {author} {\bibfnamefont {H.~E.}\ \bibnamefont {Swanson}},\ }\bibfield
  {title} {\enquote {\bibinfo {title} {{Sub-millimeter tests of the
  gravitational inverse-square law}},}\ }\href {\doibase
  10.1103/PhysRevD.70.042004} {\bibfield  {journal} {\bibinfo  {journal} {Phys.
  Rev.}\ }\textbf {\bibinfo {volume} {D70}},\ \bibinfo {pages} {042004}
  (\bibinfo {year} {2004}{\natexlab{a}})},\ \Eprint
  {http://arxiv.org/abs/hep-ph/0405262} {arXiv:hep-ph/0405262 [hep-ph]}
  \BibitemShut {NoStop}%
%%CITATION = HEP-PH/0405262;%%
\bibitem [{\citenamefont {Hoyle}\ \emph {et~al.}(2001)\citenamefont {Hoyle},
  \citenamefont {Schmidt}, \citenamefont {Heckel}, \citenamefont {Adelberger},
  \citenamefont {Gundlach}, \citenamefont {Kapner},\ and\ \citenamefont
  {Swanson}}]{Hoyle:2000cv-washington1}%
  \BibitemOpen
  \bibfield  {author} {\bibinfo {author} {\bibfnamefont {C.~D.}\ \bibnamefont
  {Hoyle}}, \bibinfo {author} {\bibfnamefont {U.}~\bibnamefont {Schmidt}},
  \bibinfo {author} {\bibfnamefont {Blayne~R.}\ \bibnamefont {Heckel}},
  \bibinfo {author} {\bibfnamefont {E.~G.}\ \bibnamefont {Adelberger}},
  \bibinfo {author} {\bibfnamefont {J.~H.}\ \bibnamefont {Gundlach}}, \bibinfo
  {author} {\bibfnamefont {D.~J.}\ \bibnamefont {Kapner}}, \ and\ \bibinfo
  {author} {\bibfnamefont {H.~E.}\ \bibnamefont {Swanson}},\ }\bibfield
  {title} {\enquote {\bibinfo {title} {{Submillimeter tests of the
  gravitational inverse square law: a search for 'large' extra dimensions}},}\
  }\href {\doibase 10.1103/PhysRevLett.86.1418} {\bibfield  {journal} {\bibinfo
   {journal} {Phys. Rev. Lett.}\ }\textbf {\bibinfo {volume} {86}},\ \bibinfo
  {pages} {1418--1421} (\bibinfo {year} {2001})},\ \Eprint
  {http://arxiv.org/abs/hep-ph/0011014} {arXiv:hep-ph/0011014 [hep-ph]}
  \BibitemShut {NoStop}%
%%CITATION = HEP-PH/0011014;%%
\bibitem [{\citenamefont {Kapner}\ \emph
  {et~al.}(2007{\natexlab{b}})\citenamefont {Kapner}, \citenamefont {Cook},
  \citenamefont {Adelberger}, \citenamefont {Gundlach}, \citenamefont {Heckel},
  \citenamefont {Hoyle},\ and\ \citenamefont {Swanson}}]{Kapner:2006si}%
  \BibitemOpen
  \bibfield  {author} {\bibinfo {author} {\bibfnamefont {D.~J.}\ \bibnamefont
  {Kapner}}, \bibinfo {author} {\bibfnamefont {T.~S.}\ \bibnamefont {Cook}},
  \bibinfo {author} {\bibfnamefont {E.~G.}\ \bibnamefont {Adelberger}},
  \bibinfo {author} {\bibfnamefont {J.~H.}\ \bibnamefont {Gundlach}}, \bibinfo
  {author} {\bibfnamefont {Blayne~R.}\ \bibnamefont {Heckel}}, \bibinfo
  {author} {\bibfnamefont {C.~D.}\ \bibnamefont {Hoyle}}, \ and\ \bibinfo
  {author} {\bibfnamefont {H.~E.}\ \bibnamefont {Swanson}},\ }\bibfield
  {title} {\enquote {\bibinfo {title} {{Tests of the gravitational
  inverse-square law below the dark-energy length scale}},}\ }\href {\doibase
  10.1103/PhysRevLett.98.021101} {\bibfield  {journal} {\bibinfo  {journal}
  {Phys. Rev. Lett.}\ }\textbf {\bibinfo {volume} {98}},\ \bibinfo {pages}
  {021101} (\bibinfo {year} {2007}{\natexlab{b}})},\ \Eprint
  {http://arxiv.org/abs/hep-ph/0611184} {arXiv:hep-ph/0611184 [hep-ph]}
  \BibitemShut {NoStop}%
%%CITATION = HEP-PH/0611184;%%
\bibitem [{\citenamefont {Tu}\ \emph {et~al.}(2007)\citenamefont {Tu},
  \citenamefont {Guan}, \citenamefont {Luo}, \citenamefont {Shao},\ and\
  \citenamefont {Liu}}]{Tu:2007zz}%
  \BibitemOpen
  \bibfield  {author} {\bibinfo {author} {\bibfnamefont {Liang-Cheng}\
  \bibnamefont {Tu}}, \bibinfo {author} {\bibfnamefont {Sheng-Guo}\
  \bibnamefont {Guan}}, \bibinfo {author} {\bibfnamefont {Jun}\ \bibnamefont
  {Luo}}, \bibinfo {author} {\bibfnamefont {Cheng-Gang}\ \bibnamefont {Shao}},
  \ and\ \bibinfo {author} {\bibfnamefont {Lin-Xia}\ \bibnamefont {Liu}},\
  }\bibfield  {title} {\enquote {\bibinfo {title} {{Null Test of Newtonian
  Inverse-Square Law at Submillimeter Range with a Dual-Modulation Torsion
  Pendulum}},}\ }\href {\doibase 10.1103/PhysRevLett.98.201101} {\bibfield
  {journal} {\bibinfo  {journal} {Phys. Rev. Lett.}\ }\textbf {\bibinfo
  {volume} {98}},\ \bibinfo {pages} {201101} (\bibinfo {year}
  {2007})}\BibitemShut {NoStop}%
%%CITATION = PRLTA,98,201101;%%
\bibitem [{\citenamefont {Yang}\ \emph {et~al.}(2012)\citenamefont {Yang},
  \citenamefont {Zhan}, \citenamefont {Wang}, \citenamefont {Shao},
  \citenamefont {Tu}, \citenamefont {Tan},\ and\ \citenamefont
  {Luo}}]{Yang:2012zzb}%
  \BibitemOpen
  \bibfield  {author} {\bibinfo {author} {\bibfnamefont {Shan-Qing}\
  \bibnamefont {Yang}}, \bibinfo {author} {\bibfnamefont {Bi-Fu}\ \bibnamefont
  {Zhan}}, \bibinfo {author} {\bibfnamefont {Qing-Lan}\ \bibnamefont {Wang}},
  \bibinfo {author} {\bibfnamefont {Cheng-Gang}\ \bibnamefont {Shao}}, \bibinfo
  {author} {\bibfnamefont {Liang-Cheng}\ \bibnamefont {Tu}}, \bibinfo {author}
  {\bibfnamefont {Wen-Hai}\ \bibnamefont {Tan}}, \ and\ \bibinfo {author}
  {\bibfnamefont {Jun}\ \bibnamefont {Luo}},\ }\bibfield  {title} {\enquote
  {\bibinfo {title} {{Test of the Gravitational Inverse Square Law at
  Millimeter Ranges}},}\ }\href {\doibase 10.1103/PhysRevLett.108.081101}
  {\bibfield  {journal} {\bibinfo  {journal} {Phys. Rev. Lett.}\ }\textbf
  {\bibinfo {volume} {108}},\ \bibinfo {pages} {081101} (\bibinfo {year}
  {2012})}\BibitemShut {NoStop}%
%%CITATION = PRLTA,108,081101;%%
\bibitem [{\citenamefont {et. al.}(2014)}]{RikkyoMurata1}%
  \BibitemOpen
  \bibfield  {author} {\bibinfo {author} {\bibfnamefont {Murata~J.}\
  \bibnamefont {et. al.}},\ }\href@noop {} {\bibfield  {journal} {\bibinfo
  {journal} {Hyperfine Interaction}\ }\textbf {\bibinfo {volume} {255}},\
  \bibinfo {pages} {193--196} (\bibinfo {year} {2014})}\BibitemShut {NoStop}%
\bibitem [{\citenamefont {Hoskins}\ \emph {et~al.}(1985)\citenamefont
  {Hoskins}, \citenamefont {Newman}, \citenamefont {Spero},\ and\ \citenamefont
  {Schultz}}]{Hoskins:1985tn}%
  \BibitemOpen
  \bibfield  {author} {\bibinfo {author} {\bibfnamefont {J.~K.}\ \bibnamefont
  {Hoskins}}, \bibinfo {author} {\bibfnamefont {R.~D.}\ \bibnamefont {Newman}},
  \bibinfo {author} {\bibfnamefont {R.}~\bibnamefont {Spero}}, \ and\ \bibinfo
  {author} {\bibfnamefont {J.}~\bibnamefont {Schultz}},\ }\bibfield  {title}
  {\enquote {\bibinfo {title} {{Experimental tests of the gravitational inverse
  square law for mass separations from 2-cm to 105-cm}},}\ }\href {\doibase
  10.1103/PhysRevD.32.3084} {\bibfield  {journal} {\bibinfo  {journal} {Phys.
  Rev.}\ }\textbf {\bibinfo {volume} {D32}},\ \bibinfo {pages} {3084--3095}
  (\bibinfo {year} {1985})}\BibitemShut {NoStop}%
%%CITATION = PHRVA,D32,3084;%%
\bibitem [{\citenamefont {Spero}\ \emph {et~al.}(1980)\citenamefont {Spero},
  \citenamefont {Hoskins}, \citenamefont {Newman}, \citenamefont {Pellam},\
  and\ \citenamefont {Schultz}}]{Spero:1980zz}%
  \BibitemOpen
  \bibfield  {author} {\bibinfo {author} {\bibfnamefont {R.}~\bibnamefont
  {Spero}}, \bibinfo {author} {\bibfnamefont {J.~K.}\ \bibnamefont {Hoskins}},
  \bibinfo {author} {\bibfnamefont {R.}~\bibnamefont {Newman}}, \bibinfo
  {author} {\bibfnamefont {J.}~\bibnamefont {Pellam}}, \ and\ \bibinfo {author}
  {\bibfnamefont {J.}~\bibnamefont {Schultz}},\ }\bibfield  {title} {\enquote
  {\bibinfo {title} {{Test of the Gravitational Inverse-Square Law at
  Laboratory Distances}},}\ }\href {\doibase 10.1103/PhysRevLett.44.1645}
  {\bibfield  {journal} {\bibinfo  {journal} {Phys. Rev. Lett.}\ }\textbf
  {\bibinfo {volume} {44}},\ \bibinfo {pages} {1645--1648} (\bibinfo {year}
  {1980})}\BibitemShut {NoStop}%
%%CITATION = PRLTA,44,1645;%%
\bibitem [{\citenamefont {Milyukov}(1985)}]{Milyukov1}%
  \BibitemOpen
  \bibfield  {author} {\bibinfo {author} {\bibfnamefont {V.~K.}\ \bibnamefont
  {Milyukov}},\ }\href@noop {} {\bibfield  {journal} {\bibinfo  {journal} {Sov.
  Phys. JETP}\ }\textbf {\bibinfo {volume} {61}},\ \bibinfo {pages} {187--191}
  (\bibinfo {year} {1985})}\BibitemShut {NoStop}%
\bibitem [{\citenamefont {Panov}\ and\ \citenamefont {Frontov}(1979)}]{Panov1}%
  \BibitemOpen
  \bibfield  {author} {\bibinfo {author} {\bibfnamefont {V.~I.}\ \bibnamefont
  {Panov}}\ and\ \bibinfo {author} {\bibfnamefont {V.~N.}\ \bibnamefont
  {Frontov}},\ }\href@noop {} {\bibfield  {journal} {\bibinfo  {journal} {Sov.
  Phys. JETP}\ }\textbf {\bibinfo {volume} {50}},\ \bibinfo {pages} {852--856}
  (\bibinfo {year} {1979})}\BibitemShut {NoStop}%
\bibitem [{\citenamefont {Moody}\ and\ \citenamefont
  {Paik}(1993)}]{Moody:1993ir}%
  \BibitemOpen
  \bibfield  {author} {\bibinfo {author} {\bibfnamefont {M.~V.}\ \bibnamefont
  {Moody}}\ and\ \bibinfo {author} {\bibfnamefont {H.~J.}\ \bibnamefont
  {Paik}},\ }\bibfield  {title} {\enquote {\bibinfo {title} {{Gauss's law test
  of gravity at short range}},}\ }\href {\doibase 10.1103/PhysRevLett.70.1195}
  {\bibfield  {journal} {\bibinfo  {journal} {Phys. Rev. Lett.}\ }\textbf
  {\bibinfo {volume} {70}},\ \bibinfo {pages} {1195--1198} (\bibinfo {year}
  {1993})}\BibitemShut {NoStop}%
%%CITATION = PRLTA,70,1195;%%
\bibitem [{\citenamefont {H.}\ \emph {et~al.}(1980)\citenamefont {H.},
  \citenamefont {K.},\ and\ \citenamefont {K.}}]{Hirakawa1}%
  \BibitemOpen
  \bibfield  {author} {\bibinfo {author} {\bibfnamefont {Hirakawa}\
  \bibnamefont {H.}}, \bibinfo {author} {\bibfnamefont {Tsubono}\ \bibnamefont
  {K.}}, \ and\ \bibinfo {author} {\bibfnamefont {Oide}\ \bibnamefont {K.}},\
  }\href@noop {} {\bibfield  {journal} {\bibinfo  {journal} {Nature}\ }\textbf
  {\bibinfo {volume} {283}},\ \bibinfo {pages} {184--185} (\bibinfo {year}
  {1980})}\BibitemShut {NoStop}%
\bibitem [{\citenamefont {Y.}\ \emph {et~al.}(1982)\citenamefont {Y.},
  \citenamefont {K.},\ and\ \citenamefont {H.}}]{Ogawa1}%
  \BibitemOpen
  \bibfield  {author} {\bibinfo {author} {\bibfnamefont {Ogawa}\ \bibnamefont
  {Y.}}, \bibinfo {author} {\bibfnamefont {Tsubono}\ \bibnamefont {K.}}, \ and\
  \bibinfo {author} {\bibfnamefont {Hirakawa}\ \bibnamefont {H.}},\ }\href@noop
  {} {\bibfield  {journal} {\bibinfo  {journal} {Phys. Rev. D}\ }\textbf
  {\bibinfo {volume} {32}},\ \bibinfo {pages} {342--346} (\bibinfo {year}
  {1982})}\BibitemShut {NoStop}%
\bibitem [{\citenamefont {K.}\ and\ \citenamefont {H.}(1985)}]{Kuroda1}%
  \BibitemOpen
  \bibfield  {author} {\bibinfo {author} {\bibfnamefont {Kuroda}\ \bibnamefont
  {K.}}\ and\ \bibinfo {author} {\bibfnamefont {Hirakawa}\ \bibnamefont {H.}},\
  }\href@noop {} {\bibfield  {journal} {\bibinfo  {journal} {Phys. Rev. D}\
  }\textbf {\bibinfo {volume} {32}},\ \bibinfo {pages} {342--346} (\bibinfo
  {year} {1985})}\BibitemShut {NoStop}%
\bibitem [{\citenamefont {N.}\ \emph {et~al.}(1987)\citenamefont {N.},
  \citenamefont {K.},\ and\ \citenamefont {H.}}]{Mio1}%
  \BibitemOpen
  \bibfield  {author} {\bibinfo {author} {\bibfnamefont {Mio}\ \bibnamefont
  {N.}}, \bibinfo {author} {\bibfnamefont {Tsubono}\ \bibnamefont {K.}}, \ and\
  \bibinfo {author} {\bibfnamefont {Hirakawa}\ \bibnamefont {H.}},\ }\href@noop
  {} {\bibfield  {journal} {\bibinfo  {journal} {Phys. Rev. D}\ }\textbf
  {\bibinfo {volume} {36}},\ \bibinfo {pages} {2321–2326} (\bibinfo {year}
  {1987})}\BibitemShut {NoStop}%
\bibitem [{\citenamefont {Brax}\ \emph {et~al.}(2019)\citenamefont {Brax},
  \citenamefont {Valageas},\ and\ \citenamefont {Vanhove}}]{Brax:2019iut}%
  \BibitemOpen
  \bibfield  {author} {\bibinfo {author} {\bibfnamefont {Philippe}\
  \bibnamefont {Brax}}, \bibinfo {author} {\bibfnamefont {Patrick}\
  \bibnamefont {Valageas}}, \ and\ \bibinfo {author} {\bibfnamefont {Pierre}\
  \bibnamefont {Vanhove}},\ }\bibfield  {title} {\enquote {\bibinfo {title}
  {{New Bounds on Dark Energy Induced Fifth Forces}},}\ }\href {\doibase
  10.1103/PhysRevD.99.064010} {\bibfield  {journal} {\bibinfo  {journal} {Phys.
  Rev.}\ }\textbf {\bibinfo {volume} {D99}},\ \bibinfo {pages} {064010}
  (\bibinfo {year} {2019})},\ \Eprint {http://arxiv.org/abs/1902.07555}
  {arXiv:1902.07555 [astro-ph.CO]} \BibitemShut {NoStop}%
%%CITATION = ARXIV:1902.07555;%%
\bibitem [{\citenamefont
  {Perivolaropoulos}(2010)}]{Perivolaropoulos:2009ak-massive-bd-ppn}%
  \BibitemOpen
  \bibfield  {author} {\bibinfo {author} {\bibfnamefont {L.}~\bibnamefont
  {Perivolaropoulos}},\ }\bibfield  {title} {\enquote {\bibinfo {title} {{PPN
  Parameter gamma and Solar System Constraints of Massive Brans-Dicke
  Theories}},}\ }\href {\doibase 10.1103/PhysRevD.81.047501} {\bibfield
  {journal} {\bibinfo  {journal} {Phys. Rev.}\ }\textbf {\bibinfo {volume}
  {D81}},\ \bibinfo {pages} {047501} (\bibinfo {year} {2010})},\ \Eprint
  {http://arxiv.org/abs/0911.3401} {arXiv:0911.3401 [gr-qc]} \BibitemShut
  {NoStop}%
%%CITATION = ARXIV:0911.3401;%%
\bibitem [{\citenamefont {Hohmann}\ \emph {et~al.}(2013)\citenamefont
  {Hohmann}, \citenamefont {Jarv}, \citenamefont {Kuusk},\ and\ \citenamefont
  {Randla}}]{Hohmann:2013rba}%
  \BibitemOpen
  \bibfield  {author} {\bibinfo {author} {\bibfnamefont {Manuel}\ \bibnamefont
  {Hohmann}}, \bibinfo {author} {\bibfnamefont {Laur}\ \bibnamefont {Jarv}},
  \bibinfo {author} {\bibfnamefont {Piret}\ \bibnamefont {Kuusk}}, \ and\
  \bibinfo {author} {\bibfnamefont {Erik}\ \bibnamefont {Randla}},\ }\bibfield
  {title} {\enquote {\bibinfo {title} {{Post-Newtonian parameters $\gamma$ and
  $\beta$ of scalar-tensor gravity with a general potential}},}\ }\href
  {\doibase 10.1103/PhysRevD.89.069901, 10.1103/PhysRevD.88.084054} {\bibfield
  {journal} {\bibinfo  {journal} {Phys. Rev.}\ }\textbf {\bibinfo {volume}
  {D88}},\ \bibinfo {pages} {084054} (\bibinfo {year} {2013})},\ \bibinfo
  {note} {[Erratum: Phys. Rev.D89,no.6,069901(2014)]},\ \Eprint
  {http://arxiv.org/abs/1309.0031} {arXiv:1309.0031 [gr-qc]} \BibitemShut
  {NoStop}%
%%CITATION = ARXIV:1309.0031;%%
\bibitem [{\citenamefont {Järv}\ \emph {et~al.}(2015)\citenamefont {Järv},
  \citenamefont {Kuusk}, \citenamefont {Saal},\ and\ \citenamefont
  {Vilson}}]{Jarv:2014hma}%
  \BibitemOpen
  \bibfield  {author} {\bibinfo {author} {\bibfnamefont {Laur}\ \bibnamefont
  {Järv}}, \bibinfo {author} {\bibfnamefont {Piret}\ \bibnamefont {Kuusk}},
  \bibinfo {author} {\bibfnamefont {Margus}\ \bibnamefont {Saal}}, \ and\
  \bibinfo {author} {\bibfnamefont {Ott}\ \bibnamefont {Vilson}},\ }\bibfield
  {title} {\enquote {\bibinfo {title} {{Invariant quantities in the
  scalar-tensor theories of gravitation}},}\ }\href {\doibase
  10.1103/PhysRevD.91.024041} {\bibfield  {journal} {\bibinfo  {journal} {Phys.
  Rev.}\ }\textbf {\bibinfo {volume} {D91}},\ \bibinfo {pages} {024041}
  (\bibinfo {year} {2015})},\ \Eprint {http://arxiv.org/abs/1411.1947}
  {arXiv:1411.1947 [gr-qc]} \BibitemShut {NoStop}%
%%CITATION = ARXIV:1411.1947;%%
\bibitem [{\citenamefont {Esposito-Farese}\ and\ \citenamefont
  {Polarski}(2001)}]{EspositoFarese:2000ij}%
  \BibitemOpen
  \bibfield  {author} {\bibinfo {author} {\bibfnamefont {Gilles}\ \bibnamefont
  {Esposito-Farese}}\ and\ \bibinfo {author} {\bibfnamefont {D.}~\bibnamefont
  {Polarski}},\ }\bibfield  {title} {\enquote {\bibinfo {title} {{Scalar tensor
  gravity in an accelerating universe}},}\ }\href {\doibase
  10.1103/PhysRevD.63.063504} {\bibfield  {journal} {\bibinfo  {journal} {Phys.
  Rev.}\ }\textbf {\bibinfo {volume} {D63}},\ \bibinfo {pages} {063504}
  (\bibinfo {year} {2001})},\ \Eprint {http://arxiv.org/abs/gr-qc/0009034}
  {arXiv:gr-qc/0009034 [gr-qc]} \BibitemShut {NoStop}%
%%CITATION = GR-QC/0009034;%%
\bibitem [{\citenamefont {Gannouji}\ \emph {et~al.}(2006)\citenamefont
  {Gannouji}, \citenamefont {Polarski}, \citenamefont {Ranquet},\ and\
  \citenamefont {Starobinsky}}]{Gannouji:2006jm}%
  \BibitemOpen
  \bibfield  {author} {\bibinfo {author} {\bibfnamefont {Radouane}\
  \bibnamefont {Gannouji}}, \bibinfo {author} {\bibfnamefont {David}\
  \bibnamefont {Polarski}}, \bibinfo {author} {\bibfnamefont {Andre}\
  \bibnamefont {Ranquet}}, \ and\ \bibinfo {author} {\bibfnamefont {Alexei~A.}\
  \bibnamefont {Starobinsky}},\ }\bibfield  {title} {\enquote {\bibinfo {title}
  {{Scalar-Tensor Models of Normal and Phantom Dark Energy}},}\ }\href
  {\doibase 10.1088/1475-7516/2006/09/016} {\bibfield  {journal} {\bibinfo
  {journal} {JCAP}\ }\textbf {\bibinfo {volume} {0609}},\ \bibinfo {pages}
  {016} (\bibinfo {year} {2006})},\ \Eprint
  {http://arxiv.org/abs/astro-ph/0606287} {arXiv:astro-ph/0606287 [astro-ph]}
  \BibitemShut {NoStop}%
%%CITATION = ASTRO-PH/0606287;%%
\bibitem [{\citenamefont {Faraoni}(2004)}]{Faraoni:2004pi}%
  \BibitemOpen
  \bibfield  {author} {\bibinfo {author} {\bibfnamefont {Valerio}\ \bibnamefont
  {Faraoni}},\ }\href {\doibase 10.1007/978-1-4020-1989-0} {\emph {\bibinfo
  {title} {{Cosmology in scalar tensor gravity}}}},\ Vol.\ \bibinfo {volume}
  {139}\ (\bibinfo {year} {2004})\BibitemShut {NoStop}%
%%CITATION = FTPHD,139,;%%
\bibitem [{\citenamefont {Chiba}(2003{\natexlab{a}})}]{Chiba:2003ir}%
  \BibitemOpen
  \bibfield  {author} {\bibinfo {author} {\bibfnamefont {Takeshi}\ \bibnamefont
  {Chiba}},\ }\bibfield  {title} {\enquote {\bibinfo {title} {{1/R gravity and
  scalar - tensor gravity}},}\ }\href {\doibase 10.1016/j.physletb.2003.09.033}
  {\bibfield  {journal} {\bibinfo  {journal} {Phys. Lett.}\ }\textbf {\bibinfo
  {volume} {B575}},\ \bibinfo {pages} {1--3} (\bibinfo {year}
  {2003}{\natexlab{a}})},\ \Eprint {http://arxiv.org/abs/astro-ph/0307338}
  {arXiv:astro-ph/0307338 [astro-ph]} \BibitemShut {NoStop}%
%%CITATION = ASTRO-PH/0307338;%%
\bibitem [{\citenamefont {Berry}\ and\ \citenamefont
  {Gair}(2011)}]{Berry:2011pb-weak-field-fr-incl-osc}%
  \BibitemOpen
  \bibfield  {author} {\bibinfo {author} {\bibfnamefont {Christopher P.~L.}\
  \bibnamefont {Berry}}\ and\ \bibinfo {author} {\bibfnamefont {Jonathan~R.}\
  \bibnamefont {Gair}},\ }\bibfield  {title} {\enquote {\bibinfo {title}
  {{Linearized f(R) Gravity: Gravitational Radiation and Solar System
  Tests}},}\ }\href {\doibase 10.1103/PhysRevD.85.089906,
  10.1103/PhysRevD.83.104022} {\bibfield  {journal} {\bibinfo  {journal} {Phys.
  Rev.}\ }\textbf {\bibinfo {volume} {D83}},\ \bibinfo {pages} {104022}
  (\bibinfo {year} {2011})},\ \bibinfo {note} {[Erratum: Phys.
  Rev.D85,089906(2012)]},\ \Eprint {http://arxiv.org/abs/1104.0819}
  {arXiv:1104.0819 [gr-qc]} \BibitemShut {NoStop}%
%%CITATION = ARXIV:1104.0819;%%
\bibitem [{\citenamefont {Capozziello}\ \emph {et~al.}(2009)\citenamefont
  {Capozziello}, \citenamefont {Stabile},\ and\ \citenamefont
  {Troisi}}]{Capozziello:2009vr}%
  \BibitemOpen
  \bibfield  {author} {\bibinfo {author} {\bibfnamefont {S.}~\bibnamefont
  {Capozziello}}, \bibinfo {author} {\bibfnamefont {A.}~\bibnamefont
  {Stabile}}, \ and\ \bibinfo {author} {\bibfnamefont {A.}~\bibnamefont
  {Troisi}},\ }\bibfield  {title} {\enquote {\bibinfo {title} {{A General
  solution in the Newtonian limit of f(R)- gravity}},}\ }\href {\doibase
  10.1142/S0217732309030382} {\bibfield  {journal} {\bibinfo  {journal} {Mod.
  Phys. Lett.}\ }\textbf {\bibinfo {volume} {A24}},\ \bibinfo {pages}
  {659--665} (\bibinfo {year} {2009})},\ \Eprint
  {http://arxiv.org/abs/0901.0448} {arXiv:0901.0448 [gr-qc]} \BibitemShut
  {NoStop}%
%%CITATION = ARXIV:0901.0448;%%
\bibitem [{\citenamefont {Schellstede}(2016)}]{Schellstede:2016ldu}%
  \BibitemOpen
  \bibfield  {author} {\bibinfo {author} {\bibfnamefont {Gerold~Oltman}\
  \bibnamefont {Schellstede}},\ }\bibfield  {title} {\enquote {\bibinfo {title}
  {{On the Newtonian limit of metric f(R) gravity}},}\ }\href {\doibase
  10.1007/s10714-016-2111-9} {\bibfield  {journal} {\bibinfo  {journal} {Gen.
  Rel. Grav.}\ }\textbf {\bibinfo {volume} {48}},\ \bibinfo {pages} {118}
  (\bibinfo {year} {2016})}\BibitemShut {NoStop}%
%%CITATION = GRGVA,48,118;%%
\bibitem [{\citenamefont {Perivolaropoulos}(2003)}]{Perivolaropoulos:2003we}%
  \BibitemOpen
  \bibfield  {author} {\bibinfo {author} {\bibfnamefont {L.}~\bibnamefont
  {Perivolaropoulos}},\ }\bibfield  {title} {\enquote {\bibinfo {title}
  {{Equation of state of oscillating Brans-Dicke scalar and extra
  dimensions}},}\ }\href {\doibase 10.1103/PhysRevD.67.123516} {\bibfield
  {journal} {\bibinfo  {journal} {Phys. Rev.}\ }\textbf {\bibinfo {volume}
  {D67}},\ \bibinfo {pages} {123516} (\bibinfo {year} {2003})},\ \Eprint
  {http://arxiv.org/abs/hep-ph/0301237} {arXiv:hep-ph/0301237 [hep-ph]}
  \BibitemShut {NoStop}%
%%CITATION = HEP-PH/0301237;%%
\bibitem [{\citenamefont {Perivolaropoulos}(2017)}]{Perivolaropoulos:2016ucs}%
  \BibitemOpen
  \bibfield  {author} {\bibinfo {author} {\bibfnamefont {Leandros}\
  \bibnamefont {Perivolaropoulos}},\ }\bibfield  {title} {\enquote {\bibinfo
  {title} {{Submillimeter spatial oscillations of Newton’s constant:
  Theoretical models and laboratory tests}},}\ }\href {\doibase
  10.1103/PhysRevD.95.084050} {\bibfield  {journal} {\bibinfo  {journal} {Phys.
  Rev.}\ }\textbf {\bibinfo {volume} {D95}},\ \bibinfo {pages} {084050}
  (\bibinfo {year} {2017})},\ \Eprint {http://arxiv.org/abs/1611.07293}
  {arXiv:1611.07293 [gr-qc]} \BibitemShut {NoStop}%
%%CITATION = ARXIV:1611.07293;%%
\bibitem [{\citenamefont {Perivolaropoulos}\ and\ \citenamefont
  {Sourdis}(2002)}]{Perivolaropoulos:2002pn}%
  \BibitemOpen
  \bibfield  {author} {\bibinfo {author} {\bibfnamefont {L.}~\bibnamefont
  {Perivolaropoulos}}\ and\ \bibinfo {author} {\bibfnamefont {C.}~\bibnamefont
  {Sourdis}},\ }\bibfield  {title} {\enquote {\bibinfo {title} {{Cosmological
  effects of radion oscillations}},}\ }\href {\doibase
  10.1103/PhysRevD.66.084018} {\bibfield  {journal} {\bibinfo  {journal} {Phys.
  Rev.}\ }\textbf {\bibinfo {volume} {D66}},\ \bibinfo {pages} {084018}
  (\bibinfo {year} {2002})},\ \Eprint {http://arxiv.org/abs/hep-ph/0204155}
  {arXiv:hep-ph/0204155 [hep-ph]} \BibitemShut {NoStop}%
%%CITATION = HEP-PH/0204155;%%
\bibitem [{\citenamefont {Donini}\ and\ \citenamefont
  {Marimón}(2016)}]{Donini:2016kgu}%
  \BibitemOpen
  \bibfield  {author} {\bibinfo {author} {\bibfnamefont {A.}~\bibnamefont
  {Donini}}\ and\ \bibinfo {author} {\bibfnamefont {S.~G.}\ \bibnamefont
  {Marimón}},\ }\bibfield  {title} {\enquote {\bibinfo {title} {{Micro-orbits
  in a many-brane model and deviations from Newton’s $1/r^2$ law}},}\ }\href
  {\doibase 10.1140/epjc/s10052-016-4537-3} {\bibfield  {journal} {\bibinfo
  {journal} {Eur. Phys. J.}\ }\textbf {\bibinfo {volume} {C76}},\ \bibinfo
  {pages} {696} (\bibinfo {year} {2016})},\ \Eprint
  {http://arxiv.org/abs/1609.05654} {arXiv:1609.05654 [hep-ph]} \BibitemShut
  {NoStop}%
%%CITATION = ARXIV:1609.05654;%%
\bibitem [{\citenamefont {Benichou}\ and\ \citenamefont
  {Estes}(2012)}]{Benichou:2011dx}%
  \BibitemOpen
  \bibfield  {author} {\bibinfo {author} {\bibfnamefont {Raphael}\ \bibnamefont
  {Benichou}}\ and\ \bibinfo {author} {\bibfnamefont {John}\ \bibnamefont
  {Estes}},\ }\bibfield  {title} {\enquote {\bibinfo {title} {{The Fate of
  Newton's Law in Brane-World Scenarios}},}\ }\href {\doibase
  10.1016/j.physletb.2012.05.031} {\bibfield  {journal} {\bibinfo  {journal}
  {Phys. Lett.}\ }\textbf {\bibinfo {volume} {B712}},\ \bibinfo {pages}
  {456--459} (\bibinfo {year} {2012})},\ \Eprint
  {http://arxiv.org/abs/1112.0565} {arXiv:1112.0565 [hep-th]} \BibitemShut
  {NoStop}%
%%CITATION = ARXIV:1112.0565;%%
\bibitem [{\citenamefont {Bronnikov}\ \emph {et~al.}(2006)\citenamefont
  {Bronnikov}, \citenamefont {Kononogov},\ and\ \citenamefont
  {Melnikov}}]{Bronnikov:2006jy}%
  \BibitemOpen
  \bibfield  {author} {\bibinfo {author} {\bibfnamefont {K.~A.}\ \bibnamefont
  {Bronnikov}}, \bibinfo {author} {\bibfnamefont {S.~A.}\ \bibnamefont
  {Kononogov}}, \ and\ \bibinfo {author} {\bibfnamefont {V.~N.}\ \bibnamefont
  {Melnikov}},\ }\bibfield  {title} {\enquote {\bibinfo {title} {{Brane world
  corrections to Newton's law}},}\ }\href {\doibase 10.1007/s10714-006-0300-7}
  {\bibfield  {journal} {\bibinfo  {journal} {Gen. Rel. Grav.}\ }\textbf
  {\bibinfo {volume} {38}},\ \bibinfo {pages} {1215--1232} (\bibinfo {year}
  {2006})},\ \Eprint {http://arxiv.org/abs/gr-qc/0601114} {arXiv:gr-qc/0601114
  [gr-qc]} \BibitemShut {NoStop}%
%%CITATION = GR-QC/0601114;%%
\bibitem [{\citenamefont {Nojiri}\ and\ \citenamefont
  {Odintsov}(2002)}]{Nojiri:2002wn-newtonptl-brane}%
  \BibitemOpen
  \bibfield  {author} {\bibinfo {author} {\bibfnamefont {Shin'ichi}\
  \bibnamefont {Nojiri}}\ and\ \bibinfo {author} {\bibfnamefont {Sergei~D.}\
  \bibnamefont {Odintsov}},\ }\bibfield  {title} {\enquote {\bibinfo {title}
  {{Newton potential in deSitter brane world}},}\ }\href {\doibase
  10.1016/S0370-2693(02)02859-9} {\bibfield  {journal} {\bibinfo  {journal}
  {Phys. Lett.}\ }\textbf {\bibinfo {volume} {B548}},\ \bibinfo {pages}
  {215--223} (\bibinfo {year} {2002})},\ \Eprint
  {http://arxiv.org/abs/hep-th/0209066} {arXiv:hep-th/0209066 [hep-th]}
  \BibitemShut {NoStop}%
%%CITATION = HEP-TH/0209066;%%
\bibitem [{\citenamefont {Faraoni}(2006{\natexlab{a}})}]{Faraoni:2006sy}%
  \BibitemOpen
  \bibfield  {author} {\bibinfo {author} {\bibfnamefont {Valerio}\ \bibnamefont
  {Faraoni}},\ }\bibfield  {title} {\enquote {\bibinfo {title} {{Matter
  instability in modified gravity}},}\ }\href {\doibase
  10.1103/PhysRevD.74.104017} {\bibfield  {journal} {\bibinfo  {journal} {Phys.
  Rev.}\ }\textbf {\bibinfo {volume} {D74}},\ \bibinfo {pages} {104017}
  (\bibinfo {year} {2006}{\natexlab{a}})},\ \Eprint
  {http://arxiv.org/abs/astro-ph/0610734} {arXiv:astro-ph/0610734 [astro-ph]}
  \BibitemShut {NoStop}%
%%CITATION = ASTRO-PH/0610734;%%
\bibitem [{\citenamefont {Dolgov}\ and\ \citenamefont
  {Kawasaki}(2003)}]{Dolgov:2003px}%
  \BibitemOpen
  \bibfield  {author} {\bibinfo {author} {\bibfnamefont {A.~D.}\ \bibnamefont
  {Dolgov}}\ and\ \bibinfo {author} {\bibfnamefont {Masahiro}\ \bibnamefont
  {Kawasaki}},\ }\bibfield  {title} {\enquote {\bibinfo {title} {{Can modified
  gravity explain accelerated cosmic expansion?}}}\ }\href {\doibase
  10.1016/j.physletb.2003.08.039} {\bibfield  {journal} {\bibinfo  {journal}
  {Phys. Lett.}\ }\textbf {\bibinfo {volume} {B573}},\ \bibinfo {pages} {1--4}
  (\bibinfo {year} {2003})},\ \Eprint {http://arxiv.org/abs/astro-ph/0307285}
  {arXiv:astro-ph/0307285 [astro-ph]} \BibitemShut {NoStop}%
%%CITATION = ASTRO-PH/0307285;%%
\bibitem [{\citenamefont {Edholm}\ \emph {et~al.}(2016)\citenamefont {Edholm},
  \citenamefont {Koshelev},\ and\ \citenamefont
  {Mazumdar}}]{Edholm:2016hbt-nonlocal-potential-stable-spatial-oscillations}%
  \BibitemOpen
  \bibfield  {author} {\bibinfo {author} {\bibfnamefont {James}\ \bibnamefont
  {Edholm}}, \bibinfo {author} {\bibfnamefont {Alexey~S.}\ \bibnamefont
  {Koshelev}}, \ and\ \bibinfo {author} {\bibfnamefont {Anupam}\ \bibnamefont
  {Mazumdar}},\ }\bibfield  {title} {\enquote {\bibinfo {title} {{Behavior of
  the Newtonian potential for ghost-free gravity and singularity-free
  gravity}},}\ }\href {\doibase 10.1103/PhysRevD.94.104033} {\bibfield
  {journal} {\bibinfo  {journal} {Phys. Rev.}\ }\textbf {\bibinfo {volume}
  {D94}},\ \bibinfo {pages} {104033} (\bibinfo {year} {2016})},\ \Eprint
  {http://arxiv.org/abs/1604.01989} {arXiv:1604.01989 [gr-qc]} \BibitemShut
  {NoStop}%
%%CITATION = ARXIV:1604.01989;%%
\bibitem [{\citenamefont {Kehagias}\ and\ \citenamefont
  {Maggiore}(2014)}]{Kehagias:2014sda-nonlocal-oscillations}%
  \BibitemOpen
  \bibfield  {author} {\bibinfo {author} {\bibfnamefont {Alex}\ \bibnamefont
  {Kehagias}}\ and\ \bibinfo {author} {\bibfnamefont {Michele}\ \bibnamefont
  {Maggiore}},\ }\bibfield  {title} {\enquote {\bibinfo {title} {{Spherically
  symmetric static solutions in a nonlocal infrared modification of General
  Relativity}},}\ }\href {\doibase 10.1007/JHEP08(2014)029} {\bibfield
  {journal} {\bibinfo  {journal} {JHEP}\ }\textbf {\bibinfo {volume} {08}},\
  \bibinfo {pages} {029} (\bibinfo {year} {2014})},\ \Eprint
  {http://arxiv.org/abs/1401.8289} {arXiv:1401.8289 [hep-th]} \BibitemShut
  {NoStop}%
%%CITATION = ARXIV:1401.8289;%%
\bibitem [{\citenamefont {Frolov}\ and\ \citenamefont
  {Zelnikov}(2016)}]{Frolov:2015usa-newton-potential-nonlocal}%
  \BibitemOpen
  \bibfield  {author} {\bibinfo {author} {\bibfnamefont {Valeri~P.}\
  \bibnamefont {Frolov}}\ and\ \bibinfo {author} {\bibfnamefont {Andrei}\
  \bibnamefont {Zelnikov}},\ }\bibfield  {title} {\enquote {\bibinfo {title}
  {{Head-on collision of ultrarelativistic particles in ghost-free theories of
  gravity}},}\ }\href {\doibase 10.1103/PhysRevD.93.064048} {\bibfield
  {journal} {\bibinfo  {journal} {Phys. Rev.}\ }\textbf {\bibinfo {volume}
  {D93}},\ \bibinfo {pages} {064048} (\bibinfo {year} {2016})},\ \Eprint
  {http://arxiv.org/abs/1509.03336} {arXiv:1509.03336 [hep-th]} \BibitemShut
  {NoStop}%
%%CITATION = ARXIV:1509.03336;%%
\bibitem [{\citenamefont {Tomboulis}(1997)}]{Tomboulis:1997gg}%
  \BibitemOpen
  \bibfield  {author} {\bibinfo {author} {\bibfnamefont {E.~T.}\ \bibnamefont
  {Tomboulis}},\ }\bibfield  {title} {\enquote {\bibinfo {title}
  {{Superrenormalizable gauge and gravitational theories}},}\ }\href@noop {} {\
   (\bibinfo {year} {1997})},\ \Eprint {http://arxiv.org/abs/hep-th/9702146}
  {arXiv:hep-th/9702146 [hep-th]} \BibitemShut {NoStop}%
%%CITATION = HEP-TH/9702146;%%
\bibitem [{\citenamefont {Siegel}(2003)}]{Siegel:2003vt}%
  \BibitemOpen
  \bibfield  {author} {\bibinfo {author} {\bibfnamefont {W.}~\bibnamefont
  {Siegel}},\ }\bibfield  {title} {\enquote {\bibinfo {title} {{Stringy gravity
  at short distances}},}\ }\href@noop {} {\  (\bibinfo {year} {2003})},\
  \Eprint {http://arxiv.org/abs/hep-th/0309093} {arXiv:hep-th/0309093 [hep-th]}
  \BibitemShut {NoStop}%
%%CITATION = HEP-TH/0309093;%%
\bibitem [{\citenamefont {Biswas}\ \emph {et~al.}(2014)\citenamefont {Biswas},
  \citenamefont {Conroy}, \citenamefont {Koshelev},\ and\ \citenamefont
  {Mazumdar}}]{Biswas:2013cha}%
  \BibitemOpen
  \bibfield  {author} {\bibinfo {author} {\bibfnamefont {Tirthabir}\
  \bibnamefont {Biswas}}, \bibinfo {author} {\bibfnamefont {Aindriú}\
  \bibnamefont {Conroy}}, \bibinfo {author} {\bibfnamefont {Alexey~S.}\
  \bibnamefont {Koshelev}}, \ and\ \bibinfo {author} {\bibfnamefont {Anupam}\
  \bibnamefont {Mazumdar}},\ }\bibfield  {title} {\enquote {\bibinfo {title}
  {{Generalized ghost-free quadratic curvature gravity}},}\ }\href {\doibase
  10.1088/0264-9381/31/1/015022, 10.1088/0264-9381/31/15/159501} {\bibfield
  {journal} {\bibinfo  {journal} {Class. Quant. Grav.}\ }\textbf {\bibinfo
  {volume} {31}},\ \bibinfo {pages} {015022} (\bibinfo {year} {2014})},\
  \bibinfo {note} {[Erratum: Class. Quant. Grav.31,159501(2014)]},\ \Eprint
  {http://arxiv.org/abs/1308.2319} {arXiv:1308.2319 [hep-th]} \BibitemShut
  {NoStop}%
%%CITATION = ARXIV:1308.2319;%%
\bibitem [{\citenamefont {Alam}\ \emph {et~al.}(2017)\citenamefont {Alam} \emph
  {et~al.}}]{Alam:2016hwk}%
  \BibitemOpen
  \bibfield  {author} {\bibinfo {author} {\bibfnamefont {Shadab}\ \bibnamefont
  {Alam}} \emph {et~al.} (\bibinfo {collaboration} {BOSS}),\ }\bibfield
  {title} {\enquote {\bibinfo {title} {{The clustering of galaxies in the
  completed SDSS-III Baryon Oscillation Spectroscopic Survey: cosmological
  analysis of the DR12 galaxy sample}},}\ }\href@noop {} {\bibfield  {journal}
  {\bibinfo  {journal} {Mon. Not. Roy. Astron. Soc.}\ }\textbf {\bibinfo
  {volume} {470}},\ \bibinfo {pages} {2617--2652} (\bibinfo {year} {2017})},\
  \Eprint {http://arxiv.org/abs/1607.03155} {arXiv:1607.03155 [astro-ph.CO]}
  \BibitemShut {NoStop}%
%%CITATION = ARXIV:1607.03155;%%
\bibitem [{\citenamefont {Amon}\ \emph {et~al.}(2017)\citenamefont {Amon} \emph
  {et~al.}}]{Amon:2017lia}%
  \BibitemOpen
  \bibfield  {author} {\bibinfo {author} {\bibfnamefont {A.}~\bibnamefont
  {Amon}} \emph {et~al.},\ }\bibfield  {title} {\enquote {\bibinfo {title}
  {{KiDS+2dFLenS+GAMA: Testing the cosmological model with the $E_{\rm G}$
  statistic}},}\ }\href {\doibase 10.1093/mnras/sty1624} {\  (\bibinfo {year}
  {2017}),\ 10.1093/mnras/sty1624},\ \Eprint {http://arxiv.org/abs/1711.10999}
  {arXiv:1711.10999 [astro-ph.CO]} \BibitemShut {NoStop}%
%%CITATION = ARXIV:1711.10999;%%
\bibitem [{\citenamefont {Abbott}\ \emph {et~al.}(2018)\citenamefont {Abbott}
  \emph {et~al.}}]{Abbott:2018jhe}%
  \BibitemOpen
  \bibfield  {author} {\bibinfo {author} {\bibfnamefont {T.~M.~C.}\
  \bibnamefont {Abbott}} \emph {et~al.} (\bibinfo {collaboration} {DES}),\
  }\bibfield  {title} {\enquote {\bibinfo {title} {{The Dark Energy Survey Data
  Release 1}},}\ }\href@noop {} {\  (\bibinfo {year} {2018})},\ \Eprint
  {http://arxiv.org/abs/1801.03181} {arXiv:1801.03181 [astro-ph.IM]}
  \BibitemShut {NoStop}%
%%CITATION = ARXIV:1801.03181;%%
\bibitem [{\citenamefont {Ade}\ \emph {et~al.}(2016)\citenamefont {Ade} \emph
  {et~al.}}]{Ade:2015xua}%
  \BibitemOpen
  \bibfield  {author} {\bibinfo {author} {\bibfnamefont {P.~A.~R.}\
  \bibnamefont {Ade}} \emph {et~al.} (\bibinfo {collaboration} {Planck}),\
  }\bibfield  {title} {\enquote {\bibinfo {title} {{Planck 2015 results. XIII.
  Cosmological parameters}},}\ }\href {\doibase 10.1051/0004-6361/201525830}
  {\bibfield  {journal} {\bibinfo  {journal} {Astron. Astrophys.}\ }\textbf
  {\bibinfo {volume} {594}},\ \bibinfo {pages} {A13} (\bibinfo {year}
  {2016})},\ \Eprint {http://arxiv.org/abs/1502.01589} {arXiv:1502.01589
  [astro-ph.CO]} \BibitemShut {NoStop}%
%%CITATION = ARXIV:1502.01589;%%
\bibitem [{\citenamefont {Abbott}\ \emph
  {et~al.}(2017{\natexlab{b}})\citenamefont {Abbott} \emph
  {et~al.}}]{Abbott:2017wau}%
  \BibitemOpen
  \bibfield  {author} {\bibinfo {author} {\bibfnamefont {T.~M.~C.}\
  \bibnamefont {Abbott}} \emph {et~al.} (\bibinfo {collaboration} {DES}),\
  }\bibfield  {title} {\enquote {\bibinfo {title} {{Dark Energy Survey Year 1
  Results: Cosmological Constraints from Galaxy Clustering and Weak
  Lensing}},}\ }\href@noop {} {\  (\bibinfo {year} {2017}{\natexlab{b}})},\
  \Eprint {http://arxiv.org/abs/1708.01530} {arXiv:1708.01530 [astro-ph.CO]}
  \BibitemShut {NoStop}%
%%CITATION = ARXIV:1708.01530;%%
\bibitem [{\citenamefont {{Blake}}\ \emph {et~al.}(2012)\citenamefont
  {{Blake}}, \citenamefont {{Brough}}, \citenamefont {{Colless}}, \citenamefont
  {{Contreras}}, \citenamefont {{Couch}}, \citenamefont {{Croom}},
  \citenamefont {{Croton}}, \citenamefont {{Davis}}, \citenamefont
  {{Drinkwater}}, \citenamefont {{Forster}}, \citenamefont {{Gilbank}},
  \citenamefont {{Gladders}}, \citenamefont {{Glazebrook}}, \citenamefont
  {{Jelliffe}}, \citenamefont {{Jurek}}, \citenamefont {{Li}}, \citenamefont
  {{Madore}}, \citenamefont {{Martin}}, \citenamefont {{Pimbblet}},
  \citenamefont {{Poole}}, \citenamefont {{Pracy}}, \citenamefont {{Sharp}},
  \citenamefont {{Wisnioski}}, \citenamefont {{Woods}}, \citenamefont
  {{Wyder}},\ and\ \citenamefont {{Yee}}}]{Blake:2012pj}%
  \BibitemOpen
  \bibfield  {author} {\bibinfo {author} {\bibfnamefont {C.}~\bibnamefont
  {{Blake}}}, \bibinfo {author} {\bibfnamefont {S.}~\bibnamefont {{Brough}}},
  \bibinfo {author} {\bibfnamefont {M.}~\bibnamefont {{Colless}}}, \bibinfo
  {author} {\bibfnamefont {C.}~\bibnamefont {{Contreras}}}, \bibinfo {author}
  {\bibfnamefont {W.}~\bibnamefont {{Couch}}}, \bibinfo {author} {\bibfnamefont
  {S.}~\bibnamefont {{Croom}}}, \bibinfo {author} {\bibfnamefont
  {D.}~\bibnamefont {{Croton}}}, \bibinfo {author} {\bibfnamefont {T.~M.}\
  \bibnamefont {{Davis}}}, \bibinfo {author} {\bibfnamefont {M.~J.}\
  \bibnamefont {{Drinkwater}}}, \bibinfo {author} {\bibfnamefont
  {K.}~\bibnamefont {{Forster}}}, \bibinfo {author} {\bibfnamefont
  {D.}~\bibnamefont {{Gilbank}}}, \bibinfo {author} {\bibfnamefont
  {M.}~\bibnamefont {{Gladders}}}, \bibinfo {author} {\bibfnamefont
  {K.}~\bibnamefont {{Glazebrook}}}, \bibinfo {author} {\bibfnamefont
  {B.}~\bibnamefont {{Jelliffe}}}, \bibinfo {author} {\bibfnamefont {R.~J.}\
  \bibnamefont {{Jurek}}}, \bibinfo {author} {\bibfnamefont {I.-h.}\
  \bibnamefont {{Li}}}, \bibinfo {author} {\bibfnamefont {B.}~\bibnamefont
  {{Madore}}}, \bibinfo {author} {\bibfnamefont {D.~C.}\ \bibnamefont
  {{Martin}}}, \bibinfo {author} {\bibfnamefont {K.}~\bibnamefont
  {{Pimbblet}}}, \bibinfo {author} {\bibfnamefont {G.~B.}\ \bibnamefont
  {{Poole}}}, \bibinfo {author} {\bibfnamefont {M.}~\bibnamefont {{Pracy}}},
  \bibinfo {author} {\bibfnamefont {R.}~\bibnamefont {{Sharp}}}, \bibinfo
  {author} {\bibfnamefont {E.}~\bibnamefont {{Wisnioski}}}, \bibinfo {author}
  {\bibfnamefont {D.}~\bibnamefont {{Woods}}}, \bibinfo {author} {\bibfnamefont
  {T.~K.}\ \bibnamefont {{Wyder}}}, \ and\ \bibinfo {author} {\bibfnamefont
  {H.~K.~C.}\ \bibnamefont {{Yee}}},\ }\bibfield  {title} {\enquote {\bibinfo
  {title} {{The WiggleZ Dark Energy Survey: joint measurements of the expansion
  and growth history at $z 1$}},}\ }\href {\doibase
  10.1111/j.1365-2966.2012.21473.x} {\bibfield  {journal} {\bibinfo  {journal}
  {mnras}\ }\textbf {\bibinfo {volume} {425}},\ \bibinfo {pages} {405--414}
  (\bibinfo {year} {2012})},\ \Eprint {http://arxiv.org/abs/1204.3674}
  {arXiv:1204.3674} \BibitemShut {NoStop}%
\bibitem [{\citenamefont {Alam}\ \emph {et~al.}(2016)\citenamefont {Alam},
  \citenamefont {Ho},\ and\ \citenamefont {Silvestri}}]{Alam:2015rsa}%
  \BibitemOpen
  \bibfield  {author} {\bibinfo {author} {\bibfnamefont {Shadab}\ \bibnamefont
  {Alam}}, \bibinfo {author} {\bibfnamefont {Shirley}\ \bibnamefont {Ho}}, \
  and\ \bibinfo {author} {\bibfnamefont {Alessandra}\ \bibnamefont
  {Silvestri}},\ }\bibfield  {title} {\enquote {\bibinfo {title} {{Testing
  deviations from $\Lambda$CDM with growth rate measurements from six
  large-scale structure surveys at $z = $0.06–1}},}\ }\href {\doibase
  10.1093/mnras/stv2935} {\bibfield  {journal} {\bibinfo  {journal} {Mon. Not.
  Roy. Astron. Soc.}\ }\textbf {\bibinfo {volume} {456}},\ \bibinfo {pages}
  {3743--3756} (\bibinfo {year} {2016})},\ \Eprint
  {http://arxiv.org/abs/1509.05034} {arXiv:1509.05034 [astro-ph.CO]}
  \BibitemShut {NoStop}%
%%CITATION = ARXIV:1509.05034;%%
\bibitem [{\citenamefont {Hinton}(2016)}]{Hinton:2016jfq}%
  \BibitemOpen
  \bibfield  {author} {\bibinfo {author} {\bibfnamefont {Samuel}\ \bibnamefont
  {Hinton}},\ }\bibfield  {title} {\enquote {\bibinfo {title} {{Extraction of
  Cosmological Information from WiggleZ}},}\ }\href@noop {} {\  (\bibinfo
  {year} {2016})},\ \Eprint {http://arxiv.org/abs/1604.01830} {arXiv:1604.01830
  [astro-ph.CO]} \BibitemShut {NoStop}%
%%CITATION = ARXIV:1604.01830;%%
\bibitem [{\citenamefont {Ballinger}\ \emph {et~al.}(1996)\citenamefont
  {Ballinger}, \citenamefont {Peacock},\ and\ \citenamefont
  {Heavens}}]{Ballinger:1996cd}%
  \BibitemOpen
  \bibfield  {author} {\bibinfo {author} {\bibfnamefont {W.~E.}\ \bibnamefont
  {Ballinger}}, \bibinfo {author} {\bibfnamefont {J.~A.}\ \bibnamefont
  {Peacock}}, \ and\ \bibinfo {author} {\bibfnamefont {A.~F.}\ \bibnamefont
  {Heavens}},\ }\bibfield  {title} {\enquote {\bibinfo {title} {{Measuring the
  cosmological constant with redshift surveys}},}\ }\href {\doibase
  10.1093/mnras/282.3.877} {\bibfield  {journal} {\bibinfo  {journal} {Mon.
  Not. Roy. Astron. Soc.}\ }\textbf {\bibinfo {volume} {282}},\ \bibinfo
  {pages} {877--888} (\bibinfo {year} {1996})},\ \Eprint
  {http://arxiv.org/abs/astro-ph/9605017} {arXiv:astro-ph/9605017 [astro-ph]}
  \BibitemShut {NoStop}%
%%CITATION = ASTRO-PH/9605017;%%
\bibitem [{\citenamefont {Amendola}\ \emph {et~al.}(2018)\citenamefont
  {Amendola} \emph {et~al.}}]{Amendola:2016saw}%
  \BibitemOpen
  \bibfield  {author} {\bibinfo {author} {\bibfnamefont {Luca}\ \bibnamefont
  {Amendola}} \emph {et~al.},\ }\bibfield  {title} {\enquote {\bibinfo {title}
  {{Cosmology and fundamental physics with the Euclid satellite}},}\ }\href
  {\doibase 10.1007/s41114-017-0010-3} {\bibfield  {journal} {\bibinfo
  {journal} {Living Rev. Rel.}\ }\textbf {\bibinfo {volume} {21}},\ \bibinfo
  {pages} {2} (\bibinfo {year} {2018})},\ \Eprint
  {http://arxiv.org/abs/1606.00180} {arXiv:1606.00180 [astro-ph.CO]}
  \BibitemShut {NoStop}%
%%CITATION = ARXIV:1606.00180;%%
\bibitem [{\citenamefont {Saito}(2016)}]{Saito16}%
  \BibitemOpen
  \bibfield  {author} {\bibinfo {author} {\bibfnamefont {Shun}\ \bibnamefont
  {Saito}},\ }\href@noop {} {\enquote {\bibinfo {title} {Lecture notes in
  cosmology},}\ } (\bibinfo {year} {2016})\BibitemShut {NoStop}%
\bibitem [{\citenamefont {Wilson}(2016)}]{Wilson:2016ggz}%
  \BibitemOpen
  \bibfield  {author} {\bibinfo {author} {\bibfnamefont {Michael~J.}\
  \bibnamefont {Wilson}},\ }\emph {\bibinfo {title} {{Geometric and growth rate
  tests of General Relativity with recovered linear cosmological
  perturbations}}},\ \href
  {https://inspirehep.net/record/1494559/files/arXiv:1610.08362.pdf} {Ph.D.
  thesis},\ \bibinfo  {school} {Edinburgh U.} (\bibinfo {year} {2016}),\
  \Eprint {http://arxiv.org/abs/1610.08362} {arXiv:1610.08362 [astro-ph.CO]}
  \BibitemShut {NoStop}%
%%CITATION = ARXIV:1610.08362;%%
\bibitem [{\citenamefont {Kazantzidis}\ \emph {et~al.}(2019)\citenamefont
  {Kazantzidis}, \citenamefont {Perivolaropoulos},\ and\ \citenamefont
  {Skara}}]{Kazantzidis:2018jtb}%
  \BibitemOpen
  \bibfield  {author} {\bibinfo {author} {\bibfnamefont {L.}~\bibnamefont
  {Kazantzidis}}, \bibinfo {author} {\bibfnamefont {L.}~\bibnamefont
  {Perivolaropoulos}}, \ and\ \bibinfo {author} {\bibfnamefont
  {F.}~\bibnamefont {Skara}},\ }\bibfield  {title} {\enquote {\bibinfo {title}
  {{Constraining power of cosmological observables: blind redshift spots and
  optimal ranges}},}\ }\href {\doibase 10.1103/PhysRevD.99.063537} {\bibfield
  {journal} {\bibinfo  {journal} {Phys. Rev.}\ }\textbf {\bibinfo {volume}
  {D99}},\ \bibinfo {pages} {063537} (\bibinfo {year} {2019})},\ \Eprint
  {http://arxiv.org/abs/1812.05356} {arXiv:1812.05356 [astro-ph.CO]}
  \BibitemShut {NoStop}%
%%CITATION = ARXIV:1812.05356;%%
\bibitem [{\citenamefont {Nesseris}\ and\ \citenamefont
  {Perivolaropoulos}(2007)}]{Nesseris:2006hp}%
  \BibitemOpen
  \bibfield  {author} {\bibinfo {author} {\bibfnamefont {S.}~\bibnamefont
  {Nesseris}}\ and\ \bibinfo {author} {\bibfnamefont {Leandros}\ \bibnamefont
  {Perivolaropoulos}},\ }\bibfield  {title} {\enquote {\bibinfo {title} {{The
  Limits of Extended Quintessence}},}\ }\href {\doibase
  10.1103/PhysRevD.75.023517} {\bibfield  {journal} {\bibinfo  {journal} {Phys.
  Rev.}\ }\textbf {\bibinfo {volume} {D75}},\ \bibinfo {pages} {023517}
  (\bibinfo {year} {2007})},\ \Eprint {http://arxiv.org/abs/astro-ph/0611238}
  {arXiv:astro-ph/0611238 [astro-ph]} \BibitemShut {NoStop}%
%%CITATION = ASTRO-PH/0611238;%%
\bibitem [{\citenamefont {Khoury}\ and\ \citenamefont
  {Weltman}(2004)}]{Khoury:2003rn}%
  \BibitemOpen
  \bibfield  {author} {\bibinfo {author} {\bibfnamefont {Justin}\ \bibnamefont
  {Khoury}}\ and\ \bibinfo {author} {\bibfnamefont {Amanda}\ \bibnamefont
  {Weltman}},\ }\bibfield  {title} {\enquote {\bibinfo {title} {{Chameleon
  cosmology}},}\ }\href {\doibase 10.1103/PhysRevD.69.044026} {\bibfield
  {journal} {\bibinfo  {journal} {Phys. Rev.}\ }\textbf {\bibinfo {volume}
  {D69}},\ \bibinfo {pages} {044026} (\bibinfo {year} {2004})},\ \Eprint
  {http://arxiv.org/abs/astro-ph/0309411} {arXiv:astro-ph/0309411 [astro-ph]}
  \BibitemShut {NoStop}%
%%CITATION = ASTRO-PH/0309411;%%
\bibitem [{\citenamefont {Copi}\ \emph {et~al.}(2004)\citenamefont {Copi},
  \citenamefont {Davis},\ and\ \citenamefont {Krauss}}]{Copi:2003xd}%
  \BibitemOpen
  \bibfield  {author} {\bibinfo {author} {\bibfnamefont {Craig~J.}\
  \bibnamefont {Copi}}, \bibinfo {author} {\bibfnamefont {Adam~N.}\
  \bibnamefont {Davis}}, \ and\ \bibinfo {author} {\bibfnamefont {Lawrence~M.}\
  \bibnamefont {Krauss}},\ }\bibfield  {title} {\enquote {\bibinfo {title} {{A
  New nucleosynthesis constraint on the variation of G}},}\ }\href {\doibase
  10.1103/PhysRevLett.92.171301} {\bibfield  {journal} {\bibinfo  {journal}
  {Phys. Rev. Lett.}\ }\textbf {\bibinfo {volume} {92}},\ \bibinfo {pages}
  {171301} (\bibinfo {year} {2004})},\ \Eprint
  {http://arxiv.org/abs/astro-ph/0311334} {arXiv:astro-ph/0311334 [astro-ph]}
  \BibitemShut {NoStop}%
%%CITATION = ASTRO-PH/0311334;%%
\bibitem [{\citenamefont {Gannouji}\ \emph {et~al.}(2018)\citenamefont
  {Gannouji}, \citenamefont {Kazantzidis}, \citenamefont {Perivolaropoulos},\
  and\ \citenamefont {Polarski}}]{Gannouji:2018ncm}%
  \BibitemOpen
  \bibfield  {author} {\bibinfo {author} {\bibfnamefont {Radouane}\
  \bibnamefont {Gannouji}}, \bibinfo {author} {\bibfnamefont {Lavrentios}\
  \bibnamefont {Kazantzidis}}, \bibinfo {author} {\bibfnamefont {Leandros}\
  \bibnamefont {Perivolaropoulos}}, \ and\ \bibinfo {author} {\bibfnamefont
  {David}\ \bibnamefont {Polarski}},\ }\bibfield  {title} {\enquote {\bibinfo
  {title} {{Consistency of modified gravity with a decreasing $G_{\rm eff}(z)$
  in a $\Lambda$CDM background}},}\ }\href {\doibase
  10.1103/PhysRevD.98.104044} {\bibfield  {journal} {\bibinfo  {journal} {Phys.
  Rev.}\ }\textbf {\bibinfo {volume} {D98}},\ \bibinfo {pages} {104044}
  (\bibinfo {year} {2018})},\ \Eprint {http://arxiv.org/abs/1809.07034}
  {arXiv:1809.07034 [gr-qc]} \BibitemShut {NoStop}%
%%CITATION = ARXIV:1809.07034;%%
\bibitem [{\citenamefont {Boisseau}\ \emph {et~al.}(2000)\citenamefont
  {Boisseau}, \citenamefont {Esposito-Farese}, \citenamefont {Polarski},\ and\
  \citenamefont {Starobinsky}}]{Boisseau:2000pr}%
  \BibitemOpen
  \bibfield  {author} {\bibinfo {author} {\bibfnamefont {B.}~\bibnamefont
  {Boisseau}}, \bibinfo {author} {\bibfnamefont {Gilles}\ \bibnamefont
  {Esposito-Farese}}, \bibinfo {author} {\bibfnamefont {D.}~\bibnamefont
  {Polarski}}, \ and\ \bibinfo {author} {\bibfnamefont {Alexei~A.}\
  \bibnamefont {Starobinsky}},\ }\bibfield  {title} {\enquote {\bibinfo {title}
  {{Reconstruction of a scalar tensor theory of gravity in an accelerating
  universe}},}\ }\href {\doibase 10.1103/PhysRevLett.85.2236} {\bibfield
  {journal} {\bibinfo  {journal} {Phys. Rev. Lett.}\ }\textbf {\bibinfo
  {volume} {85}},\ \bibinfo {pages} {2236} (\bibinfo {year} {2000})},\ \Eprint
  {http://arxiv.org/abs/gr-qc/0001066} {arXiv:gr-qc/0001066 [gr-qc]}
  \BibitemShut {NoStop}%
%%CITATION = GR-QC/0001066;%%
\bibitem [{\citenamefont {Nesseris}\ and\ \citenamefont
  {Perivolaropoulos}(2006)}]{Nesseris:2006jc}%
  \BibitemOpen
  \bibfield  {author} {\bibinfo {author} {\bibfnamefont {S.}~\bibnamefont
  {Nesseris}}\ and\ \bibinfo {author} {\bibfnamefont {Leandros}\ \bibnamefont
  {Perivolaropoulos}},\ }\bibfield  {title} {\enquote {\bibinfo {title}
  {{Evolving newton's constant, extended gravity theories and snia data
  analysis}},}\ }\href {\doibase 10.1103/PhysRevD.73.103511} {\bibfield
  {journal} {\bibinfo  {journal} {Phys. Rev.}\ }\textbf {\bibinfo {volume}
  {D73}},\ \bibinfo {pages} {103511} (\bibinfo {year} {2006})},\ \Eprint
  {http://arxiv.org/abs/astro-ph/0602053} {arXiv:astro-ph/0602053 [astro-ph]}
  \BibitemShut {NoStop}%
%%CITATION = ASTRO-PH/0602053;%%
\bibitem [{\citenamefont {Hojjati}\ \emph {et~al.}(2011)\citenamefont
  {Hojjati}, \citenamefont {Pogosian},\ and\ \citenamefont
  {Zhao}}]{Hojjati:2011ix}%
  \BibitemOpen
  \bibfield  {author} {\bibinfo {author} {\bibfnamefont {Alireza}\ \bibnamefont
  {Hojjati}}, \bibinfo {author} {\bibfnamefont {Levon}\ \bibnamefont
  {Pogosian}}, \ and\ \bibinfo {author} {\bibfnamefont {Gong-Bo}\ \bibnamefont
  {Zhao}},\ }\bibfield  {title} {\enquote {\bibinfo {title} {{Testing gravity
  with CAMB and CosmoMC}},}\ }\href {\doibase 10.1088/1475-7516/2011/08/005}
  {\bibfield  {journal} {\bibinfo  {journal} {JCAP}\ }\textbf {\bibinfo
  {volume} {1108}},\ \bibinfo {pages} {005} (\bibinfo {year} {2011})},\ \Eprint
  {http://arxiv.org/abs/1106.4543} {arXiv:1106.4543 [astro-ph.CO]} \BibitemShut
  {NoStop}%
%%CITATION = ARXIV:1106.4543;%%
\bibitem [{\citenamefont {Hoyle}\ \emph
  {et~al.}(2004{\natexlab{b}})\citenamefont {Hoyle}, \citenamefont {Kapner},
  \citenamefont {Heckel}, \citenamefont {Adelberger}, \citenamefont {Gundlach},
  \citenamefont {Schmidt},\ and\ \citenamefont
  {Swanson}}]{Hoyle:2004cw-washington2}%
  \BibitemOpen
  \bibfield  {author} {\bibinfo {author} {\bibfnamefont {C.~D.}\ \bibnamefont
  {Hoyle}}, \bibinfo {author} {\bibfnamefont {D.~J.}\ \bibnamefont {Kapner}},
  \bibinfo {author} {\bibfnamefont {Blayne~R.}\ \bibnamefont {Heckel}},
  \bibinfo {author} {\bibfnamefont {E.~G.}\ \bibnamefont {Adelberger}},
  \bibinfo {author} {\bibfnamefont {J.~H.}\ \bibnamefont {Gundlach}}, \bibinfo
  {author} {\bibfnamefont {U.}~\bibnamefont {Schmidt}}, \ and\ \bibinfo
  {author} {\bibfnamefont {H.~E.}\ \bibnamefont {Swanson}},\ }\bibfield
  {title} {\enquote {\bibinfo {title} {{Sub-millimeter tests of the
  gravitational inverse-square law}},}\ }\href {\doibase
  10.1103/PhysRevD.70.042004} {\bibfield  {journal} {\bibinfo  {journal} {Phys.
  Rev.}\ }\textbf {\bibinfo {volume} {D70}},\ \bibinfo {pages} {042004}
  (\bibinfo {year} {2004}{\natexlab{b}})},\ \Eprint
  {http://arxiv.org/abs/hep-ph/0405262} {arXiv:hep-ph/0405262 [hep-ph]}
  \BibitemShut {NoStop}%
%%CITATION = HEP-PH/0405262;%%
\bibitem [{\citenamefont {Capozziello}\ \emph {et~al.}(2007)\citenamefont
  {Capozziello}, \citenamefont {Stabile},\ and\ \citenamefont
  {Troisi}}]{Capozziello:2007ms-fR-newtonian-limit-with-osc-good-disc}%
  \BibitemOpen
  \bibfield  {author} {\bibinfo {author} {\bibfnamefont {S.}~\bibnamefont
  {Capozziello}}, \bibinfo {author} {\bibfnamefont {A.}~\bibnamefont
  {Stabile}}, \ and\ \bibinfo {author} {\bibfnamefont {A.}~\bibnamefont
  {Troisi}},\ }\bibfield  {title} {\enquote {\bibinfo {title} {{The Newtonian
  Limit of f(R) gravity}},}\ }\href {\doibase 10.1103/PhysRevD.76.104019}
  {\bibfield  {journal} {\bibinfo  {journal} {Phys. Rev.}\ }\textbf {\bibinfo
  {volume} {D76}},\ \bibinfo {pages} {104019} (\bibinfo {year} {2007})},\
  \Eprint {http://arxiv.org/abs/0708.0723} {arXiv:0708.0723 [gr-qc]}
  \BibitemShut {NoStop}%
%%CITATION = ARXIV:0708.0723;%%
\bibitem [{\citenamefont {Starobinsky}(2007)}]{Starobinsky:2007hu}%
  \BibitemOpen
  \bibfield  {author} {\bibinfo {author} {\bibfnamefont {Alexei~A.}\
  \bibnamefont {Starobinsky}},\ }\bibfield  {title} {\enquote {\bibinfo {title}
  {{Disappearing cosmological constant in f(R) gravity}},}\ }\href {\doibase
  10.1134/S0021364007150027} {\bibfield  {journal} {\bibinfo  {journal} {JETP
  Lett.}\ }\textbf {\bibinfo {volume} {86}},\ \bibinfo {pages} {157--163}
  (\bibinfo {year} {2007})},\ \Eprint {http://arxiv.org/abs/0706.2041}
  {arXiv:0706.2041 [astro-ph]} \BibitemShut {NoStop}%
%%CITATION = ARXIV:0706.2041;%%
\bibitem [{\citenamefont {De~Felice}\ and\ \citenamefont
  {Tsujikawa}(2010)}]{DeFelice:2010aj}%
  \BibitemOpen
  \bibfield  {author} {\bibinfo {author} {\bibfnamefont {Antonio}\ \bibnamefont
  {De~Felice}}\ and\ \bibinfo {author} {\bibfnamefont {Shinji}\ \bibnamefont
  {Tsujikawa}},\ }\bibfield  {title} {\enquote {\bibinfo {title} {{f(R)
  theories}},}\ }\href {\doibase 10.12942/lrr-2010-3} {\bibfield  {journal}
  {\bibinfo  {journal} {Living Rev. Rel.}\ }\textbf {\bibinfo {volume} {13}},\
  \bibinfo {pages} {3} (\bibinfo {year} {2010})},\ \Eprint
  {http://arxiv.org/abs/1002.4928} {arXiv:1002.4928 [gr-qc]} \BibitemShut
  {NoStop}%
%%CITATION = ARXIV:1002.4928;%%
\bibitem [{\citenamefont
  {Chiba}(2003{\natexlab{b}})}]{Chiba:2003ir-gamma05-weak-field}%
  \BibitemOpen
  \bibfield  {author} {\bibinfo {author} {\bibfnamefont {Takeshi}\ \bibnamefont
  {Chiba}},\ }\bibfield  {title} {\enquote {\bibinfo {title} {{1/R gravity and
  scalar - tensor gravity}},}\ }\href {\doibase 10.1016/j.physletb.2003.09.033}
  {\bibfield  {journal} {\bibinfo  {journal} {Phys. Lett.}\ }\textbf {\bibinfo
  {volume} {B575}},\ \bibinfo {pages} {1--3} (\bibinfo {year}
  {2003}{\natexlab{b}})},\ \Eprint {http://arxiv.org/abs/astro-ph/0307338}
  {arXiv:astro-ph/0307338 [astro-ph]} \BibitemShut {NoStop}%
%%CITATION = ASTRO-PH/0307338;%%
\bibitem [{\citenamefont
  {Faraoni}(2006{\natexlab{b}})}]{Faraoni:2006hx-ill-defined-equivalency-st-fR}%
  \BibitemOpen
  \bibfield  {author} {\bibinfo {author} {\bibfnamefont {Valerio}\ \bibnamefont
  {Faraoni}},\ }\bibfield  {title} {\enquote {\bibinfo {title} {{Solar System
  experiments do not yet veto modified gravity models}},}\ }\href {\doibase
  10.1103/PhysRevD.74.023529} {\bibfield  {journal} {\bibinfo  {journal} {Phys.
  Rev.}\ }\textbf {\bibinfo {volume} {D74}},\ \bibinfo {pages} {023529}
  (\bibinfo {year} {2006}{\natexlab{b}})},\ \Eprint
  {http://arxiv.org/abs/gr-qc/0607016} {arXiv:gr-qc/0607016 [gr-qc]}
  \BibitemShut {NoStop}%
%%CITATION = GR-QC/0607016;%%
\bibitem [{\citenamefont {Dolan}(2007)}]{Dolan:2007mj}%
  \BibitemOpen
  \bibfield  {author} {\bibinfo {author} {\bibfnamefont {Sam~R.}\ \bibnamefont
  {Dolan}},\ }\bibfield  {title} {\enquote {\bibinfo {title} {{Instability of
  the massive Klein-Gordon field on the Kerr spacetime}},}\ }\href {\doibase
  10.1103/PhysRevD.76.084001} {\bibfield  {journal} {\bibinfo  {journal} {Phys.
  Rev.}\ }\textbf {\bibinfo {volume} {D76}},\ \bibinfo {pages} {084001}
  (\bibinfo {year} {2007})},\ \Eprint {http://arxiv.org/abs/0705.2880}
  {arXiv:0705.2880 [gr-qc]} \BibitemShut {NoStop}%
%%CITATION = ARXIV:0705.2880;%%
\bibitem [{\citenamefont {Kapner}(2005)}]{Kapner:2005qy-thesis}%
  \BibitemOpen
  \bibfield  {author} {\bibinfo {author} {\bibfnamefont {Daniel~J.}\
  \bibnamefont {Kapner}},\ }\emph {\bibinfo {title} {{A Short-range test of
  Newton's gravitational inverse-square law}}},\ \href
  {http://wwwlib.umi.com/dissertations/fullcit?p3183373} {Ph.D. thesis},\
  \bibinfo  {school} {Washington U., Seattle} (\bibinfo {year}
  {2005})\BibitemShut {NoStop}%
%%CITATION = UMI-31-83373;%%
\bibitem [{\citenamefont {Biswas}\ \emph {et~al.}(2012)\citenamefont {Biswas},
  \citenamefont {Gerwick}, \citenamefont {Koivisto},\ and\ \citenamefont
  {Mazumdar}}]{Biswas:2011ar}%
  \BibitemOpen
  \bibfield  {author} {\bibinfo {author} {\bibfnamefont {Tirthabir}\
  \bibnamefont {Biswas}}, \bibinfo {author} {\bibfnamefont {Erik}\ \bibnamefont
  {Gerwick}}, \bibinfo {author} {\bibfnamefont {Tomi}\ \bibnamefont
  {Koivisto}}, \ and\ \bibinfo {author} {\bibfnamefont {Anupam}\ \bibnamefont
  {Mazumdar}},\ }\bibfield  {title} {\enquote {\bibinfo {title} {{Towards
  singularity and ghost free theories of gravity}},}\ }\href {\doibase
  10.1103/PhysRevLett.108.031101} {\bibfield  {journal} {\bibinfo  {journal}
  {Phys. Rev. Lett.}\ }\textbf {\bibinfo {volume} {108}},\ \bibinfo {pages}
  {031101} (\bibinfo {year} {2012})},\ \Eprint {http://arxiv.org/abs/1110.5249}
  {arXiv:1110.5249 [gr-qc]} \BibitemShut {NoStop}%
%%CITATION = ARXIV:1110.5249;%%
\bibitem [{\citenamefont {Goroff}\ and\ \citenamefont
  {Sagnotti}(1986)}]{Goroff:1985th}%
  \BibitemOpen
  \bibfield  {author} {\bibinfo {author} {\bibfnamefont {Marc~H.}\ \bibnamefont
  {Goroff}}\ and\ \bibinfo {author} {\bibfnamefont {Augusto}\ \bibnamefont
  {Sagnotti}},\ }\bibfield  {title} {\enquote {\bibinfo {title} {{The
  Ultraviolet Behavior of Einstein Gravity}},}\ }\href {\doibase
  10.1016/0550-3213(86)90193-8} {\bibfield  {journal} {\bibinfo  {journal}
  {Nucl. Phys.}\ }\textbf {\bibinfo {volume} {B266}},\ \bibinfo {pages}
  {709--736} (\bibinfo {year} {1986})}\BibitemShut {NoStop}%
%%CITATION = NUPHA,B266,709;%%
\bibitem [{\citenamefont {Stelle}(1978)}]{Stelle:1977ry}%
  \BibitemOpen
  \bibfield  {author} {\bibinfo {author} {\bibfnamefont {K.~S.}\ \bibnamefont
  {Stelle}},\ }\bibfield  {title} {\enquote {\bibinfo {title} {{Classical
  Gravity with Higher Derivatives}},}\ }\href {\doibase 10.1007/BF00760427}
  {\bibfield  {journal} {\bibinfo  {journal} {Gen. Rel. Grav.}\ }\textbf
  {\bibinfo {volume} {9}},\ \bibinfo {pages} {353--371} (\bibinfo {year}
  {1978})}\BibitemShut {NoStop}%
%%CITATION = GRGVA,9,353;%%
\bibitem [{\citenamefont {Deser}\ and\ \citenamefont
  {Redlich}(1986)}]{Deser:1986xr}%
  \BibitemOpen
  \bibfield  {author} {\bibinfo {author} {\bibfnamefont {Stanley}\ \bibnamefont
  {Deser}}\ and\ \bibinfo {author} {\bibfnamefont {A.~N.}\ \bibnamefont
  {Redlich}},\ }\bibfield  {title} {\enquote {\bibinfo {title} {{String Induced
  Gravity and Ghost Freedom}},}\ }\href {\doibase 10.1016/0370-2693(86)90177-2}
  {\bibfield  {journal} {\bibinfo  {journal} {Phys. Lett.}\ }\textbf {\bibinfo
  {volume} {B176}},\ \bibinfo {pages} {350} (\bibinfo {year} {1986})},\
  \bibinfo {note} {[Erratum: Phys. Lett.B186,461(1987)]}\BibitemShut {NoStop}%
%%CITATION = PHLTA,B176,350;%%
\bibitem [{\citenamefont {Modesto}(2012)}]{Modesto:2011kw}%
  \BibitemOpen
  \bibfield  {author} {\bibinfo {author} {\bibfnamefont {Leonardo}\
  \bibnamefont {Modesto}},\ }\bibfield  {title} {\enquote {\bibinfo {title}
  {{Super-renormalizable Quantum Gravity}},}\ }\href {\doibase
  10.1103/PhysRevD.86.044005} {\bibfield  {journal} {\bibinfo  {journal} {Phys.
  Rev.}\ }\textbf {\bibinfo {volume} {D86}},\ \bibinfo {pages} {044005}
  (\bibinfo {year} {2012})},\ \Eprint {http://arxiv.org/abs/1107.2403}
  {arXiv:1107.2403 [hep-th]} \BibitemShut {NoStop}%
%%CITATION = ARXIV:1107.2403;%%
\bibitem [{\citenamefont
  {Frolov}(2008)}]{Frolov:2008uf-singularity-problem-forfR}%
  \BibitemOpen
  \bibfield  {author} {\bibinfo {author} {\bibfnamefont {Andrei~V.}\
  \bibnamefont {Frolov}},\ }\bibfield  {title} {\enquote {\bibinfo {title} {{A
  Singularity Problem with f(R) Dark Energy}},}\ }\href {\doibase
  10.1103/PhysRevLett.101.061103} {\bibfield  {journal} {\bibinfo  {journal}
  {Phys. Rev. Lett.}\ }\textbf {\bibinfo {volume} {101}},\ \bibinfo {pages}
  {061103} (\bibinfo {year} {2008})},\ \Eprint {http://arxiv.org/abs/0803.2500}
  {arXiv:0803.2500 [astro-ph]} \BibitemShut {NoStop}%
%%CITATION = ARXIV:0803.2500;%%
\bibitem [{\citenamefont {Maggiore}\ and\ \citenamefont
  {Mancarella}(2014)}]{Maggiore:2014sia-nonlocal-gravity-oscil}%
  \BibitemOpen
  \bibfield  {author} {\bibinfo {author} {\bibfnamefont {Michele}\ \bibnamefont
  {Maggiore}}\ and\ \bibinfo {author} {\bibfnamefont {Michele}\ \bibnamefont
  {Mancarella}},\ }\bibfield  {title} {\enquote {\bibinfo {title} {{Nonlocal
  gravity and dark energy}},}\ }\href {\doibase 10.1103/PhysRevD.90.023005}
  {\bibfield  {journal} {\bibinfo  {journal} {Phys. Rev.}\ }\textbf {\bibinfo
  {volume} {D90}},\ \bibinfo {pages} {023005} (\bibinfo {year} {2014})},\
  \Eprint {http://arxiv.org/abs/1402.0448} {arXiv:1402.0448 [hep-th]}
  \BibitemShut {NoStop}%
%%CITATION = ARXIV:1402.0448;%%
\bibitem [{\citenamefont {Park}\ and\ \citenamefont
  {Dodelson}(2013)}]{Park:2012cp}%
  \BibitemOpen
  \bibfield  {author} {\bibinfo {author} {\bibfnamefont {Sohyun}\ \bibnamefont
  {Park}}\ and\ \bibinfo {author} {\bibfnamefont {Scott}\ \bibnamefont
  {Dodelson}},\ }\bibfield  {title} {\enquote {\bibinfo {title} {{Structure
  formation in a nonlocally modified gravity model}},}\ }\href {\doibase
  10.1103/PhysRevD.87.024003} {\bibfield  {journal} {\bibinfo  {journal} {Phys.
  Rev.}\ }\textbf {\bibinfo {volume} {D87}},\ \bibinfo {pages} {024003}
  (\bibinfo {year} {2013})},\ \Eprint {http://arxiv.org/abs/1209.0836}
  {arXiv:1209.0836 [astro-ph.CO]} \BibitemShut {NoStop}%
%%CITATION = ARXIV:1209.0836;%%
\bibitem [{\citenamefont {Calcagni}\ and\ \citenamefont
  {Nardelli}(2010)}]{Calcagni:2010ab}%
  \BibitemOpen
  \bibfield  {author} {\bibinfo {author} {\bibfnamefont {Gianluca}\
  \bibnamefont {Calcagni}}\ and\ \bibinfo {author} {\bibfnamefont {Giuseppe}\
  \bibnamefont {Nardelli}},\ }\bibfield  {title} {\enquote {\bibinfo {title}
  {{Non-local gravity and the diffusion equation}},}\ }\href {\doibase
  10.1103/PhysRevD.82.123518} {\bibfield  {journal} {\bibinfo  {journal} {Phys.
  Rev.}\ }\textbf {\bibinfo {volume} {D82}},\ \bibinfo {pages} {123518}
  (\bibinfo {year} {2010})},\ \Eprint {http://arxiv.org/abs/1004.5144}
  {arXiv:1004.5144 [hep-th]} \BibitemShut {NoStop}%
%%CITATION = ARXIV:1004.5144;%%
\bibitem [{\citenamefont {Barvinsky}(2012)}]{Barvinsky:2011hd}%
  \BibitemOpen
  \bibfield  {author} {\bibinfo {author} {\bibfnamefont {A.~O.}\ \bibnamefont
  {Barvinsky}},\ }\bibfield  {title} {\enquote {\bibinfo {title} {{Dark energy
  and dark matter from nonlocal ghost-free gravity theory}},}\ }\href {\doibase
  10.1016/j.physletb.2012.02.075} {\bibfield  {journal} {\bibinfo  {journal}
  {Phys. Lett.}\ }\textbf {\bibinfo {volume} {B710}},\ \bibinfo {pages}
  {12--16} (\bibinfo {year} {2012})},\ \Eprint {http://arxiv.org/abs/1107.1463}
  {arXiv:1107.1463 [hep-th]} \BibitemShut {NoStop}%
%%CITATION = ARXIV:1107.1463;%%
\bibitem [{\citenamefont {Dirian}\ \emph {et~al.}(2016)\citenamefont {Dirian},
  \citenamefont {Foffa}, \citenamefont {Kunz}, \citenamefont {Maggiore},\ and\
  \citenamefont {Pettorino}}]{Dirian:2016puz}%
  \BibitemOpen
  \bibfield  {author} {\bibinfo {author} {\bibfnamefont {Yves}\ \bibnamefont
  {Dirian}}, \bibinfo {author} {\bibfnamefont {Stefano}\ \bibnamefont {Foffa}},
  \bibinfo {author} {\bibfnamefont {Martin}\ \bibnamefont {Kunz}}, \bibinfo
  {author} {\bibfnamefont {Michele}\ \bibnamefont {Maggiore}}, \ and\ \bibinfo
  {author} {\bibfnamefont {Valeria}\ \bibnamefont {Pettorino}},\ }\bibfield
  {title} {\enquote {\bibinfo {title} {{Non-local gravity and comparison with
  observational datasets. II. Updated results and Bayesian model comparison
  with $\Lambda$CDM}},}\ }\href {\doibase 10.1088/1475-7516/2016/05/068}
  {\bibfield  {journal} {\bibinfo  {journal} {JCAP}\ }\textbf {\bibinfo
  {volume} {1605}},\ \bibinfo {pages} {068} (\bibinfo {year} {2016})},\ \Eprint
  {http://arxiv.org/abs/1602.03558} {arXiv:1602.03558 [astro-ph.CO]}
  \BibitemShut {NoStop}%
%%CITATION = ARXIV:1602.03558;%%
\bibitem [{\citenamefont {Edholm}\ and\ \citenamefont
  {Conroy}(2017)}]{Conroy:2017nkc}%
  \BibitemOpen
  \bibfield  {author} {\bibinfo {author} {\bibfnamefont {James}\ \bibnamefont
  {Edholm}}\ and\ \bibinfo {author} {\bibfnamefont {Aindriú}\ \bibnamefont
  {Conroy}},\ }\bibfield  {title} {\enquote {\bibinfo {title} {{Newtonian
  Potential and Geodesic Completeness in Infinite Derivative Gravity}},}\
  }\href {\doibase 10.1103/PhysRevD.96.044012} {\bibfield  {journal} {\bibinfo
  {journal} {Phys. Rev.}\ }\textbf {\bibinfo {volume} {D96}},\ \bibinfo {pages}
  {044012} (\bibinfo {year} {2017})},\ \Eprint
  {http://arxiv.org/abs/1705.02382} {arXiv:1705.02382 [gr-qc]} \BibitemShut
  {NoStop}%
%%CITATION = ARXIV:1705.02382;%%
\bibitem [{\citenamefont {Song}\ and\ \citenamefont
  {Percival}(2009)}]{Song:2008qt}%
  \BibitemOpen
  \bibfield  {author} {\bibinfo {author} {\bibfnamefont {Yong-Seon}\
  \bibnamefont {Song}}\ and\ \bibinfo {author} {\bibfnamefont {Will~J.}\
  \bibnamefont {Percival}},\ }\bibfield  {title} {\enquote {\bibinfo {title}
  {{Reconstructing the history of structure formation using Redshift
  Distortions}},}\ }\href {\doibase 10.1088/1475-7516/2009/10/004} {\bibfield
  {journal} {\bibinfo  {journal} {JCAP}\ }\textbf {\bibinfo {volume} {0910}},\
  \bibinfo {pages} {004} (\bibinfo {year} {2009})},\ \Eprint
  {http://arxiv.org/abs/0807.0810} {arXiv:0807.0810 [astro-ph]} \BibitemShut
  {NoStop}%
%%CITATION = ARXIV:0807.0810;%%
\bibitem [{\citenamefont {Tegmark}\ \emph {et~al.}(2006)\citenamefont {Tegmark}
  \emph {et~al.}}]{Tegmark:2006az}%
  \BibitemOpen
  \bibfield  {author} {\bibinfo {author} {\bibfnamefont {Max}\ \bibnamefont
  {Tegmark}} \emph {et~al.} (\bibinfo {collaboration} {SDSS}),\ }\bibfield
  {title} {\enquote {\bibinfo {title} {{Cosmological Constraints from the SDSS
  Luminous Red Galaxies}},}\ }\href {\doibase 10.1103/PhysRevD.74.123507}
  {\bibfield  {journal} {\bibinfo  {journal} {Phys. Rev.}\ }\textbf {\bibinfo
  {volume} {D74}},\ \bibinfo {pages} {123507} (\bibinfo {year} {2006})},\
  \Eprint {http://arxiv.org/abs/astro-ph/0608632} {arXiv:astro-ph/0608632
  [astro-ph]} \BibitemShut {NoStop}%
%%CITATION = ASTRO-PH/0608632;%%
\bibitem [{\citenamefont {Davis}\ \emph {et~al.}(2011)\citenamefont {Davis},
  \citenamefont {Nusser}, \citenamefont {Masters}, \citenamefont {Springob},
  \citenamefont {Huchra},\ and\ \citenamefont {Lemson}}]{Davis:2010sw}%
  \BibitemOpen
  \bibfield  {author} {\bibinfo {author} {\bibfnamefont {Marc}\ \bibnamefont
  {Davis}}, \bibinfo {author} {\bibfnamefont {Adi}\ \bibnamefont {Nusser}},
  \bibinfo {author} {\bibfnamefont {Karen}\ \bibnamefont {Masters}}, \bibinfo
  {author} {\bibfnamefont {Christopher}\ \bibnamefont {Springob}}, \bibinfo
  {author} {\bibfnamefont {John~P.}\ \bibnamefont {Huchra}}, \ and\ \bibinfo
  {author} {\bibfnamefont {Gerard}\ \bibnamefont {Lemson}},\ }\bibfield
  {title} {\enquote {\bibinfo {title} {{Local Gravity versus Local Velocity:
  Solutions for $\beta$ and nonlinear bias}},}\ }\href {\doibase
  10.1111/j.1365-2966.2011.18362.x} {\bibfield  {journal} {\bibinfo  {journal}
  {Mon. Not. Roy. Astron. Soc.}\ }\textbf {\bibinfo {volume} {413}},\ \bibinfo
  {pages} {2906} (\bibinfo {year} {2011})},\ \Eprint
  {http://arxiv.org/abs/1011.3114} {arXiv:1011.3114 [astro-ph.CO]} \BibitemShut
  {NoStop}%
%%CITATION = ARXIV:1011.3114;%%
\bibitem [{\citenamefont {Hudson}\ and\ \citenamefont
  {Turnbull}(2012)}]{Hudson:2012gt}%
  \BibitemOpen
  \bibfield  {author} {\bibinfo {author} {\bibfnamefont {Michael~J.}\
  \bibnamefont {Hudson}}\ and\ \bibinfo {author} {\bibfnamefont {Stephen~J.}\
  \bibnamefont {Turnbull}},\ }\bibfield  {title} {\enquote {\bibinfo {title}
  {The growth rate of cosmic structure from peculiar velocities at low and high
  redshifts},}\ }\href {http://stacks.iop.org/2041-8205/751/i=2/a=L30}
  {\bibfield  {journal} {\bibinfo  {journal} {The Astrophysical Journal
  Letters}\ }\textbf {\bibinfo {volume} {751}},\ \bibinfo {pages} {L30}
  (\bibinfo {year} {2012})}\BibitemShut {NoStop}%
\bibitem [{\citenamefont {Turnbull}\ \emph {et~al.}(2012)\citenamefont
  {Turnbull}, \citenamefont {Hudson}, \citenamefont {Feldman}, \citenamefont
  {Hicken}, \citenamefont {Kirshner},\ and\ \citenamefont
  {Watkins}}]{Turnbull:2011ty}%
  \BibitemOpen
  \bibfield  {author} {\bibinfo {author} {\bibfnamefont {Stephen~J.}\
  \bibnamefont {Turnbull}}, \bibinfo {author} {\bibfnamefont {Michael~J.}\
  \bibnamefont {Hudson}}, \bibinfo {author} {\bibfnamefont {Hume~A.}\
  \bibnamefont {Feldman}}, \bibinfo {author} {\bibfnamefont {Malcolm}\
  \bibnamefont {Hicken}}, \bibinfo {author} {\bibfnamefont {Robert~P.}\
  \bibnamefont {Kirshner}}, \ and\ \bibinfo {author} {\bibfnamefont {Richard}\
  \bibnamefont {Watkins}},\ }\bibfield  {title} {\enquote {\bibinfo {title}
  {Cosmic flows in the nearby universe from type ia supernovae},}\ }\href@noop
  {} {\bibfield  {journal} {\bibinfo  {journal} {Monthly Notices of the Royal
  Astronomical Society}\ }\textbf {\bibinfo {volume} {420}},\ \bibinfo {pages}
  {447--454} (\bibinfo {year} {2012})}\BibitemShut {NoStop}%
\bibitem [{\citenamefont {Samushia}\ \emph {et~al.}(2012)\citenamefont
  {Samushia}, \citenamefont {Percival},\ and\ \citenamefont
  {Raccanelli}}]{Samushia:2011cs}%
  \BibitemOpen
  \bibfield  {author} {\bibinfo {author} {\bibfnamefont {L.}~\bibnamefont
  {Samushia}}, \bibinfo {author} {\bibfnamefont {W.~J.}\ \bibnamefont
  {Percival}}, \ and\ \bibinfo {author} {\bibfnamefont {A.}~\bibnamefont
  {Raccanelli}},\ }\bibfield  {title} {\enquote {\bibinfo {title} {Interpreting
  large-scale redshift-space distortion measurements},}\ }\href {\doibase
  10.1111/j.1365-2966.2011.20169.x} {\bibfield  {journal} {\bibinfo  {journal}
  {Monthly Notices of the Royal Astronomical Society}\ }\textbf {\bibinfo
  {volume} {420}},\ \bibinfo {pages} {2102--2119} (\bibinfo {year}
  {2012})}\BibitemShut {NoStop}%
\bibitem [{\citenamefont {Beutler}\ \emph {et~al.}(2012)\citenamefont
  {Beutler}, \citenamefont {Blake}, \citenamefont {Colless}, \citenamefont
  {Jones}, \citenamefont {Staveley-Smith}, \citenamefont {Poole}, \citenamefont
  {Campbell}, \citenamefont {Parker}, \citenamefont {Saunders},\ and\
  \citenamefont {Watson}}]{Beutler:2012px}%
  \BibitemOpen
  \bibfield  {author} {\bibinfo {author} {\bibfnamefont {Florian}\ \bibnamefont
  {Beutler}}, \bibinfo {author} {\bibfnamefont {Chris}\ \bibnamefont {Blake}},
  \bibinfo {author} {\bibfnamefont {Matthew}\ \bibnamefont {Colless}}, \bibinfo
  {author} {\bibfnamefont {D.~Heath}\ \bibnamefont {Jones}}, \bibinfo {author}
  {\bibfnamefont {Lister}\ \bibnamefont {Staveley-Smith}}, \bibinfo {author}
  {\bibfnamefont {Gregory~B.}\ \bibnamefont {Poole}}, \bibinfo {author}
  {\bibfnamefont {Lachlan}\ \bibnamefont {Campbell}}, \bibinfo {author}
  {\bibfnamefont {Quentin}\ \bibnamefont {Parker}}, \bibinfo {author}
  {\bibfnamefont {Will}\ \bibnamefont {Saunders}}, \ and\ \bibinfo {author}
  {\bibfnamefont {Fred}\ \bibnamefont {Watson}},\ }\bibfield  {title} {\enquote
  {\bibinfo {title} {The 6df galaxy survey: $z \approx 0$ measurements of the
  growth rate and $\sigma 8$},}\ }\href {\doibase
  10.1111/j.1365-2966.2012.21136.x} {\bibfield  {journal} {\bibinfo  {journal}
  {Monthly Notices of the Royal Astronomical Society}\ }\textbf {\bibinfo
  {volume} {423}},\ \bibinfo {pages} {3430--3444} (\bibinfo {year}
  {2012})}\BibitemShut {NoStop}%
\bibitem [{\citenamefont {Tojeiro}\ \emph {et~al.}(2012)\citenamefont
  {Tojeiro}, \citenamefont {Percival}, \citenamefont {Brinkmann}, \citenamefont
  {Brownstein}, \citenamefont {Eisenstein}, \citenamefont {Manera},
  \citenamefont {Maraston}, \citenamefont {McBride}, \citenamefont {Muna},
  \citenamefont {Reid}, \citenamefont {Ross}, \citenamefont {Ross},
  \citenamefont {Samushia}, \citenamefont {Padmanabhan}, \citenamefont
  {Schneider}, \citenamefont {Skibba}, \citenamefont {Sánchez}, \citenamefont
  {Swanson}, \citenamefont {Thomas}, \citenamefont {Tinker}, \citenamefont
  {Verde}, \citenamefont {Wake}, \citenamefont {Weaver},\ and\ \citenamefont
  {Zhao}}]{Tojeiro:2012rp}%
  \BibitemOpen
  \bibfield  {author} {\bibinfo {author} {\bibfnamefont {Rita}\ \bibnamefont
  {Tojeiro}}, \bibinfo {author} {\bibfnamefont {Will~J.}\ \bibnamefont
  {Percival}}, \bibinfo {author} {\bibfnamefont {Jon}\ \bibnamefont
  {Brinkmann}}, \bibinfo {author} {\bibfnamefont {Joel~R.}\ \bibnamefont
  {Brownstein}}, \bibinfo {author} {\bibfnamefont {Daniel~J.}\ \bibnamefont
  {Eisenstein}}, \bibinfo {author} {\bibfnamefont {Marc}\ \bibnamefont
  {Manera}}, \bibinfo {author} {\bibfnamefont {Claudia}\ \bibnamefont
  {Maraston}}, \bibinfo {author} {\bibfnamefont {Cameron~K.}\ \bibnamefont
  {McBride}}, \bibinfo {author} {\bibfnamefont {Demitri}\ \bibnamefont {Muna}},
  \bibinfo {author} {\bibfnamefont {Beth}\ \bibnamefont {Reid}}, \bibinfo
  {author} {\bibfnamefont {Ashley~J.}\ \bibnamefont {Ross}}, \bibinfo {author}
  {\bibfnamefont {Nicholas~P.}\ \bibnamefont {Ross}}, \bibinfo {author}
  {\bibfnamefont {Lado}\ \bibnamefont {Samushia}}, \bibinfo {author}
  {\bibfnamefont {Nikhil}\ \bibnamefont {Padmanabhan}}, \bibinfo {author}
  {\bibfnamefont {Donald~P.}\ \bibnamefont {Schneider}}, \bibinfo {author}
  {\bibfnamefont {Ramin}\ \bibnamefont {Skibba}}, \bibinfo {author}
  {\bibfnamefont {Ariel~G.}\ \bibnamefont {Sánchez}}, \bibinfo {author}
  {\bibfnamefont {Molly E.~C.}\ \bibnamefont {Swanson}}, \bibinfo {author}
  {\bibfnamefont {Daniel}\ \bibnamefont {Thomas}}, \bibinfo {author}
  {\bibfnamefont {Jeremy~L.}\ \bibnamefont {Tinker}}, \bibinfo {author}
  {\bibfnamefont {Licia}\ \bibnamefont {Verde}}, \bibinfo {author}
  {\bibfnamefont {David~A.}\ \bibnamefont {Wake}}, \bibinfo {author}
  {\bibfnamefont {Benjamin~A.}\ \bibnamefont {Weaver}}, \ and\ \bibinfo
  {author} {\bibfnamefont {Gong-Bo}\ \bibnamefont {Zhao}},\ }\bibfield  {title}
  {\enquote {\bibinfo {title} {The clustering of galaxies in the sdss-iii
  baryon oscillation spectroscopic survey: measuring structure growth using
  passive galaxies},}\ }\href {\doibase 10.1111/j.1365-2966.2012.21404.x}
  {\bibfield  {journal} {\bibinfo  {journal} {Monthly Notices of the Royal
  Astronomical Society}\ }\textbf {\bibinfo {volume} {424}},\ \bibinfo {pages}
  {2339--2344} (\bibinfo {year} {2012})}\BibitemShut {NoStop}%
\bibitem [{\citenamefont {de~la Torre}\ \emph {et~al.}(2013)\citenamefont
  {de~la Torre} \emph {et~al.}}]{delaTorre:2013rpa}%
  \BibitemOpen
  \bibfield  {author} {\bibinfo {author} {\bibfnamefont {S.}~\bibnamefont
  {de~la Torre}} \emph {et~al.},\ }\bibfield  {title} {\enquote {\bibinfo
  {title} {{The VIMOS Public Extragalactic Redshift Survey (VIPERS). Galaxy
  clustering and redshift-space distortions at z=0.8 in the first data
  release}},}\ }\href {\doibase 10.1051/0004-6361/201321463} {\bibfield
  {journal} {\bibinfo  {journal} {Astron. Astrophys.}\ }\textbf {\bibinfo
  {volume} {557}},\ \bibinfo {pages} {A54} (\bibinfo {year} {2013})},\ \Eprint
  {http://arxiv.org/abs/1303.2622} {arXiv:1303.2622 [astro-ph.CO]} \BibitemShut
  {NoStop}%
%%CITATION = ARXIV:1303.2622;%%
\bibitem [{\citenamefont {Chuang}\ and\ \citenamefont
  {Wang}(2013)}]{Chuang:2012qt}%
  \BibitemOpen
  \bibfield  {author} {\bibinfo {author} {\bibfnamefont {Chia-Hsun}\
  \bibnamefont {Chuang}}\ and\ \bibinfo {author} {\bibfnamefont {Yun}\
  \bibnamefont {Wang}},\ }\bibfield  {title} {\enquote {\bibinfo {title}
  {Modelling the anisotropic two-point galaxy correlation function on small
  scales and single-probe measurements of $h(z)$, $da(z)$ and $f(z)\sigma 8(z)$
  from the sloan digital sky survey dr7 luminous red galaxies},}\ }\href
  {\doibase 10.1093/mnras/stt1290} {\bibfield  {journal} {\bibinfo  {journal}
  {Monthly Notices of the Royal Astronomical Society}\ }\textbf {\bibinfo
  {volume} {435}},\ \bibinfo {pages} {255--262} (\bibinfo {year}
  {2013})}\BibitemShut {NoStop}%
\bibitem [{\citenamefont {Komatsu}\ \emph {et~al.}(2011)\citenamefont {Komatsu}
  \emph {et~al.}}]{Komatsu:2010fb}%
  \BibitemOpen
  \bibfield  {author} {\bibinfo {author} {\bibfnamefont {E.}~\bibnamefont
  {Komatsu}} \emph {et~al.} (\bibinfo {collaboration} {WMAP}),\ }\bibfield
  {title} {\enquote {\bibinfo {title} {{Seven-Year Wilkinson Microwave
  Anisotropy Probe (WMAP) Observations: Cosmological Interpretation}},}\ }\href
  {\doibase 10.1088/0067-0049/192/2/18} {\bibfield  {journal} {\bibinfo
  {journal} {Astrophys. J. Suppl.}\ }\textbf {\bibinfo {volume} {192}},\
  \bibinfo {pages} {18} (\bibinfo {year} {2011})},\ \Eprint
  {http://arxiv.org/abs/1001.4538} {arXiv:1001.4538 [astro-ph.CO]} \BibitemShut
  {NoStop}%
%%CITATION = ARXIV:1001.4538;%%
\bibitem [{\citenamefont {Blake}\ \emph {et~al.}(2013)\citenamefont {Blake}
  \emph {et~al.}}]{Blake:2013nif}%
  \BibitemOpen
  \bibfield  {author} {\bibinfo {author} {\bibfnamefont {Chris}\ \bibnamefont
  {Blake}} \emph {et~al.},\ }\bibfield  {title} {\enquote {\bibinfo {title}
  {{Galaxy And Mass Assembly (GAMA): improved cosmic growth measurements using
  multiple tracers of large-scale structure}},}\ }\href {\doibase
  10.1093/mnras/stt1791} {\bibfield  {journal} {\bibinfo  {journal} {Mon. Not.
  Roy. Astron. Soc.}\ }\textbf {\bibinfo {volume} {436}},\ \bibinfo {pages}
  {3089} (\bibinfo {year} {2013})},\ \Eprint {http://arxiv.org/abs/1309.5556}
  {arXiv:1309.5556 [astro-ph.CO]} \BibitemShut {NoStop}%
%%CITATION = ARXIV:1309.5556;%%
\bibitem [{\citenamefont {Sanchez}\ \emph {et~al.}(2014)\citenamefont {Sanchez}
  \emph {et~al.}}]{Sanchez:2013tga}%
  \BibitemOpen
  \bibfield  {author} {\bibinfo {author} {\bibfnamefont {Ariel~G.}\
  \bibnamefont {Sanchez}} \emph {et~al.},\ }\bibfield  {title} {\enquote
  {\bibinfo {title} {{The clustering of galaxies in the SDSS-III Baryon
  Oscillation Spectroscopic Survey: cosmological implications of the full shape
  of the clustering wedges in the data release 10 and 11 galaxy samples}},}\
  }\href {\doibase 10.1093/mnras/stu342} {\bibfield  {journal} {\bibinfo
  {journal} {Mon. Not. Roy. Astron. Soc.}\ }\textbf {\bibinfo {volume} {440}},\
  \bibinfo {pages} {2692--2713} (\bibinfo {year} {2014})},\ \Eprint
  {http://arxiv.org/abs/1312.4854} {arXiv:1312.4854 [astro-ph.CO]} \BibitemShut
  {NoStop}%
%%CITATION = ARXIV:1312.4854;%%
\bibitem [{\citenamefont {Anderson}\ \emph {et~al.}(2014)\citenamefont
  {Anderson} \emph {et~al.}}]{Anderson:2013zyy}%
  \BibitemOpen
  \bibfield  {author} {\bibinfo {author} {\bibfnamefont {Lauren}\ \bibnamefont
  {Anderson}} \emph {et~al.} (\bibinfo {collaboration} {BOSS}),\ }\bibfield
  {title} {\enquote {\bibinfo {title} {{The clustering of galaxies in the
  SDSS-III Baryon Oscillation Spectroscopic Survey: baryon acoustic
  oscillations in the Data Releases 10 and 11 Galaxy samples}},}\ }\href
  {\doibase 10.1093/mnras/stu523} {\bibfield  {journal} {\bibinfo  {journal}
  {Mon. Not. Roy. Astron. Soc.}\ }\textbf {\bibinfo {volume} {441}},\ \bibinfo
  {pages} {24--62} (\bibinfo {year} {2014})},\ \Eprint
  {http://arxiv.org/abs/1312.4877} {arXiv:1312.4877 [astro-ph.CO]} \BibitemShut
  {NoStop}%
%%CITATION = ARXIV:1312.4877;%%
\bibitem [{\citenamefont {Howlett}\ \emph {et~al.}(2015)\citenamefont
  {Howlett}, \citenamefont {Ross}, \citenamefont {Samushia}, \citenamefont
  {Percival},\ and\ \citenamefont {Manera}}]{Howlett:2014opa}%
  \BibitemOpen
  \bibfield  {author} {\bibinfo {author} {\bibfnamefont {Cullan}\ \bibnamefont
  {Howlett}}, \bibinfo {author} {\bibfnamefont {Ashley}\ \bibnamefont {Ross}},
  \bibinfo {author} {\bibfnamefont {Lado}\ \bibnamefont {Samushia}}, \bibinfo
  {author} {\bibfnamefont {Will}\ \bibnamefont {Percival}}, \ and\ \bibinfo
  {author} {\bibfnamefont {Marc}\ \bibnamefont {Manera}},\ }\bibfield  {title}
  {\enquote {\bibinfo {title} {{The clustering of the SDSS main galaxy sample
  – II. Mock galaxy catalogues and a measurement of the growth of structure
  from redshift space distortions at $z = 0.15$}},}\ }\href {\doibase
  10.1093/mnras/stu2693} {\bibfield  {journal} {\bibinfo  {journal} {Mon. Not.
  Roy. Astron. Soc.}\ }\textbf {\bibinfo {volume} {449}},\ \bibinfo {pages}
  {848--866} (\bibinfo {year} {2015})},\ \Eprint
  {http://arxiv.org/abs/1409.3238} {arXiv:1409.3238 [astro-ph.CO]} \BibitemShut
  {NoStop}%
%%CITATION = ARXIV:1409.3238;%%
\bibitem [{\citenamefont {Feix}\ \emph {et~al.}(2015)\citenamefont {Feix},
  \citenamefont {Nusser},\ and\ \citenamefont {Branchini}}]{Feix:2015dla}%
  \BibitemOpen
  \bibfield  {author} {\bibinfo {author} {\bibfnamefont {Martin}\ \bibnamefont
  {Feix}}, \bibinfo {author} {\bibfnamefont {Adi}\ \bibnamefont {Nusser}}, \
  and\ \bibinfo {author} {\bibfnamefont {Enzo}\ \bibnamefont {Branchini}},\
  }\bibfield  {title} {\enquote {\bibinfo {title} {{Growth Rate of Cosmological
  Perturbations at $z \approx 0.1$ from a New Observational Test}},}\ }\href
  {\doibase 10.1103/PhysRevLett.115.011301} {\bibfield  {journal} {\bibinfo
  {journal} {Phys. Rev. Lett.}\ }\textbf {\bibinfo {volume} {115}},\ \bibinfo
  {pages} {011301} (\bibinfo {year} {2015})},\ \Eprint
  {http://arxiv.org/abs/1503.05945} {arXiv:1503.05945 [astro-ph.CO]}
  \BibitemShut {NoStop}%
%%CITATION = ARXIV:1503.05945;%%
\bibitem [{\citenamefont {Tegmark}\ \emph {et~al.}(2004)\citenamefont {Tegmark}
  \emph {et~al.}}]{Tegmark:2003uf}%
  \BibitemOpen
  \bibfield  {author} {\bibinfo {author} {\bibfnamefont {Max}\ \bibnamefont
  {Tegmark}} \emph {et~al.} (\bibinfo {collaboration} {SDSS}),\ }\bibfield
  {title} {\enquote {\bibinfo {title} {{The 3-D power spectrum of galaxies from
  the SDSS}},}\ }\href {\doibase 10.1086/382125} {\bibfield  {journal}
  {\bibinfo  {journal} {Astrophys. J.}\ }\textbf {\bibinfo {volume} {606}},\
  \bibinfo {pages} {702--740} (\bibinfo {year} {2004})},\ \Eprint
  {http://arxiv.org/abs/astro-ph/0310725} {arXiv:astro-ph/0310725 [astro-ph]}
  \BibitemShut {NoStop}%
%%CITATION = ASTRO-PH/0310725;%%
\bibitem [{\citenamefont {Okumura}\ \emph {et~al.}(2016)\citenamefont {Okumura}
  \emph {et~al.}}]{Okumura:2015lvp}%
  \BibitemOpen
  \bibfield  {author} {\bibinfo {author} {\bibfnamefont {Teppei}\ \bibnamefont
  {Okumura}} \emph {et~al.},\ }\bibfield  {title} {\enquote {\bibinfo {title}
  {{The Subaru FMOS galaxy redshift survey (FastSound). IV. New constraint on
  gravity theory from redshift space distortions at $z\sim 1.4$}},}\ }\href
  {\doibase 10.1093/pasj/psw029} {\bibfield  {journal} {\bibinfo  {journal}
  {Publ. Astron. Soc. Jap.}\ }\textbf {\bibinfo {volume} {68}},\ \bibinfo
  {pages} {24} (\bibinfo {year} {2016})},\ \Eprint
  {http://arxiv.org/abs/1511.08083} {arXiv:1511.08083 [astro-ph.CO]}
  \BibitemShut {NoStop}%
%%CITATION = ARXIV:1511.08083;%%
\bibitem [{\citenamefont {Hinshaw}\ \emph {et~al.}(2013)\citenamefont {Hinshaw}
  \emph {et~al.}}]{Hinshaw:2012aka}%
  \BibitemOpen
  \bibfield  {author} {\bibinfo {author} {\bibfnamefont {G.}~\bibnamefont
  {Hinshaw}} \emph {et~al.} (\bibinfo {collaboration} {WMAP}),\ }\bibfield
  {title} {\enquote {\bibinfo {title} {{Nine-Year Wilkinson Microwave
  Anisotropy Probe (WMAP) Observations: Cosmological Parameter Results}},}\
  }\href {\doibase 10.1088/0067-0049/208/2/19} {\bibfield  {journal} {\bibinfo
  {journal} {Astrophys. J. Suppl.}\ }\textbf {\bibinfo {volume} {208}},\
  \bibinfo {pages} {19} (\bibinfo {year} {2013})},\ \Eprint
  {http://arxiv.org/abs/1212.5226} {arXiv:1212.5226 [astro-ph.CO]} \BibitemShut
  {NoStop}%
%%CITATION = ARXIV:1212.5226;%%
\bibitem [{\citenamefont {Chuang}\ \emph {et~al.}(2016)\citenamefont {Chuang}
  \emph {et~al.}}]{Chuang:2013wga}%
  \BibitemOpen
  \bibfield  {author} {\bibinfo {author} {\bibfnamefont {Chia-Hsun}\
  \bibnamefont {Chuang}} \emph {et~al.},\ }\bibfield  {title} {\enquote
  {\bibinfo {title} {{The clustering of galaxies in the SDSS-III Baryon
  Oscillation Spectroscopic Survey: single-probe measurements from CMASS
  anisotropic galaxy clustering}},}\ }\href {\doibase 10.1093/mnras/stw1535}
  {\bibfield  {journal} {\bibinfo  {journal} {Mon. Not. Roy. Astron. Soc.}\
  }\textbf {\bibinfo {volume} {461}},\ \bibinfo {pages} {3781--3793} (\bibinfo
  {year} {2016})},\ \Eprint {http://arxiv.org/abs/1312.4889} {arXiv:1312.4889
  [astro-ph.CO]} \BibitemShut {NoStop}%
%%CITATION = ARXIV:1312.4889;%%
\bibitem [{\citenamefont {Beutler}\ \emph {et~al.}(2017)\citenamefont {Beutler}
  \emph {et~al.}}]{Beutler:2016arn}%
  \BibitemOpen
  \bibfield  {author} {\bibinfo {author} {\bibfnamefont {Florian}\ \bibnamefont
  {Beutler}} \emph {et~al.} (\bibinfo {collaboration} {BOSS}),\ }\bibfield
  {title} {\enquote {\bibinfo {title} {{The clustering of galaxies in the
  completed SDSS-III Baryon Oscillation Spectroscopic Survey: Anisotropic
  galaxy clustering in Fourier-space}},}\ }\href {\doibase
  10.1093/mnras/stw3298} {\bibfield  {journal} {\bibinfo  {journal} {Mon. Not.
  Roy. Astron. Soc.}\ }\textbf {\bibinfo {volume} {466}},\ \bibinfo {pages}
  {2242--2260} (\bibinfo {year} {2017})},\ \Eprint
  {http://arxiv.org/abs/1607.03150} {arXiv:1607.03150 [astro-ph.CO]}
  \BibitemShut {NoStop}%
%%CITATION = ARXIV:1607.03150;%%
\bibitem [{\citenamefont {Gil-Marín}\ \emph {et~al.}(2017)\citenamefont
  {Gil-Marín}, \citenamefont {Percival}, \citenamefont {Verde}, \citenamefont
  {Brownstein}, \citenamefont {Chuang}, \citenamefont {Kitaura}, \citenamefont
  {Rodríguez-Torres},\ and\ \citenamefont {Olmstead}}]{Gil-Marin:2016wya}%
  \BibitemOpen
  \bibfield  {author} {\bibinfo {author} {\bibfnamefont {Héctor}\ \bibnamefont
  {Gil-Marín}}, \bibinfo {author} {\bibfnamefont {Will~J.}\ \bibnamefont
  {Percival}}, \bibinfo {author} {\bibfnamefont {Licia}\ \bibnamefont {Verde}},
  \bibinfo {author} {\bibfnamefont {Joel~R.}\ \bibnamefont {Brownstein}},
  \bibinfo {author} {\bibfnamefont {Chia-Hsun}\ \bibnamefont {Chuang}},
  \bibinfo {author} {\bibfnamefont {Francisco-Shu}\ \bibnamefont {Kitaura}},
  \bibinfo {author} {\bibfnamefont {Sergio~A.}\ \bibnamefont
  {Rodríguez-Torres}}, \ and\ \bibinfo {author} {\bibfnamefont {Matthew~D.}\
  \bibnamefont {Olmstead}},\ }\bibfield  {title} {\enquote {\bibinfo {title}
  {{The clustering of galaxies in the SDSS-III Baryon Oscillation Spectroscopic
  Survey: RSD measurement from the power spectrum and bispectrum of the DR12
  BOSS galaxies}},}\ }\href {\doibase 10.1093/mnras/stw2679} {\bibfield
  {journal} {\bibinfo  {journal} {Mon. Not. Roy. Astron. Soc.}\ }\textbf
  {\bibinfo {volume} {465}},\ \bibinfo {pages} {1757--1788} (\bibinfo {year}
  {2017})},\ \Eprint {http://arxiv.org/abs/1606.00439} {arXiv:1606.00439
  [astro-ph.CO]} \BibitemShut {NoStop}%
%%CITATION = ARXIV:1606.00439;%%
\bibitem [{\citenamefont {Hawken}\ \emph {et~al.}(2017)\citenamefont {Hawken}
  \emph {et~al.}}]{Hawken:2016qcy}%
  \BibitemOpen
  \bibfield  {author} {\bibinfo {author} {\bibfnamefont {A.~J.}\ \bibnamefont
  {Hawken}} \emph {et~al.},\ }\bibfield  {title} {\enquote {\bibinfo {title}
  {{The VIMOS Public Extragalactic Redshift Survey: Measuring the growth rate
  of structure around cosmic voids}},}\ }\href {\doibase
  10.1051/0004-6361/201629678} {\bibfield  {journal} {\bibinfo  {journal}
  {Astron. Astrophys.}\ }\textbf {\bibinfo {volume} {607}},\ \bibinfo {pages}
  {A54} (\bibinfo {year} {2017})},\ \Eprint {http://arxiv.org/abs/1611.07046}
  {arXiv:1611.07046 [astro-ph.CO]} \BibitemShut {NoStop}%
%%CITATION = ARXIV:1611.07046;%%
\bibitem [{\citenamefont {Huterer}\ \emph {et~al.}(2017)\citenamefont
  {Huterer}, \citenamefont {Shafer}, \citenamefont {Scolnic},\ and\
  \citenamefont {Schmidt}}]{Huterer:2016uyq}%
  \BibitemOpen
  \bibfield  {author} {\bibinfo {author} {\bibfnamefont {Dragan}\ \bibnamefont
  {Huterer}}, \bibinfo {author} {\bibfnamefont {Daniel}\ \bibnamefont
  {Shafer}}, \bibinfo {author} {\bibfnamefont {Daniel}\ \bibnamefont
  {Scolnic}}, \ and\ \bibinfo {author} {\bibfnamefont {Fabian}\ \bibnamefont
  {Schmidt}},\ }\bibfield  {title} {\enquote {\bibinfo {title} {{Testing
  $\Lambda$CDM at the lowest redshifts with SN Ia and galaxy velocities}},}\
  }\href {\doibase 10.1088/1475-7516/2017/05/015} {\bibfield  {journal}
  {\bibinfo  {journal} {JCAP}\ }\textbf {\bibinfo {volume} {1705}},\ \bibinfo
  {pages} {015} (\bibinfo {year} {2017})},\ \Eprint
  {http://arxiv.org/abs/1611.09862} {arXiv:1611.09862 [astro-ph.CO]}
  \BibitemShut {NoStop}%
%%CITATION = ARXIV:1611.09862;%%
\bibitem [{\citenamefont {de~la Torre}\ \emph {et~al.}(2017)\citenamefont
  {de~la Torre} \emph {et~al.}}]{delaTorre:2016rxm}%
  \BibitemOpen
  \bibfield  {author} {\bibinfo {author} {\bibfnamefont {S.}~\bibnamefont
  {de~la Torre}} \emph {et~al.},\ }\bibfield  {title} {\enquote {\bibinfo
  {title} {{The VIMOS Public Extragalactic Redshift Survey (VIPERS). Gravity
  test from the combination of redshift-space distortions and galaxy-galaxy
  lensing at $0.5 < z < 1.2$}},}\ }\href {\doibase 10.1051/0004-6361/201630276}
  {\bibfield  {journal} {\bibinfo  {journal} {Astron. Astrophys.}\ }\textbf
  {\bibinfo {volume} {608}},\ \bibinfo {pages} {A44} (\bibinfo {year}
  {2017})},\ \Eprint {http://arxiv.org/abs/1612.05647} {arXiv:1612.05647
  [astro-ph.CO]} \BibitemShut {NoStop}%
%%CITATION = ARXIV:1612.05647;%%
\bibitem [{\citenamefont {Pezzotta}\ \emph {et~al.}(2017)\citenamefont
  {Pezzotta} \emph {et~al.}}]{Pezzotta:2016gbo}%
  \BibitemOpen
  \bibfield  {author} {\bibinfo {author} {\bibfnamefont {A.}~\bibnamefont
  {Pezzotta}} \emph {et~al.},\ }\bibfield  {title} {\enquote {\bibinfo {title}
  {{The VIMOS Public Extragalactic Redshift Survey (VIPERS): The growth of
  structure at $0.5 < z < 1.2$ from redshift-space distortions in the
  clustering of the PDR-2 final sample}},}\ }\href {\doibase
  10.1051/0004-6361/201630295} {\bibfield  {journal} {\bibinfo  {journal}
  {Astron. Astrophys.}\ }\textbf {\bibinfo {volume} {604}},\ \bibinfo {pages}
  {A33} (\bibinfo {year} {2017})},\ \Eprint {http://arxiv.org/abs/1612.05645}
  {arXiv:1612.05645 [astro-ph.CO]} \BibitemShut {NoStop}%
%%CITATION = ARXIV:1612.05645;%%
\bibitem [{\citenamefont {Feix}\ \emph {et~al.}(2017)\citenamefont {Feix},
  \citenamefont {Branchini},\ and\ \citenamefont {Nusser}}]{Feix:2016qhh}%
  \BibitemOpen
  \bibfield  {author} {\bibinfo {author} {\bibfnamefont {Martin}\ \bibnamefont
  {Feix}}, \bibinfo {author} {\bibfnamefont {Enzo}\ \bibnamefont {Branchini}},
  \ and\ \bibinfo {author} {\bibfnamefont {Adi}\ \bibnamefont {Nusser}},\
  }\bibfield  {title} {\enquote {\bibinfo {title} {{Speed from light: growth
  rate and bulk flow at $z \approx 0.1$ from improved SDSS DR13 photometry}},}\
  }\href {\doibase 10.1093/mnras/stx566} {\bibfield  {journal} {\bibinfo
  {journal} {Mon. Not. Roy. Astron. Soc.}\ }\textbf {\bibinfo {volume} {468}},\
  \bibinfo {pages} {1420--1425} (\bibinfo {year} {2017})},\ \Eprint
  {http://arxiv.org/abs/1612.07809} {arXiv:1612.07809 [astro-ph.CO]}
  \BibitemShut {NoStop}%
%%CITATION = ARXIV:1612.07809;%%
\bibitem [{\citenamefont {Howlett}\ \emph {et~al.}(2017)\citenamefont
  {Howlett}, \citenamefont {Staveley-Smith}, \citenamefont {Elahi},
  \citenamefont {Hong}, \citenamefont {Jarrett}, \citenamefont {Jones},
  \citenamefont {Koribalski}, \citenamefont {Macri}, \citenamefont {Masters},\
  and\ \citenamefont {Springob}}]{Howlett:2017asq}%
  \BibitemOpen
  \bibfield  {author} {\bibinfo {author} {\bibfnamefont {Cullan}\ \bibnamefont
  {Howlett}}, \bibinfo {author} {\bibfnamefont {Lister}\ \bibnamefont
  {Staveley-Smith}}, \bibinfo {author} {\bibfnamefont {Pascal~J.}\ \bibnamefont
  {Elahi}}, \bibinfo {author} {\bibfnamefont {Tao}\ \bibnamefont {Hong}},
  \bibinfo {author} {\bibfnamefont {Tom~H.}\ \bibnamefont {Jarrett}}, \bibinfo
  {author} {\bibfnamefont {D.~Heath}\ \bibnamefont {Jones}}, \bibinfo {author}
  {\bibfnamefont {Bärbel~S.}\ \bibnamefont {Koribalski}}, \bibinfo {author}
  {\bibfnamefont {Lucas~M.}\ \bibnamefont {Macri}}, \bibinfo {author}
  {\bibfnamefont {Karen~L.}\ \bibnamefont {Masters}}, \ and\ \bibinfo {author}
  {\bibfnamefont {Christopher~M.}\ \bibnamefont {Springob}},\ }\bibfield
  {title} {\enquote {\bibinfo {title} {{2MTF VI. Measuring the velocity power
  spectrum}},}\ }\href {\doibase 10.1093/mnras/stx1521} {\bibfield  {journal}
  {\bibinfo  {journal} {Mon. Not. Roy. Astron. Soc.}\ }\textbf {\bibinfo
  {volume} {471}},\ \bibinfo {pages} {3135} (\bibinfo {year} {2017})},\ \Eprint
  {http://arxiv.org/abs/1706.05130} {arXiv:1706.05130 [astro-ph.CO]}
  \BibitemShut {NoStop}%
%%CITATION = ARXIV:1706.05130;%%
\bibitem [{\citenamefont {Mohammad}\ \emph {et~al.}(2017)\citenamefont
  {Mohammad} \emph {et~al.}}]{Mohammad:2017lzz}%
  \BibitemOpen
  \bibfield  {author} {\bibinfo {author} {\bibfnamefont {F.~G.}\ \bibnamefont
  {Mohammad}} \emph {et~al.},\ }\bibfield  {title} {\enquote {\bibinfo {title}
  {{The VIMOS Public Extragalactic Redshift Survey (VIPERS): An unbiased
  estimate of the growth rate of structure at $\mathbf{\left<z\right>=0.85}$
  using the clustering of luminous blue galaxies}},}\ }\href@noop {} {\
  (\bibinfo {year} {2017})},\ \Eprint {http://arxiv.org/abs/1708.00026}
  {arXiv:1708.00026 [astro-ph.CO]} \BibitemShut {NoStop}%
%%CITATION = ARXIV:1708.00026;%%
\bibitem [{\citenamefont {Wang}\ \emph {et~al.}(2017)\citenamefont {Wang},
  \citenamefont {Zhao}, \citenamefont {Chuang}, \citenamefont
  {Pellejero-Ibanez}, \citenamefont {Zhao}, \citenamefont {Kitaura},\ and\
  \citenamefont {Rodriguez-Torres}}]{Wang:2017wia}%
  \BibitemOpen
  \bibfield  {author} {\bibinfo {author} {\bibfnamefont {Yuting}\ \bibnamefont
  {Wang}}, \bibinfo {author} {\bibfnamefont {Gong-Bo}\ \bibnamefont {Zhao}},
  \bibinfo {author} {\bibfnamefont {Chia-Hsun}\ \bibnamefont {Chuang}},
  \bibinfo {author} {\bibfnamefont {Marcos}\ \bibnamefont {Pellejero-Ibanez}},
  \bibinfo {author} {\bibfnamefont {Cheng}\ \bibnamefont {Zhao}}, \bibinfo
  {author} {\bibfnamefont {Francisco-Shu}\ \bibnamefont {Kitaura}}, \ and\
  \bibinfo {author} {\bibfnamefont {Sergio}\ \bibnamefont {Rodriguez-Torres}},\
  }\bibfield  {title} {\enquote {\bibinfo {title} {{The clustering of galaxies
  in the completed SDSS-III Baryon Oscillation Spectroscopic Survey: a
  tomographic analysis of structure growth and expansion rate from anisotropic
  galaxy clustering}},}\ }\href@noop {} {\  (\bibinfo {year} {2017})},\ \Eprint
  {http://arxiv.org/abs/1709.05173} {arXiv:1709.05173 [astro-ph.CO]}
  \BibitemShut {NoStop}%
%%CITATION = ARXIV:1709.05173;%%
\bibitem [{\citenamefont {Shi}\ \emph {et~al.}(2017)\citenamefont {Shi} \emph
  {et~al.}}]{Shi:2017qpr}%
  \BibitemOpen
  \bibfield  {author} {\bibinfo {author} {\bibfnamefont {Feng}\ \bibnamefont
  {Shi}} \emph {et~al.},\ }\bibfield  {title} {\enquote {\bibinfo {title}
  {{Mapping the Real Space Distributions of Galaxies in SDSS DR7: II. Measuring
  the growth rate, linear mass variance and biases of galaxies at redshift
  0.1}},}\ }\href@noop {} {\  (\bibinfo {year} {2017})},\ \Eprint
  {http://arxiv.org/abs/1712.04163} {arXiv:1712.04163 [astro-ph.CO]}
  \BibitemShut {NoStop}%
%%CITATION = ARXIV:1712.04163;%%
\bibitem [{\citenamefont {Gil-Marín}\ \emph {et~al.}(2018)\citenamefont
  {Gil-Marín} \emph {et~al.}}]{Gil-Marin:2018cgo}%
  \BibitemOpen
  \bibfield  {author} {\bibinfo {author} {\bibfnamefont {Héctor}\ \bibnamefont
  {Gil-Marín}} \emph {et~al.},\ }\bibfield  {title} {\enquote {\bibinfo
  {title} {{The clustering of the SDSS-IV extended Baryon Oscillation
  Spectroscopic Survey DR14 quasar sample: structure growth rate measurement
  from the anisotropic quasar power spectrum in the redshift range
  0.8$<$z$<$2.2}},}\ }\href@noop {} {\  (\bibinfo {year} {2018})},\ \Eprint
  {http://arxiv.org/abs/1801.02689} {arXiv:1801.02689 [astro-ph.CO]}
  \BibitemShut {NoStop}%
%%CITATION = ARXIV:1801.02689;%%
\bibitem [{\citenamefont {Hou}\ \emph {et~al.}(2018)\citenamefont {Hou} \emph
  {et~al.}}]{Hou:2018yny}%
  \BibitemOpen
  \bibfield  {author} {\bibinfo {author} {\bibfnamefont {Jiamin}\ \bibnamefont
  {Hou}} \emph {et~al.},\ }\bibfield  {title} {\enquote {\bibinfo {title} {{The
  clustering of the SDSS-IV extended Baryon Oscillation Spectroscopic Survey
  DR14 quasar sample: anisotropic clustering analysis in
  configuration-space}},}\ }\href@noop {} {\  (\bibinfo {year} {2018})},\
  \Eprint {http://arxiv.org/abs/1801.02656} {arXiv:1801.02656 [astro-ph.CO]}
  \BibitemShut {NoStop}%
%%CITATION = ARXIV:1801.02656;%%
\bibitem [{\citenamefont {Zhao}\ \emph {et~al.}(2018)\citenamefont {Zhao} \emph
  {et~al.}}]{Zhao:2018jxv}%
  \BibitemOpen
  \bibfield  {author} {\bibinfo {author} {\bibfnamefont {Gong-Bo}\ \bibnamefont
  {Zhao}} \emph {et~al.},\ }\bibfield  {title} {\enquote {\bibinfo {title}
  {{The clustering of the SDSS-IV extended Baryon Oscillation Spectroscopic
  Survey DR14 quasar sample: a tomographic measurement of cosmic structure
  growth and expansion rate based on optimal redshift weights}},}\ }\href@noop
  {} {\  (\bibinfo {year} {2018})},\ \Eprint {http://arxiv.org/abs/1801.03043}
  {arXiv:1801.03043 [astro-ph.CO]} \BibitemShut {NoStop}%
%%CITATION = ARXIV:1801.03043;%%
\end{thebibliography}%

\end{document}